\documentclass[12pt, preprint]{aastex}

\usepackage{lineno}
\usepackage{lscape}
\linenumbers

\usepackage{amsmath}
\usepackage{graphicx}
\usepackage{color}
\usepackage{url}
\usepackage{subfigure}

\usepackage{enumitem}
\setlist{leftmargin=2cm}

\usepackage{todonotes}

\usepackage{color}

\usepackage{xspace}

\newcommand{\pointlike}{\ensuremath{\mathtt{pointlike}}\xspace}
\newcommand{\gtlike}{\ensuremath{\mathtt{gtlike}}\xspace}

\graphicspath{{./figures/}}

\begin{document}

\title{CONSTRAINTS ON THE GALACTIC POPULATION OF TEV PULSAR WIND NEBULAE USING \emph{Fermi} LARGE AREA TELESCOPE OBSERVATIONS}

\keywords{
catalogs;
gamma rays: general; 
methods: data analysis
}

\sloppy

\author{
F.~Acero\altaffilmark{1}, 
M.~Ackermann\altaffilmark{2}, 
M.~Ajello\altaffilmark{3}, 
A.~Allafort\altaffilmark{4}, 
L.~Baldini\altaffilmark{5}, 
J.~Ballet\altaffilmark{6}, 
G.~Barbiellini\altaffilmark{7,8}, 
D.~Bastieri\altaffilmark{9,10}, 
K.~Bechtol\altaffilmark{4}, 
R.~Bellazzini\altaffilmark{11}
% ...
}
\altaffiltext{1}{NASA Goddard Space Flight Center, Greenbelt, MD 20771, USA}
\altaffiltext{2}{Deutsches Elektronen Synchrotron DESY, D-15738 Zeuthen, Germany}
\altaffiltext{3}{Space Sciences Laboratory, 7 Gauss Way, University of California, Berkeley, CA 94720-7450, USA, , USA}
\altaffiltext{4}{W. W. Hansen Experimental Physics Laboratory, Kavli Institute for Particle Astrophysics and Cosmology, Department of Physics and SLAC National Accelerator Laboratory, Stanford University, Stanford, CA 94305, USA}
\altaffiltext{5}{Universit\`a  di Pisa and Istituto Nazionale di Fisica Nucleare, Sezione di Pisa I-56127 Pisa, Italy}
\altaffiltext{6}{Laboratoire AIM, CEA-IRFU/CNRS/Universit\'e Paris Diderot, Service d'Astrophysique, CEA Saclay, 91191 Gif sur Yvette, France}
\altaffiltext{7}{Istituto Nazionale di Fisica Nucleare, Sezione di Trieste, I-34127 Trieste, Italy}
\altaffiltext{8}{Dipartimento di Fisica, Universit\`a di Trieste, I-34127 Trieste, Italy}
\altaffiltext{9}{Istituto Nazionale di Fisica Nucleare, Sezione di Padova, I-35131 Padova, Italy}
\altaffiltext{10}{Dipartimento di Fisica e Astronomia "G. Galilei", Universit\`a di Padova, I-35131 Padova, Italy}
\altaffiltext{11}{Istituto Nazionale di Fisica Nucleare, Sezione di Pisa, I-56127 Pisa, Italy}
% ...

\begin{abstract}
  Pulsar wind nebulae (PWNe) have been established as the most populous class of TeV $\gamma$-ray emitters. Since launch, the \emph{Fermi} Large Area Telescope (LAT) identified five high-energy (100$\,$MeV $<$E$<$ 100$\,$GeV) $\gamma$-ray sources as PWNe, and detected a large number of PWNe candidates, all powered by young and energetic pulsars. The wealth of multi-wavelength data available and the new results provided by \emph{Fermi}-LAT give us an opportunity to find new PWNe and to explore the radiative processes taking place in known ones. The TeV $\gamma$-ray unidentified sources (UNIDs) are the best candidates for finding new PWNe. Using 45 months of \emph{Fermi}-LAT data for energies above 10$\,$GeV, an analysis was performed near the position of 58$\,$TeV PWNe and UNIDs within 5$\degr$ of the Galactic Plane to establish new constraints on PWNe properties and find new clues on the nature of UNIDs. Of the 58 sources, 30 were detected, and this work provides their $\gamma$-ray fluxes for energies above 10$\,$GeV. The spectral energy distributions (SED) and upper limits, in the multi-wavelength context, also provide new information on the source nature and can help distinguish between emission scenarios, i.e. between classification as a pulsar candidate or as a PWN candidate. Six new GeV PWNe candidates are described in detail and compared with existing models. A population study of GeV PWNe candidates as a function of the pulsar/PWN system characteristics is presented.
\end{abstract}

\section{INTRODUCTION}

Since 2003, the extensive observations of the Galactic Plane by Cherenkov telescopes have detected more than 80 Galactic TeV sources \citep{2009ARAA..47..523H}. Pulsar wind nebulae (PWNe) are the dominant class with more than 30 firm identifications. A similar number of Galactic sources cannot be associated with a counterpart at any other wavelength; they form the unidentified (UNID) source class. The third largest class of Galactic sources are the supernova remnants (SNRs). 

The Large Area Telescope (LAT) on board the \emph{Fermi Gamma-ray Space Telescope} provides all-sky coverage of the $\gamma$-ray sky at energies from 20$\,$MeV to more than 300$\,$GeV. With 2 years of observations, the \emph{Fermi}-LAT Second Source Catalog \citep[2FGL,][]{2012ApJS..199...31N} reports the detection of 1873 sources, 1298 being identified and 575 without clear identification. 400 of them lie within 5$\degr$ of the Galactic Plane.

Most of the LAT UNID sources are expected to be pulsars, SNRs, binary systems or PWNe. The $\gamma$-ray emission from these sources is expected to be either hadronic or leptonic. In the leptonic scenario, $\gamma$-ray photons are created by inverse Compton (IC) scattering of highly relativistic leptons from the source on the ambient photon fields such as Cosmic Microwave Background (CMB), stellar radiation or infrared emission from dust \citep[e.g.][]{2009arXiv0906.2644D,2009ASSL..357..451D}. This leptonic approach could well explain several UNIDs \citep[e.g.][]{2011ICRC....6..197T,2012AIPC.1505..349T} and, moreover, one of its biggest advantages is that it provides a natural explanation for the UNIDs that lack a lower energy (radio and X-ray) counterpart \citep[e.g.][]{2009arXiv0906.2644D,dalton1303}, i.e. the so-called ``dark sources". In the hadronic scenario, hadrons accelerated by a source collide with the nuclei in the ambient medium (e.g. molecular cloud) and secondary neutral pions decay to $\gamma$-rays \citep[e.g.][]{2009MNRAS.396.1629G}.

The leptonic PWN scenario requires an energetic and young pulsar to be present. Pulsars are the largest class of Galactic sources detected above 100$\,$MeV with the LAT. These pulsars could make up part of the LAT UNID population. In the LAT energy range, pulsars are point-like sources and exhibit power-law spectra with exponential cutoffs between 0.5 and 6$\,$GeV \citep{2010ApJS..187..460A}, while PWNe have hard power-law spectra without cutoffs in the GeV energy range and might be spatially resolved by the LAT. Middle-aged SNRs, interacting with molecular clouds, detected by the LAT are generally bright and exhibit a break at $\sim$2\,GeV \citep{2011arXiv1104.1197U}. Radio, X-ray and $\gamma$-ray photons probe the non-thermal particle populations and therefore provide information to discriminate between scenarios in which the $\gamma$-ray emission is dominated by leptonic or hadronic processes.

Here we report on the analysis of 58 PWNe and Galactic UNIDs detected at TeV energies, using 45 months of \emph{Fermi}-LAT data above 10$\,$GeV. A complementary search for PWNe in the off-peak phase ranges of pulsars for which the LAT sees $\gamma$-ray pulsations (henceforth ``LAT-detected pulsars"), updating the analysis of \cite{2011ApJ...726...35A}, is presented in the second \emph{Fermi}-LAT 
catalog of $\gamma$-ray pulsars \citep{2PC}, henceforth ``2PC".

The objective of this work is to constrain some general characteristics of PWNe such as their $\gamma$-ray efficiency. This study of TeV sources might also increase the number of PWNe candidates detected at GeV energies by the LAT. \cite{2010ApJ...720..266S}, \cite{2011ApJ...738...42G} and \cite{Rousseau1857} have demonstrated the potential of LAT observations to study PWNe candidates. With the exception of Vela$-$X \citep{2010ApJ...713..146A}, the five PWNe firmly identified by \emph{Fermi} are associated with TeV counterparts. Furthermore, their spectra are consistent with predictions from a leptonic PWN scenario where the IC spectra peak above 100\,GeV \citep{2010ApJ...708.1254A,2011ApJ...738...42G}.

In Section 2, we establish a list of TeV sources potentially associated with PWNe. In Section 3, we describe the methods and tools we used to analyze LAT data. In Section 4, we describe the spectral and spatial analysis of the LAT data, and in Section 5, we perform a population study based on these new $\gamma$-ray results.

\section{TEV $\gamma$-RAY SOURCE SAMPLE}
\label{candidatelist}

We selected our PWNe candidates from the online catalog of TeV $\gamma$-ray sources, TeVCat\footnote{TeVCat is developed by the University of Chicago and can be obtained from: \url{http://tevcat.uchicago.edu}}. As of 2013 January 1, the catalog contained 143 sources observed with the Very High Energy (VHE) experiments: H.E.S.S. \citep{2006AA...457..899A}, VERITAS \citep{2002APh....17..221W}, MAGIC \citep{2012APh....35..435A}, Milagro \citep{2003ApJ...595..803A} and others.

Except for N157~B \citep{2012arXiv1201.0639K}, all PWNe detected by VHE experiments lie inside our Galaxy. A survey of the Galactic Plane was performed by H.E.S.S. and the current version \citep{2012arXiv1204.5860G} covers $\pm4\degr$ in latitude and longitudes between $l=65\degr$ and $l=250\degr$. Another survey, specific to the Cygnus region, was performed by VERITAS \citep{2009arXiv0912.4492W}. Milagro surveyed the Northern hemisphere, covering the Galactic Plane from $l=30\degr$ to $l=220\degr$  \citep{2007ApJ...664L..91A}.  To be conservative compared to the H.E.S.S. survey, we selected the 84 sources that lie within 5$\degr$ of the Galactic Plane. The Galactic center is a complex region to investigate with \emph{Fermi}-LAT, due to the confusion by the large density of sources and by the diffuse emission, so we removed the three VHE sources within 2$\degr$ of the Galactic center from our list. These sources, HESS~J1745$-$303 \citep{2008AA...483..509A}, HESS~J1741$-$302 \citep{2008AIPC.1085..249T} and SNR G0.9$+$0.1 \citep{2005AA...432L..25A}, will be presented separately.

We also excluded from our list the 21 TeV $\gamma$-ray sources associated with radio-detected SNRs. LAT observations of these objects will be presented in the \emph{Fermi}-LAT catalog of SNRs, henceforth ``SNR catalog". Finally, we excluded the Crab Nebula and Vela$-$X, both already studied in detail \citep{2010ApJ...708.1254A,2012ApJ...749...26B,VelaX}. The final list of 58 TeV $\gamma$-ray sources that we selected is presented in Table~\ref{tab:TeV_sources} along with their best-fit morphologies measured by VHE experiments.

\begin{deluxetable}{llrrcccl}
\tabletypesize{\scriptsize}
\tablecaption{List of analyzed VHE sources
\label{tab:TeV_sources}}
\tablehead{\colhead{Name} & \colhead{Class} & \colhead{$l$} & \colhead{$b$} & \colhead{TeV morphology} & \colhead{$\sigma_1$} & \colhead{$\sigma_2$} &  \colhead{Reference}\\\colhead{} & \colhead{} & \colhead{(deg)} & \colhead{(deg)} & \colhead{} & \colhead{(deg)} & \colhead{(deg)}& \colhead{}}
\startdata
VER~J0006$+$727 & PWN & 119.58 & 10.20 & PS & \nodata & \nodata & \cite{2011arXiv1111.2591M}\\
MGRO~J0631$+$105 & PWN & 201.30 & 0.51 & PS & \nodata & \nodata & \cite{2009ApJ...700L.127A}\\
MGRO~J0632$+$17 & PWN & 195.34 & 3.78 & G & 1.30 & \nodata & \cite{2009ApJ...700L.127A} \\
HESS~J1018$-$589 & UNID & 284.23 & $-1.72$ & PS & \nodata & \nodata & \cite{2012AA...541A...5H} \\
HESS~J1023$-$575 & MSC & 284.22 & $-0.40$ & G & 0.18 & \nodata & \cite{2011AA...525A..46H}\\
HESS~J1026$-$582 & PWN & 284.80 & $-0.52$ & G & 0.14 & \nodata & \cite{2011AA...525A..46H} \\
HESS~J1119$-$614 & PWN & 292.10 & $-0.49$ & G & 0.05 & \nodata & Presentation\tablenotemark{a}\\
HESS~J1303$-$631 & PWN & 304.24 & $-0.36$ & G & 0.16 & \nodata & \cite{2005AA...439.1013A}\\
HESS~J1356$-$645 & PWN & 309.81 & $-2.49$ & G & 0.20 & \nodata & \cite{2011AA...533A.103H}\\
HESS~J1418$-$609 & PWN & 313.25 & 0.15 & EG & 0.08 & 0.06 &\cite{2006AA...456..245A}\\
HESS~J1420$-$607 & PWN & 313.56 & 0.27 & G & 0.06 & \nodata & \cite{2006AA...456..245A}\\
HESS~J1427$-$608 & UNID & 314.41 & $-0.14$ & EG & 0.04 & 0.08 & \cite{2008AA...477..353A}\\
HESS~J1458$-$608 & PWN & 317.75 &  $-1.70$ & G & 0.17 & \nodata & \cite{2012arXiv1205.0719D}\\
HESS~J1503$-$582 & UNID & 319.62 & 0.29 & G & 0.26 & \nodata & \cite{2008AIPC.1085..281R}\\
HESS~J1507$-$622 & UNID & 317.95 & $-3.49$ & G & 0.15 & \nodata & \cite{2011AA...525A..45H}\\
HESS~J1514$-$591 & PWN & 320.33 & $-1.19$ & EG & 0.11 & 0.04 & \cite{2005AA...435L..17A}\\
HESS~J1554$-$550 & PWN & 327.16 & $-1.07$ & PS & \nodata & \nodata & \cite{2012arXiv1201.0481A}\\
HESS~J1614$-$518 & MSC & 331.52 & $-0.58$ & EG & 0.23 & 0.15 & \cite{2006ApJ...636..777A}\\
HESS~J1616$-$508 & PWN & 332.39 & $-0.14$ & G & 0.14 & \nodata & \cite{2006ApJ...636..777A}\\
HESS~J1626$-$490 & UNID & 334.77 & 0.05 & EG & 0.07 & 0.10 & \cite{2008AA...477..353A}\\
HESS~J1632$-$478 & PWN & 336.38 & 0.19 & EG & 0.21 & 0.06 &\cite{2006ApJ...636..777A}\\
HESS~J1634$-$472 & UNID & 337.11 & 0.22 & G & 0.11 & \nodata &\cite{2006ApJ...636..777A}\\
HESS~J1640$-$465 & PWN & 338.32 & $-0.02$ & G & 0.04 & \nodata &\cite{2006ApJ...636..777A}\\
HESS~J1646$-$458A & MSC & 339.57 & $-0.02$ & G & 0.35 & \nodata & \cite{2012AA...537A.114A}\\
HESS~J1646$-$458B & MSC & 339.01 & $-0.79$ & G & 0.25 & \nodata & \cite{2012AA...537A.114A}\\
HESS~J1702$-$420 & UNID & 344.30 & $-0.18$ & EG & 0.30 & 0.15 & \cite{2006ApJ...636..777A}\\
HESS~J1708$-$443 & PWN & 343.06 & $-2.38$ & G & 0.29 & \nodata & \cite{2011AA...528A.143H}\\
HESS~J1718$-$385 & PWN & 348.83 & $-0.49$ & EG & 0.15 & 0.07 & \cite{2007AA...472..489A}\\
HESS~J1729$-$345 & UNID & 353.44 & $-0.13$ & G & 0.14 & \nodata & \cite{2011AA...531A..81H}\\
HESS~J1804$-$216 & UNID & 8.40 & $-0.03$ & EG & 0.16 & 0.27 & \cite{2006ApJ...636..777A} \\
HESS~J1809$-$193 & PWN & 11.18 & $-0.09$ & EG & 0.53 & 0.25 &\cite{2007AA...472..489A}\\
HESS~J1813$-$178 & PWN & 12.81 & $-0.03$ & G & 0.04 & \nodata & \cite{2006ApJ...636..777A}\\
HESS~J1818$-$154 & PWN & 15.41 & 0.17 & G & 0.14 & \nodata & \cite{2011arXiv1112.2901H} \\
HESS~J1825$-$137 & PWN & 17.71 & $-0.70$ & EG & 0.13 & 0.12 &\cite{2006AA...460..365A}\\
HESS~J1831$-$098 & PWN & 21.85 & $-0.11$ &G & 0.15 & \nodata & \cite{2011ICRC....7..243S}\\
HESS~J1833$-$105 & PWN & 21.51 & $-0.88$ & PS & \nodata & \nodata & \cite{2008ICRC....2..823D}\\
HESS~J1834$-$087 & UNID & 23.24 & $-0.31$ & G & 0.09 & \nodata & \cite{2006ApJ...636..777A}\\
HESS~J1837$-$069 & UNID & 25.18 & $-0.12$ & EG & 0.12 & 0.05 &\cite{2006ApJ...636..777A}\\
HESS~J1841$-$055 & UNID & 26.80 & $-0.20$ & EG & 0.41 & 0.25 & \cite{2008AA...477..353A}\\
HESS~J1843$-$033 & UNID & 29.30 & 0.51 & PS & \nodata & \nodata & \cite{2008ICRC....2..579H}\\
MGRO~J1844$-$035 & UNID & 28.91 & $-0.02$ & PS & \nodata & \nodata & \cite{2009ApJ...700L.127A}\\
HESS~J1846$-$029 & PWN & 29.70 & $-0.24$ & PS & \nodata & \nodata & \cite{2008ICRC....2..823D}\\
HESS~J1848$-$018 & UNID & 31.00 & $-0.16$ & G & 0.32 & \nodata & \cite{2008AIPC.1085..372C}\\
HESS~J1849$-$000 & PWN & 32.64 & 0.53 & PS & \nodata & \nodata & \cite{2008AIPC.1085..312T}\\
HESS~J1857$+$026 & UNID & 35.96 & $-0.06$ & EG & 0.11 & 0.08 &\cite{2008AA...477..353A}\\
HESS~J1858$+$020 & UNID & 35.58 & $-0.58$ & EG & 0.08 & 0.02 &\cite{2008AA...477..353A}\\
MGRO~J1900$+$039 & UNID & 37.42 & $-0.11$ & PS & \nodata & \nodata & \cite{2009ApJ...700L.127A}\\
MGRO~J1908$+$06 & UNID & 40.39 & $-0.79$ & G & 0.34 & \nodata & \cite{2009AA...499..723A}\\
HESS~J1912$+$101 & PWN & 44.39 & $-0.07$ & G & 0.26 & \nodata & \cite{2008AA...484..435A}\\
VER~J1930$+$188 & PWN & 54.10 & 0.26 & PS & \nodata & \nodata & \cite{2010ApJ...719L..69A} \\
MGRO~J1958$+$2848 & PWN & 65.85 & $-0.23$ & PS & \nodata & \nodata & \cite{2009ApJ...700L.127A}\\
VER~J1959$+$208 & PSR & 59.20 & $-4.70$ & PS & \nodata & \nodata & \cite{2003ApJ...583..853H}\\
VER~J2016$+$372 & UNID & 74.94 & 1.15 & PS & \nodata & \nodata & \cite{2011arXiv1110.4656A}\\
MGRO~J2019$+$37 & PWN & 75.00 & 0.39 & G & 0.55 & \nodata & \cite{2007ApJ...664L..91A}\\
MGRO~J2031$+$41A& UNID & 79.53 & 0.64 & G & 1.50 & \nodata &\cite{2007ApJ...664L..91A}\\
MGRO~J2031$+$41B& UNID & 80.25 & 1.07 & G & 0.10 & \nodata & \cite{2012ApJ...745L..22B}\\
MGRO~J2228$+$61 & PWN & 106.57 & 2.91 & PS & \nodata & \nodata & \cite{2009ApJ...700L.127A}\\
W49A & SFR & 43.27 & $-0.00$ & PS & \nodata & \nodata & \cite{2011arXiv1104.5003B}\\
\enddata

\begin{flushleft}
a- This work was presented at the "Supernova Remnants and Pulsar Wind
Nebulae in the Chandra Era", 2009. See \url{http://cxc.harvard.edu/cdo/snr09/pres/DjannatiAtai\_Arache\_v2.pdf}.
\end{flushleft}
\tablecomments{VHE sources analyzed with LAT observations. The first two columns list the VHE source names and classifications as defined in the TeV catalog (see Section 2): PWN for Pulsar Wind Nebulae, PSR for Pulsars, UNID for Unidentified sources, MSC for Massive Star Clusters and SFR for Star Forming Regions. The third and fourth column give the Galactic longitude and latitude for each source. The fifth column presents the best-fit morphology of the source when observed by VHE experiments: PS, G and EG respectively stand for point-source, Gaussian and elliptical Gaussian. The sixth and seventh columns present the Gaussian and elliptical Gaussian extension. A reference is cited in the eighth column.}
\end{deluxetable}
\clearpage

\section{CONVENTIONS AND METHODS}

The LAT is a $\gamma$-ray telescope that detects photons by conversion into electron-positron pairs. It operates in the energy range between 20$\,$MeV and more than 300$\,$GeV. Details of the instrument and data processing are given in \cite{2009ApJ...697.1071A}. The on-orbit calibration is described in \cite{2009ApJ...696.1084A} and \cite{2012ApJS..203....4A}. This Section will present the data set and the method used to analyze LAT data.

\subsection{Dataset}

This paper uses 45 months of data collected from 2008 August 4 to 2012 April 18 (Mission Elapsed Time: 239557440-356439741 s) for regions centered on the positions of each VHE source. We excluded $\gamma$-rays coming from a zenith angle larger than 100$\degr$. We used the Pass 7 Clean event class that has substantially less instrumental background above 10 GeV with only a marginal loss in effective area, compared to the Pass 7 Source event class \citep{2012ApJS..203....4A}.

We analyzed LAT data only between 10 and 316$\,$GeV to avoid systematics associated with the modeling of adjacent sources with soft spectra. It also reduces systematics associated with imperfect modeling of the Galactic diffuse emission. The maximum energy of 316\,GeV increases the overlap between the energy range covered by the VHE experiments and the LAT. The 100 to 316\,GeV energy range was also shown to be crucial in previous analyses like \cite{Rousseau1857}.

\begin{figure}[h!]
\centering
\includegraphics[width=\textwidth]{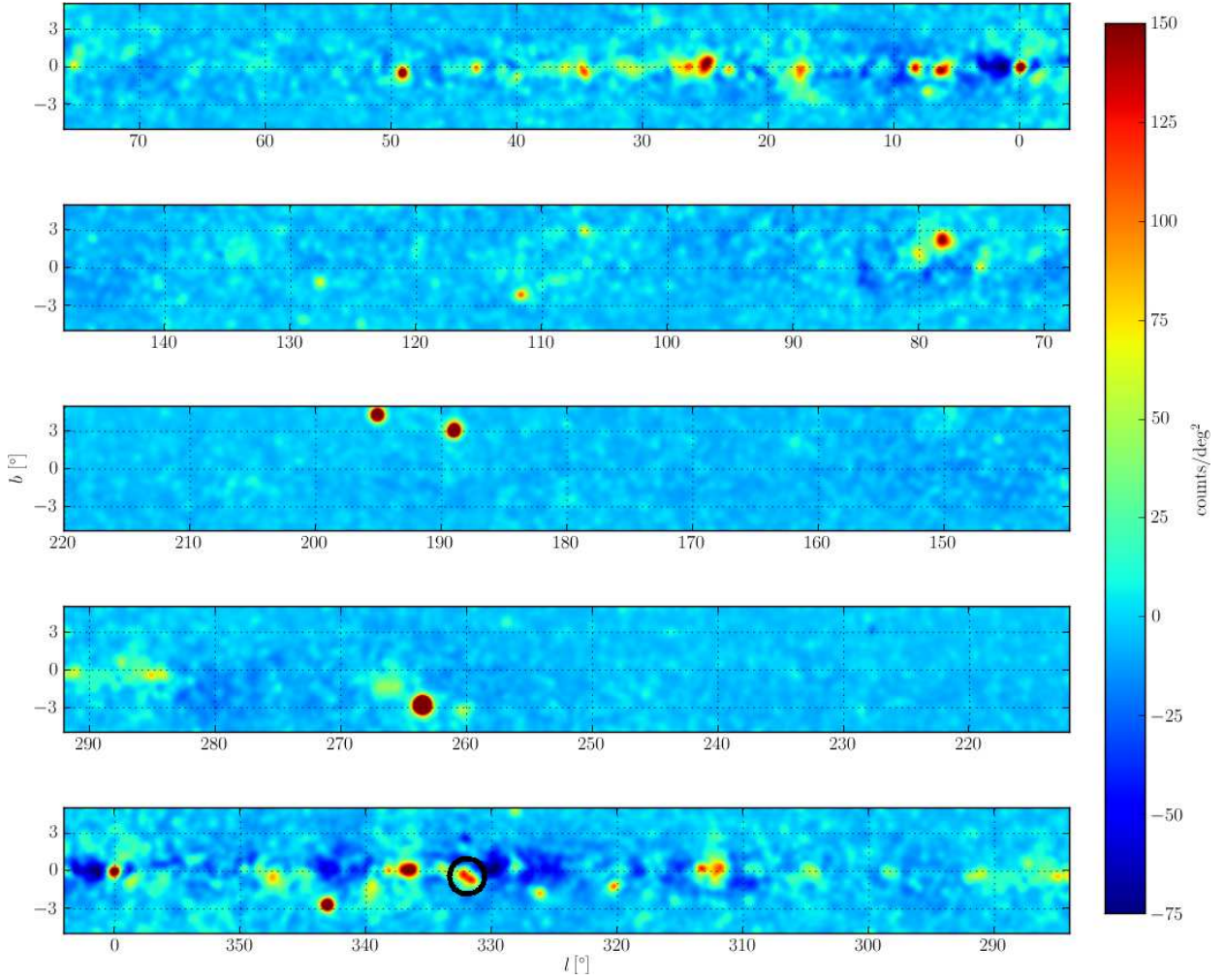}
\caption{Count map of the Galactic plane in the energy range between 10 GeV and 316 GeV with the contribution of the Galactic diffuse, isotropic diffuse, and identified blazars subtracted. All sources associated to blazars are subtracted using the spectral parameters listed in 1FHL. Most of the large positive residuals in the map are Galactic sources, while a few could be unassociated blazars. The count map is smoothed with a Gaussian of $0 \fdg 27$. The black circle shows the positions of HESS~J1614$-$518 and HESS~J1616$-$508.
\label{fig:PlanGal}}
\end{figure}

Figure~\ref{fig:PlanGal} shows a background-subtracted count map of the Galactic Plane observed by the LAT above 10$\,$GeV. The bright Vela ($l,b=263 \fdg 55,-2 \fdg 79$) and Geminga ($l,b$=$195 \fdg 13,4 \fdg 27$) pulsars , and the SNR IC 443 ($l,b=189 \fdg 06 ,3 \fdg 23$) clearly stand out. In addition to these well-known objects, a large number of other sources are apparent. Several are coincident with sources detected by VHE experiments, such as HESS~J1614$-$518 and HESS~J1616$-$508 \citep{2012arXiv1207.0027L}, and will be discussed in Section~\ref{res}. The large number of other sources visible in the map highlights the LAT sensitivity at high energies.

\subsection{Modeling the Regions of Interest}
\label{Model}
Two different software packages for maximum-likelihood fitting were used to analyze LAT data: \gtlike and \pointlike. These tools fit LAT data with a parametrized model of the sky, including models for the instrumental, extragalactic and Galactic components of the background.

The first one, \gtlike, is a maximum-likelihood method distributed in the \emph{Fermi} Science Tools by the FSSC\footnote{\emph{Fermi} Science Support Center: \url{http://fermi.gsfc.nasa.gov/ssc/}}. We used version 09-28-00 of the package in binned mode. \pointlike is an alternate software package that we used to fit the positions of point-like sources and fit the spatial parameters of spatially-extended sources. \cite{2011arXiv1101.6072K} describes the implementation of \pointlike, and \cite{2012arXiv1207.0027L} validates \pointlike's extension-fitting functionality. In \pointlike, the data are binned spatially, using a HEALPix pixellization\footnote{The HEALPix libraries can be obtained from: \url{http://healpix.jpl.nasa.gov/}} \citep{2005ApJ...622..759G}, and spectrally, and the likelihood is maximized over all bins in a region.

We used \pointlike to evaluate a position and extension estimate in the LAT data for each source of our sample. Using those morphologies, we used \gtlike to obtain the best-fit spectral parameters and statistical significances. \gtlike makes fewer approximations in calculating the likelihood for spectra than \pointlike. Both methods agree with each other within 10$\%$ for all derived quantities, but all spectral parameters quoted in the following were obtained using \gtlike.

Since \pointlike and \gtlike use two different shapes for the regions of the sky modeled for each source, we included photons within a radius of 5$\degr$ centered on our source of interest when using \pointlike and within a 7$\degr\,\times\,7\degr$ square region aligned with Galactic coordinates when using \gtlike. We tried to keep the two methods as close as possible by using the same conventions, i.e. same energy binning of 8 energy bins per decade between 10$\,$GeV and 316$\,$GeV, and the same optimizer: MINUIT \citep{JamesRoos1975}.

The Galactic diffuse emission was modeled by the standard LAT interstellar emission model \texttt{ring\_2yearp7v6\_v0.fits}. The residual cosmic-ray background and extragalactic radiation are described by a single isotropic component with a spectral shape described by the file \texttt{isotrop\_2year\_P76\_clean\_v0.txt}. The models have been released and described by the \emph{Fermi}-LAT Collaboration through the FSSC\footnote{Background models are available at: \url{http://fermi.gsfc.nasa.gov/ssc/data/access/lat/BackgroundModels.html}}. In the following, we fit the Galactic diffuse normalization. Since the isotropic diffuse component is not well constrained over the small regions of interest used in this work, we fixed its normalization to 1.

We included in our sky model all cataloged LAT sources within a radius of 10$\degr$ of each source of interest and listed in the hard source list \citep{1FHL}, henceforth ``1FHL". The 1FHL catalog is a forthcoming catalog of sources using three years of LAT data above 10~GeV. The data selection used clean events as done in this work. The spectral parameters of sources closer than 2$\degr$ to the source of interest were fit, while the spectra of all other 1FHL sources were fixed.

\begin{deluxetable}{llllcc}
\tabletypesize{\scriptsize}
\tablecaption{VHE sources with a LAT detected pulsar within $0 \fdg 5$
\label{tab:pulsars}}
\tablehead{\colhead{Name} & \colhead{Pulsar Name} & \colhead{distance} & \colhead{2FGL name} & \colhead{In the model} &  \colhead{Justification}\\ \colhead{} & \colhead{} & \colhead{(deg)} & \colhead{} & \colhead{} & \colhead{}}
\startdata
VER~J0006+727 & PSR~J0007+7303 & 0.26 & 2FGL~J0007.0+7303 & N & a \\
MGRO~J0631+105 & PSR~J0631+1036 & 0.10 & 2FGL~J0631.5+ 1035 & N & a \\
MGRO~J0632+17 & PSR~J0633+1746 & 0.00 & 2FGL~J0633.9+1746 & N & a \\
HESS~J1018$-$589 & PSR~J1016$-$5857 & 0.22 & 2FGL~J1016.5$-$5858 & N & a \\
HESS~J1023$-$575 & PSR~J1023$-$5746 & 0.05 & 2FGL~J1022.7$-$5741 & N & b \\
HESS~J1026$-$582 & PSR~J1028$-$5819 & 0.27 & 2FGL~J1028.5$-$5819 & Y & \nodata \\
HESS~J1119$-$614 & PSR~J1119$-$6127 & 0.07 & 2FGL~J1118.8$-$6128 & N & a \\
HESS~J1356$-$645 & PSR~J1357$-$6429 & 0.12 & 2FGL~J1356.0$-$6436 & N & a \\
HESS~J1418$-$609 & PSR~J1418$-$6058 & 0.05 & 2FGL~J1418.7$-$6058 & N & b \\
HESS~J1420$-$607 & PSR~J1420$-$6048 & 0.05 & 2FGL~J1420.1$-$6047 & N & b \\
HESS~J1458$-$608 & PSR~J1459$-$6053 & 0.17 & 2FGL~J1459.4$-$6054 & N & a \\
HESS~J1514$-$591 & PSR~J1513$-$5908 & 0.03 & \nodata & N & b\\
HESS~J1646$-$458B & PSR~J1648$-$4611 & 0.42 & 2FGL~J1646$-$4611 & N & c \\
HESS~J1702$-$420 & PSR~J1702$-$4128 & 0.53 & \nodata & N & c\\
HESS~J1708$-$443 & PSR~J1709$-$4429 & 0.25 & 2FGL~J1709.7$-$4429 & N & b \\
HESS~J1718$-$385 & PSR~J1718$-$3825 & 0.13 & 2FGL~J1718.3$-$3827 & N & a \\
HESS~J1804$-$216 & PSR~J1803$-$2149 & 0.27 & 2FGL~J1803.3$-$2148 & N & b \\
HESS~J1833$-$105 & PSR~J1833$-$1034 & 0.01 & 2FGL~J1833.6$-$1032 & N & a \\
HESS~J1841$-$055 & PSR~J1838$-$0537 & 0.48 & 2FGL~J1839.0$-$0539 & Y & \nodata \\
MGRO~J1908+06 & PSR~J1907+0602 & 0.23 & 2FGL~J1907.9+0602 & N & a \\
MGRO~J1958+2848 & PSR~J1958+2846 & 0.12 & 2FGL~J1958.6+2845 & N & a \\
VER~J1959+208 & PSR~J1959+2048 & 0.02 & 2FGL~J1959.5+2047 & N & a \\
MGRO~J2019+37 & PSR~J2021+3651 & 0.36 & 2FGL~J2021.0+3651 & N & b \\
MGRO~J2031+41B & PSR~J2032+4127 & 0.05 & 2FGL~J2032.2+4126 & N & b \\  
MGRO~J2228+61 & PSR~J2229+6114 & 0.09 & 2FGL~J2229+6114 & N & a \\
\enddata\\
\begin{flushleft}
a- The distance between the pulsar and the source is closer than $0 \fdg 27$.\\
b- The pulsar is located inside the edge of the shape observed by VHE experiments.\\
c- Not a and not b, but no significant excess above 10 GeV at the position of the pulsar.
\end{flushleft}
\tablecomments{$_{ }$Sources with a LAT-detected $\gamma$-ray pulsar within $0 \fdg 5$. The first two columns list the names of the VHE sources and their associated pulsars. The third column is the angular distance between the center of the VHE source and the LAT pulsar. The pulsar position comes from 2PC. The fourth column gives the pulsar 2FGL name. The fifth column says if the emission from the pulsar was included in the model of the background emission. ``Y" means the emission was included and ``N" means that it was not. The sixth column gives the justification when the pulsar is not included in the model.}
\end{deluxetable}

$\gamma$-ray pulsars are found near many of the VHE sources. Table~\ref{tab:pulsars} summarizes the sources analyzed in this paper that are near a LAT-detected pulsar. In the LAT energy range, pulsars tend to be brighter than their associated PWNe \citep{2011ApJ...726...35A}. During the fit, the observed photon sample is shared by all modeled sources. The high energy range used in this work prevents a reasonable fit of a pulsar component modeled by a power law with an exponential cut-off spectrum. If a pulsar's model overestimates its emission above 10 GeV, a putative PWN emission would be underestimated. This effect could obscure the detection of a faint PWN. On the contrary, not including an existing source, such as a pulsar, would artificially increase the flux attributed to a putative PWN at low energy. Therefore, we decided to keep as separate sources in the model all pulsars located outside the VHE experiment template and more than $0 \fdg 27$ away from the source of interest. The $0 \fdg 27$ radius corresponds to the 68\% containment radius of the Point Spread Function (PSF) averaged over energies above 10$\,$GeV \citep{2012ApJS..203....4A}. For VHE sources less than $0 \fdg 27$ from a LAT-detected pulsar, we analyzed the sources twice. Removing the pulsar from the background model amounts to neglecting it, while leaving it in the background model amounts to subtracting the underlying pulsar contribution from each putative PWN. In all cases the parameters of the pulsar models have been fixed to those obtained in the 2FGL catalog \citep{2012ApJS..199...31N}. If contamination from the pulsar is a concern, it is possible to phase-fold photons and analyze the data in the off-peak phase intervals of the pulsar. This will be performed in 2PC. Here we analyze data at all phases to have the largest possible statistics and therefore better sensitivity to faint PWNe.

Due to the 45-months integration time of our analysis compared to the 36 months of the forthcoming 1FHL catalog, we expected to find new statistically-significant background sources. To prevent bias from these sources, we included in our model any nearby background sources with a significance above 4$\,\sigma$ (TS$\,>\,25$ with four degrees of freedom, d.o.f.). We fit their spectra with power laws. The locations and spectra of the six such sources found in our analysis are described in Table~\ref{tab:newsources}. 

\tabletypesize{\scriptsize}
\begin{deluxetable}{llrlll}
\tablewidth{0pt}
\tablecaption{Additional background sources 
\label{tab:newsources}}
\tablehead{\colhead{Name} & \colhead{$l$} & \colhead{$b$} & \colhead{$\text{TS}$} & \colhead{Prefactor} & \colhead{Spectral index}\\\colhead{} & \colhead{(deg)} & \colhead{(deg)} & \colhead{} & \colhead{$\left(\text{cm}^{-2}\,\text{s}^{-1}\,\text{MeV}^{-1}\right)$} & \colhead{}}
\startdata
2FGL~J1405.5$-$6121 & 311.81 & 0.30 & 31 & (1.2 $\pm$ 0.4) $\times 10^{-15}$ & 1.8 $\pm$ 0.3 \\
Background Source 1 & 333.59 & $-0.31$ & 29 & (6.5 $\pm$ 2.5) $\times 10^{-17}$ & 4.3 $\pm$ 0.9\\
Background Source 2 & 336.96 & $-0.07$ & 25 & (1.2 $\pm$ 0.4) $\times 10^{-15}$ & 1.9 $\pm$ 0.4\\
2FGL~J1823.1$-$1338c & 17.51 & $-0.12$ & 30 & (4.9 $\pm$ 1.9) $\times 10^{-15}$  & 2.9 $\pm$ 0.7\\
2FGL~J1836.8$-$0623c & 25.41 & 0.42 & 25 & ( 9.4 $\pm$ 3.9 ) $\times 10^{-16}$ & 2.0 $\pm$ 0.4 \\
PSR~J1838$-$0536 &  26.28 & 0.62 & 16 & (5.0 $\pm$ 1.8) $\times 10^{-17}$ & 4.1 $\pm$ 1.0 \\
\enddata
\tablecomments{New background sources found in our analysis for energies above 10 GeV that are not included in 1FHL. The first three columns are their name, Galactic longitude, and Galactic latitude. The TS values for the sources are provided in the fourth column. The spectral results are presented in columns 5 and 6 assuming a power-law spectral model (Equation~\ref{PL}) with a scale parameter $E_0 = 56.2$ GeV (corresponding to the mid-value of the energy range in logarithmic scale). PSR~J1838$-$0536 improves the morphology fit of the diffuse source HESS~J1841$-$055. The spectral fit is consistent with the pulsar component (spectral index of $\sim$4). Additional sources appearing to be spatially consistent with a 2FGL source are labeled with the name of the associated 2FGL source.}
\end{deluxetable}
\normalsize
\noindent

\subsection{Analysis Procedure}

We developed a uniform procedure for analyzing any potential emission in the 58 regions listed in Table~\ref{tab:TeV_sources}. The small statistics above 10 GeV and the narrow energy range (1.5 decade) prevent any spectral curvature to be significant as will be discussed in the 1FHL catalog. Therefore, in the following we derived the spectra assuming a power-law spectral model:
\begin{equation}
\label{PL}
\frac{dF}{dE}=N_0 \left( \frac{E}{E_0} \right)^{-\Gamma}
\end{equation}
where $N_0$ is the normalization,  $\Gamma$ is the spectral index, and $E_0$ is the scale parameter. To minimize the covariance between $N_0$ and $\Gamma$, we ran the analysis twice. In the first iteration, we fitted the source assuming a power-law model depending on the integral flux $F$ and $\Gamma$,

\begin{equation}
\label{PLFlux}
\frac{dF}{dE}=\frac{F (-\Gamma +1) E^{-\Gamma}}{E_{\text{max}}^{-\Gamma+1}-E_{\text{min}}^{-\Gamma+1}}.
\end{equation}

Using the covariance matrix of the fit parameters, we derived the pivot energy $E_p$, computed as the energy at which the relative uncertainty on the normalization $N_0$ was minimal \citep{2012ApJS..199...31N}. In a second iteration, we refitted the spectrum of the source assuming a power-law spectral model (Eq. \ref{PL}) with the scale parameter $E_0$ fixed to $E_p$. For sources not significantly detected, we computed a 99\% confidence level (c.l.) Bayesian upper limit on the flux of the source assuming the published VHE morphology and a power-law photon spectral index of 2.

We performed the spatial analysis in two steps. As a first step, we assumed that the LAT emission originates from the same population of emitting particles cooling by the same radiation process as the VHE emission. These assumptions mean that there should be a correlation between the spatial morphology of a source at LAT and VHE energies. While it is possible for the $\gamma$-ray emission from PWNe to have both the synchrotron and IC components visible in the LAT energy range \citep[as in the case of the Crab nebula,][]{2010ApJ...708.1254A}, this scenario is unlikely for observations above 10\,GeV since electrons of energy around 1\,PeV should radiate in a magnetic field $\sim 0.3$\,G to emit photons above 10\,GeV. This value is unrealistic for a PWN. A second exception would appear if the LAT and VHE emissions originate from two different populations of electrons \citep[as in the case of Vela$-$X,][]{2010ApJ...713..146A}. However, most PWNe observed by the LAT show only IC emission from the same population responsible for the VHE emission \citep[e.g.,][]{2010ApJ...714..927A,2011ApJ...738...42G,Rousseau1857} supporting our assumption in this first step.

We assumed that any LAT emission would have the same morphology as the best-fit VHE morphology. Therefore, we modeled the spatial distribution of the emission as Gaussian with a position and extension fixed at the value of the spatial best-fit performed in the VHE energy range. To homogenize the analysis, if the source was modeled with an elliptical Gaussian at VHE, the Gaussian was fixed to the averaged extension. We tested for the significance of the source at LAT energies using a likelihood-ratio test: $\text{TS}=2\times\log (\mathcal{L}_1/\mathcal{L}_0)$ where $\mathcal{L}$ is the Poisson likelihood of obtaining the observed data given the assumed model, $\mathcal{L}_1$ corresponds to the likelihood obtained by fitting a model of the source of interest and the background model, and $\mathcal{L}_0$ corresponds to the likelihood obtained by fitting the background model only. In the following, we refer to the TS assuming the VHE shape as $\text{TS}_{\text{TeV}}$. By assuming a fixed spatial model, our test has fewer d.o.f. which makes our test more sensitive to the LAT emission, assuming that the VHE spatial model reproduces the LAT observation. 

The formal statistical significance of this test can be obtained from Wilks' theorem \citep{STMAZ.03029575}. In the null hypothesis, TS follows a $\chi^2$ distribution with $n$ d.o.f. where $n$ is the number of additional parameters in the model. We consider a source to be significantly detected when $\text{TS}_{\text{TeV}} \geq 16$. Our test has only two d.o.f. (the flux and the spectral index) so our threshold corresponds to a formal significance of $3.6\,\sigma$. For significantly detected sources, $\text{TS}_{\text{TeV}}$ is presented in Table~\ref{tab:Spat_results}.

\thispagestyle{empty}
\tabletypesize{\scriptsize}
\begin{deluxetable}{lllllrrll}
\tablewidth{0pt}
\tablecaption{Spatial results for detected sources
\label{tab:Spat_results}}
\tablehead{ \colhead{Name} & \colhead{ID} & \colhead{$\text{TS}_{\text{TeV}}$} & \colhead{$\text{TS}_{\text{GeV}}$} & \colhead{$\text{TS}_{\text{ext}}$} & \colhead{$l$} & \colhead{$b$} & \colhead{$\text{Unc}_{\text{GeV}}$} & \colhead{$\sigma$}\\ 
 \colhead{} & \colhead{} & \colhead{} & \colhead{} & \colhead{} & \colhead{(deg)} & \colhead{(deg)} & \colhead{(deg)}}
\startdata
VER~J0006+727 & PSR & 655 & 1206 & 0 & $119.68$ & $10.47$ & $0.01,0.01$ & $< 0.07$ \\ 
\tablenotemark{a} & & 2 & \nodata & \nodata & \nodata & \nodata & \nodata & \nodata\\
MGRO~J0632+17 & PSR  & 699 & 2056 & 1 & $195.13$ & $4.28$ & $0.01,0.01$ & $< 0.08$ \\ 
\tablenotemark{a} & & 5 & \nodata & \nodata & \nodata & \nodata & \nodata & \nodata \\
HESS~J1018$-$589 & O  & 29 & 29 & 0 & $284.33$ & $-1.66$ & $0.04,0.02$ & $< 0.87$ \\ 
\tablenotemark{a} &   & 25 & 25 & 2 & $284.34$ & $-1.65$ & $0.04,0.02$ & $< 0.87$ \\
HESS~J1023$-$575 & PWNc  & 52 & 58 & 8 & $284.13$ & $-0.45$ & $0.03,0.02$ & $< 0.77$ \\ 
\tablenotemark{a} &   & 52 & 58 & 8 & $284.13$ & $-0.45$ & $0.03,0.02$ & $< 0.77$ \\
HESS~J1119$-$614 & PWNc  & 27 & 27 & 9 & $292.16$ & $-0.56$ & $ 0.05,0.02 $ & $< 0.31$ \\
\tablenotemark{a} &  & 16 & 16 & 9 & $292.18$ & $-0.57$ & $0.05,0.02$ & $< 0.32$ \\ 
HESS~J1303$-$631 & PWNc & 37 & 58 & 29 & $304.56$ & $-0.11$ & $0.04,0.03$ & $0.45 \pm 0.09 \pm 0.10$\\
HESS~J1356$-$645 & PWN  & 24 & 26 & 3 & $309.71$ & $-2.33$ & $0.05,0.01$ & $< 0.39$ \\ 
\tablenotemark{a} &   & 24 & 26 & 3 & $309.71$ & $-2.32$ & $0.05,0.01$ & $< 0.39$ \\
HESS~J1418$-$609 & PSR  & 31 & 32 & 0 & $313.28$ & $0.13$ & $0.03,0.01$ & $< 0.32$ \\ 
\tablenotemark{a} & & 15 & \nodata & \nodata & \nodata & \nodata & \nodata & \nodata\\
HESS~J1420$-$607 & PWNc  & 42 & 42 & 0 & $313.55$ & $0.27$ & $0.04,0.02$ & $< 0.38$ \\
\tablenotemark{a} & & 36 & 36 & 0 & $313.55$ & $0.28$ & $0.04,0.02$ & $< 0.39$ \\ 
HESS~J1507$-$622 & O  & 21 & 23 & 7 & $317.77$ & $-3.60$ & $0.05,0.03$ & $< 1.04$ \\ 
\tablenotemark{a} & & 21 & 23 & 7 & $317.76$ & $-3.61$ & $0.05,0.03$ & $< 1.04$ \\
HESS~J1514$-$591 & PWN  & 156 & 147 & 10 & $320.35$ & $-1.25$ & $0.03,0.01$ & $< 0.16$ \\ 
\tablenotemark{a} & & 156 & 147 & 10 & $320.35$ & $-1.25$ & $0.03,0.01$ & $< 0.16$ \\
HESS~J1614$-$518 & O  & 110 & 133 & 47 & $331.62$ & $-0.74$ & $0.04,0.03$ & $0.28 \pm 0.03 \pm 0.05$\\
HESS~J1616$-$508 & PWNc  & 75 & 94 & 31 & $332.39$ & $-0.27$ & $0.04,0.02$ & $0.25 \pm 0.03 \pm 0.05$\\
HESS~J1632$-$478 & PWNc  & 137 & 161 & 56 & $336.50$ & $0.10$ & $0.03,0.02$ & $0.30 \pm 0.06 \pm 0.06$\\
HESS~J1634$-$472 & O  & 33 & 34 & 1 & $337.23$ & $0.35$ & $0.03,0.01$ &$< 1.21$ \\ 
HESS~J1640$-$465 & PWNc  & 47 & 42 & 9 & $338.33$ & $0.05$ & $0.05,0.01$ & $< 1.17$ \\ 
HESS~J1708$-$443 & PSR  & 722 & 1153 & 0 & $343.11$ & $-2.70$ & $0.01,0.01$ & $< 0.05$ \\
\tablenotemark{a} & & 33 & 64 & 0 & $343.12$ & $-2.70$ & $0.01,0.01$ & $< 0.09$ \\ 
HESS~J1804$-$216 & O  & 138 & 141 & 37 & $8.40$ & $-0.09$ & $0.04,0.01$ & $0.25 \pm 0.03 \pm 0.04$\\
\tablenotemark{a} & & 124 & 128 & 30 & $8.42$ & $-0.10$ & $0.04,0.01$ & $0.24 \pm 0.03 \pm 0.04$\\
HESS~J1825$-$137 & PWN  & 56 & 82 & 30 & $17.55$ & $-0.47$ & $0.05,0.03$ & $0.44 \pm 0.08 \pm 0.09$\\
HESS~J1834$-$087 & O  & 27 & 36 & 4 & $23.20$ & $-0.26$ & $0.05,0.01$ & $< 0.22$ \\ 
HESS~J1837$-$069 & PWNc  & 73 & 119 & 46 & $25.17$ & $0.00$  & $0.05,0.03$ & $0.36 \pm 0.06 \pm 0.04$\\
HESS~J1841$-$055 & PWNc  & 64 & 70 & 32 & $27.01$ & $-0.15$ & $0.05,0.03$ & $0.38 \pm 0.06 \pm 0.06$\\
HESS~J1848$-$018 & PWNc  & 19 & 19 & 0 & $30.90$ & $-0.20$ & $0.04,0.01$ & $< 1.50$ \\ 
HESS~J1857+026 & PWNc  & 53 & 55 & 8 & $36.08$ & $0.02$ & $0.04,0.01$ & $< 0.28$ \\ 
MGRO~J1908+06 & PSR  & 16 & 37 & 2 & $40.11$ & $-0.84$ & $0.03,0.01$ &$< 0.19$ \\ 
\tablenotemark{a} & & 9 & \nodata & \nodata & \nodata & \nodata & \nodata & \nodata \\
MGRO~J1958+2848 & PSR  & 21 & 24 & 0 & $65.88$ & $-0.34$ & $0.04,0.01$ & $< 0.56$ \\ 
\tablenotemark{a} & & 8 & \nodata & \nodata & \nodata & \nodata & \nodata & \nodata\\
VER~J2016+372 & O  & 31 & 33 & 1 & $74.86$ & $1.22$ & $0.05,0.02$ & $< 1.16$ \\ 
MGRO~J2019+37 & PSR & 31 & 100 & 1 & $75.23$ & $0.13$ & $0.02,0.01$ & $< 0.07$ \\ 
\tablenotemark{a} & & 5 & \nodata & \nodata & \nodata & \nodata & \nodata & \nodata\\
MGRO~J2031+41B & PSR  & 58 & 66 & 5 & $80.20$ & $1.03$ & $0.05,0.01$ & $< 2.47$ \\ 
\tablenotemark{a} & & 12 & \nodata & \nodata & \nodata & \nodata & \nodata & \nodata\\
MGRO~J2228+61 & PSR & 94 & 113 & 0 & $106.65$ & $2.94$ & $0.02,0.01$ & $< 0.10$ \\ 
\tablenotemark{a} & & 15 & \nodata & \nodata & \nodata & \nodata & \nodata & \nodata\\
\enddata

\tablecomments{Results of the maximum likelihood spatial fits for LAT-detected VHE sources. ``a" in the first column corresponds to the results with contribution of the pulsar associated in Table~\ref{tab:pulsars} subtracted from the emission of the source just above. Column 2 lists the classification for the LAT emission: either ``PWN" for clearly identified PWNe, ``PWNc" for PWNe candidates, ``PSR" for pulsar emission, and ``O" for anything else. Column 3 is the TS when the source is modeled with the spatial model obtained from VHE data. Column 4 is the TS when the source is modeled assuming it is point-like, and column 5 is the TS of the source assuming it is spatially-extended with a Gaussian spatial model. Columns 6 and 7 give the position of the source fit in our LAT analysis. Column 8 gives the statistical (68\% confidence radius) and systematic uncertainties on the position. The methods for determining systematic uncertainties on the spatial parameters are described in Section \ref{systext}. Column 9 gives the extension fit in the LAT energy range if $\text{TS}_{\text{ext}} > 16$ or a 99 \% c.l. upper limit on the extension otherwise.}
\end{deluxetable}
\normalsize
\noindent
\clearpage

As a second step, for significantly-detected sources, we then independently characterized the best-fit morphology obtained from the LAT emission. Following the method of \cite{2012ApJS..199...31N}, we assumed the source to be point-like and fit its position. Following the method adopted in \cite{2012arXiv1207.0027L}, we then assumed the source to be spatially-extended with a Gaussian spatial model and fit its position and extension. From this, we obtained $\text{TS}_{\text{point}}$ and $\text{TS}_{\text{Gaussian}}$. We then defined the extension significance $\text{TS}_{\text{ext}} = \text{TS}_{\text{Gaussian}} - \text{TS}_{\text{point}}$ following the method of \cite{2012arXiv1207.0027L} and set the threshold for claiming the source to be spatially extended as $\text{TS}_{\text{ext}}>16$, corresponding to a significance of $4\,\sigma$. $\text{TS}_{\text{GeV}}=\text{TS}_{\text{Gaussian}}$ when $\text{TS}_{\text{ext}}>16$ and $\text{TS}_{\text{GeV}}=\text{TS}_{\text{point}}$ otherwise. Table~\ref{tab:Spat_results} lists eight significantly extended sources and 22 point sources. If the source was not significantly extended, we derived a 99\% c.l. Bayesian upper limit on the extension.

\thispagestyle{empty}
\clearpage
\thispagestyle{empty}
\begin{landscape}
\tabletypesize{\scriptsize}
\begin{deluxetable}{l*{11}l}
\tablewidth{0pt}
\tablecaption{Spectral results for detected sources 
\label{tab:Det_results}}
\tablehead{ \colhead{Name} & \colhead{TS} & \colhead{F$_{10\,\text{GeV}}^{316\,\text{GeV}}$} & \colhead{$\Gamma$} & \colhead{TS$_{10\,\text{GeV}}^{31\,\text{GeV}}$} & \colhead{F$_{10\,\text{GeV}}^{31\,\text{GeV}}$} & \colhead{TS$_{31\,\text{GeV}}^{100\,\text{GeV}}$} & \colhead{F$_{31\,\text{GeV}}^{100\,\text{GeV}}$} & \colhead{TS$_{100\,\text{GeV}}^{316\,\text{GeV}}$} & \colhead{F$_{100\,\text{GeV}}^{316\,\text{GeV}}$}\\
\colhead{} & \colhead{} & \colhead{$\left(10^{-10}\,\text{cm}^{-2}\,\text{s}^{-1}\right)$} & \colhead{} & \colhead{} & \colhead{$\left(10^{-10}\,\text{cm}^{-2}\,\text{s}^{-1}\right)$} & \colhead{} & \colhead{$\left(10^{-10}\,\text{cm}^{-2}\,\text{s}^{-1}\right)$} & \colhead{} & \colhead{$\left(10^{-10}\,\text{cm}^{-2}\,\text{s}^{-1}\right)$}}
\startdata
VER~J0006$+$727 & 655 & $11.3 \pm 0.9 \pm 1.2$ & $3.96 \pm 0.25 \pm 0.36$& 647 & $11.2 \pm 0.9 \pm 2.0$ & 11 & $0.4 \pm 0.2 \pm 0.2$ & 0 & $<\,0.3$\\
\tablenotemark{a} & 2 & $< 1.2$ & \nodata & 3 & $<\,1.2$& 0 & $<\,0.4$& 0 & $<\,0.3$\\
MGRO~J0632$+$17 & 699 & $34.9 \pm 2.1 \pm 10.2$ & $4.53 \pm 0.25 \pm 0.51$& 695 & $36.8 \pm 2.0 \pm 10.1$ & 6 & $<\,1.6$& 1 & $<\,0.7$\\
\tablenotemark{a} & 5 & $< 5.1$ & \nodata & 9 & $<\,5.5$& 1 & $<\,1.1$& 0 & $<\,0.6$\\
HESS~J1018$-$589 & 29 & $1.7 \pm 0.5 \pm 0.7$ & $2.41 \pm 0.49 \pm 0.49$& 25 & $1.4 \pm 0.5 \pm 0.6$ & 0 & $<\,0.6$& 6 & $<\,0.6$\\
\tablenotemark{a} & 25 & $1.5 \pm 0.5 \pm 0.7$ & $2.31 \pm 0.50 \pm 0.49$& 20 & $1.3 \pm 0.4 \pm 0.6$ & 0 & $<\,0.6$& 6 & $<\,0.6$\\
HESS~J1023$-$575 & 52 & $4.6 \pm 0.9 \pm 1.2$ & $1.99 \pm 0.24 \pm 0.32$& 40 & $3.8 \pm 0.8 \pm 1.8$ & 2 & $<\,0.9$& 9 & $<\,1.2$\\
\tablenotemark{a} & 52 & $4.6 \pm 0.9 \pm 1.2$ & $1.99 \pm 0.24 \pm 0.32$& 40 & $3.8 \pm 0.8 \pm 1.8$ & 2 & $<\,0.9$& 9 & $<\,1.2$\\
HESS~J1119$-$614 & 27 & $2.1 \pm 0.6 \pm 0.8$ & $2.15 \pm 0.37 \pm 0.36$& 17 & $1.5 \pm 0.5 \pm 0.5$ & 1 & $<\,0.7$& 11 & $0.2 \pm 0.1 \pm 0.1$ \\
\tablenotemark{a} & 16 & $2.0 \pm 0.6 \pm 0.8$ & $1.83 \pm 0.41 \pm 0.36$& 5 & $<\,2.1$& 1 & $<\,0.8$& 11 & $0.2 \pm 0.1 \pm 0.1$ \\
HESS~J1303$-$631 & 37 & $3.6 \pm 0.9 \pm 2.1$ & $1.53 \pm 0.23 \pm 0.37$& 10 & $1.6 \pm 0.7 \pm 1.5$ & 25 & $1.6 \pm 0.5 \pm 0.7$ & 3 & $<\,0.7$\\
HESS~J1356$-$645 & 24 & $1.1 \pm 0.4 \pm 0.5$ & $0.95 \pm 0.40 \pm 0.40$& 0 & $<\,0.9$& 14 & $0.6 \pm 0.3 \pm 0.3$ & 10 & $0.3 \pm 0.2 \pm 0.2$ \\
\tablenotemark{a} & 24 & $1.1 \pm 0.4 \pm 0.5$ & $0.94 \pm 0.40 \pm 0.40$& 0 & $<\,0.9$& 14 & $0.6 \pm 0.3 \pm 0.3$ & 10 & $0.3 \pm 0.2 \pm 0.2$ \\
HESS~J1418$-$609 & 31 & $4.0 \pm 1.0 \pm 1.3$ & $3.52 \pm 0.81 \pm 0.61$& 29 & $3.6 \pm 0.9 \pm 1.2$ & 2 & $<\,1.0$& 0 & $<\,0.6$\\
\tablenotemark{a} & 15 & $< 4.3$ & \nodata & 13 & $2.6 \pm 0.9 \pm 1.2$ & 2 & $<\,1.0$& 0 & $<\,0.6$\\
HESS~J1420$-$607 & 42 & $3.7 \pm 0.9 \pm 1.1$ & $1.89 \pm 0.28 \pm 0.31$& 19 & $2.4 \pm 0.7 \pm 0.7$ & 13 & $0.8 \pm 0.3 \pm 0.3$ & 12 & $0.5 \pm 0.2 \pm 0.2$ \\
\tablenotemark{a} & 36 & $3.4 \pm 0.9 \pm 1.1$ & $1.81 \pm 0.29 \pm 0.31$& 15 & $2.2 \pm 0.7 \pm 0.7$ & 13 & $0.8 \pm 0.3 \pm 0.3$ & 12 & $0.5 \pm 0.2 \pm 0.2$ \\
HESS~J1507$-$622 & 21 & $1.5 \pm 0.5 \pm 0.5$ & $2.33 \pm 0.48 \pm 0.48$& 18 & $1.2 \pm 0.4 \pm 0.4$ & 3 & $<\,0.7$& 0 & $<\,0.4$\\
HESS~J1514$-$591 & 156 & $6.2 \pm 0.9 \pm 1.3$ & $1.72 \pm 0.16 \pm 0.17$& 69 & $3.9 \pm 0.7 \pm 1.1$ & 54 & $1.7 \pm 0.4 \pm 0.4$ & 36 & $0.7 \pm 0.3 \pm 0.4$ \\
HESS~J1614$-$518 & 110 & $9.9 \pm 1.4 \pm 3.1$ & $1.75 \pm 0.15 \pm 0.18$& 47 & $6.1 \pm 1.1 \pm 2.7$ & 37 & $2.6 \pm 0.6 \pm 0.8$ & 31 & $1.1 \pm 0.4 \pm 0.3$ \\
HESS~J1616$-$508 & 75 & $9.3 \pm 1.4 \pm 2.3$ & $2.18 \pm 0.19 \pm 0.20$& 46 & $6.5 \pm 1.2 \pm 2.1$ & 29 & $2.3 \pm 0.6 \pm 0.6$ & 3 & $<\,1.0$\\
HESS~J1632$-$478 & 137 & $11.8 \pm 1.5 \pm 5.3$ & $1.82 \pm 0.14 \pm 0.19$& 69 & $7.8 \pm 1.2 \pm 4.2$ & 37 & $2.6 \pm 0.6 \pm 0.9$ & 39 & $1.5 \pm 0.4 \pm 0.5$ \\
HESS~J1634$-$472 & 33 & $5.6 \pm 1.3 \pm 2.5$ & $1.96 \pm 0.25 \pm 0.29$& 20 & $3.6 \pm 1.0 \pm 2.1$ & 12 & $1.3 \pm 0.5 \pm 0.5$ & 2 & $<\,0.8$\\
HESS~J1640$-$465 & 47 & $5.0 \pm 1.0 \pm 1.7$ & $1.95 \pm 0.23 \pm 0.20$& 24 & $3.4 \pm 0.9 \pm 1.3$ & 28 & $1.7 \pm 0.5 \pm 0.5$ & 0 & $<\,0.5$\\
HESS~J1708$-$443 & 722 & $24.5 \pm 1.7 \pm 3.5$ & $3.80 \pm 0.24 \pm 0.33$& 714 & $23.9 \pm 1.6 \pm 3.0$ & 14 & $1.1 \pm 0.4 \pm 0.5$ & 6 & $<\,1.0$\\
\tablenotemark{a} & 33 & $5.5 \pm 1.3 \pm 3.5$ & $2.13 \pm 0.31 \pm 0.33$& 17 & $4.0 \pm 1.2 \pm 3.0$ & 11 & $1.0 \pm 0.4 \pm 0.5$ & 6 & $<\,1.0$\\
HESS~J1804$-$216 & 138 & $14.2 \pm 1.6 \pm 3.1$ & $2.10 \pm 0.16 \pm 0.24$& 91 & $10.1 \pm 1.4 \pm 2.3$ & 38 & $2.9 \pm 0.7 \pm 0.6$ & 21 & $1.2 \pm 0.4 \pm 0.4$ \\
\tablenotemark{a} & 124 & $13.4 \pm 1.6 \pm 3.1$ & $2.04 \pm 0.16 \pm 0.24$& 77 & $9.3 \pm 1.4 \pm 2.3$ & 36 & $2.8 \pm 0.7 \pm 0.6$ & 21 & $1.2 \pm 0.4 \pm 0.4$ \\
HESS~J1825$-$137 & 56 & $5.6 \pm 1.2 \pm 9.0$ & $1.32 \pm 0.20 \pm 0.39$& 10 & $1.7 \pm 0.9 \pm 1.5$ & 30 & $2.9 \pm 0.7 \pm 1.6$ & 17 & $0.8 \pm 0.3 \pm 0.8$ \\
HESS~J1834$-$087 & 27 & $5.5 \pm 1.2 \pm 2.5$ & $2.24 \pm 0.34 \pm 0.42$& 19 & $4.2 \pm 1.1 \pm 1.9$ & 7 & $<\,1.9$& 3 & $<\,1.0$\\
HESS~J1837$-$069 & 73 & $7.5 \pm 1.3 \pm 4.2$ & $1.47 \pm 0.18 \pm 0.30$& 28 & $4.3 \pm 1.1 \pm 3.5$ & 21 & $1.9 \pm 0.6 \pm 0.9$ & 27 & $1.4 \pm 0.5 \pm 0.5$ \\
HESS~J1841$-$055 & 64 & $10.9 \pm 0.8 \pm 4.1$ & $1.60 \pm 0.27 \pm 0.33$& 20 & $7.4 \pm 1.6 \pm 2.9$ & 13 & $2.1 \pm 0.7 \pm 1.0$ & 31 & $1.8 \pm 0.5 \pm 0.7$ \\
HESS~J1848$-$018 & 19 & $7.4 \pm 1.9 \pm 2.7$ & $2.46 \pm 0.50 \pm 0.51$& 16 & $5.8 \pm 1.6 \pm 2.9$ & 4 & $<\,2.6$& 0 & $<\,1.0$\\
HESS~J1857$+$026 & 53 & $4.2 \pm 0.3 \pm 1.3$ & $1.01 \pm 0.24 \pm 0.25$& 6 & $<\,2.9$& 19 & $1.2 \pm 0.4 \pm 0.4$ & 31 & $1.5 \pm 0.4 \pm 0.4$ \\
MGRO~J1908$+$06 & 16 & $4.6 \pm 1.3 \pm 1.5$ & $3.50 \pm 1.11 \pm 1.60$& 11 & $3.4 \pm 1.2 \pm 1.2$ & 2 & $<\,1.6$& 7 & $<\,1.5$\\
\tablenotemark{a} & 9 & $< 5.6$ & \nodata & 3 & $<\,3.9$& 2 & $<\,1.6$& 7 & $<\,1.5$\\
MGRO~J1958$+$2848 & 21 & $1.3 \pm 0.4 \pm 0.7$ & $4.36 \pm 1.09 \pm 1.21$& 19 & $1.3 \pm 0.5 \pm 0.5$ & 0 & $<\,0.3$& 0 & $<\,0.3$\\
\tablenotemark{a} & 8 & $< 1.8$ & \nodata & 15 & $1.2 \pm 0.5 \pm 0.5$ & 0 & $<\,0.3$& 0 & $<\,0.3$\\
VER~J2016$+$372 & 31 & $1.8 \pm 0.5 \pm 0.8$ & $2.45 \pm 0.44 \pm 0.49$& 26 & $1.6 \pm 0.5 \pm 0.5$ & 3 & $<\,0.6$& 3 & $<\,0.5$\\
MGRO~J2019$+$37 & 31 & $4.4 \pm 1.7 \pm 1.1$ & $6.37 \pm 3.04 \pm 1.21$& 26 & $5.4 \pm 1.2 \pm 1.8$ & 0 & $<\,0.7$& 1 & $<\,0.8$\\
\tablenotemark{a} & 5 & $< 4.7$ & \nodata & 9 & $<\,5.3$& 0 & $<\,0.8$& 1 & $<\,0.8$\\
MGRO~J2031$+$41B & 58 & $6.4 \pm 1.0 \pm 1.1$ & $3.07 \pm 0.33 \pm 0.37$& 58 & $6.2 \pm 1.0 \pm 1.3$ & 3 & $<\,1.0$& 0 & $<\,0.4$\\
\tablenotemark{a}& 12 & $< 4.4$ & \nodata & 11 & $3.0 \pm 0.9 \pm 1.0 $ & 1 & $< 0.9$ & 0 & $< 0.4$\\
MGRO~J2228$+$61 & 94 & $2.8 \pm 0.5 \pm 0.5$ & $3.06 \pm 0.41 \pm 0.42$& 87 & $2.5 \pm 0.5 \pm 0.5$ & 7 & $<\,0.8$& 0 & $<\,0.2$\\
\tablenotemark{a} & 15 & $< 2.0$ & \nodata & 9 & $<\,1.8$& 7 & $<\,0.8$& 0 & $<\,0.2$\\
\enddata
\tablecomments{Results of the maximum likelihood spectral fits for LAT-detected VHE sources. These results are obtained assuming the sources have the same morphology as was measured by VHE experiments. ``a" in the first column corresponds to the results with the contribution of pulsars associated in Table~\ref{tab:pulsars} subtracted from the SED for the source just above. Columns 2 to 4 respectively give the TS, the integrated flux and the spectral index of the source fit in the energy range from 10 GeV to 316 GeV. Columns 5 to 10 give the TS and the integrated flux fit in three logarithmically-spaced energy ranges: 10-31 GeV, 31-100 GeV, 100-316 GeV. When the TS in the energy bin is $<10$ a 99\% c.l. upper limit on the flux is given instead. The two uncertainties respectively correspond to the statistical and systematic uncertainties.}
\end{deluxetable}
\normalsize
\noindent
\end{landscape}
\thispagestyle{empty}
\clearpage
\begin{landscape}
\thispagestyle{empty}
\tabletypesize{\scriptsize}
\begin{deluxetable}{l*{11}l}
\tablewidth{0pt}
\tablecaption{Spectral results for undetected sources 
\label{tab:Not_Det_Res}}
\tablehead{ \colhead{Name} & \colhead{TS} & \colhead{F$_{10\,\text{GeV}}^{316\,\text{GeV}}$} & \colhead{TS$_{10\,\text{GeV}}^{31\,\text{GeV}}$} & \colhead{F$_{10\,\text{GeV}}^{31\,\text{GeV}}$}& \colhead{TS$_{31\,\text{GeV}}^{100\,\text{GeV}}$} & \colhead{F$_{31\,\text{GeV}}^{100\,\text{GeV}}$} & \colhead{TS$_{100\,\text{GeV}}^{316\,\text{GeV}}$} & \colhead{F$_{100\,\text{GeV}}^{316\,\text{GeV}}$}\\
\colhead{} & \colhead{} & \colhead{$\left(10^{-10}\,\text{cm}^{-2}\,\text{s}^{-1}\right)$} & \colhead{} & \colhead{$\left(10^{-10}\,\text{cm}^{-2}\,\text{s}^{-1}\right)$} & \colhead{} & \colhead{$\left(10^{-10}\,\text{cm}^{-2}\,\text{s}^{-1}\right)$} & \colhead{} & \colhead{$\left(10^{-10}\,\text{cm}^{-2}\,\text{s}^{-1}\right)$}}
\startdata
MGRO~J0631$+$105 & 6 & $<1.4$ & 4 & $<\,1.2$& 3 & $<\,0.6$& 0 & $<\,0.4$\\
\tablenotemark{a} & 2 & $< 1.0$ & 0 & $<\,0.8$& 3 & $<\,0.6$& 0 & $<\,0.4$\\
HESS~J1026$-$582 & 1 & $<1.6$ & 0 & $<\,1.6$& 0 & $<\,0.4$& 6 & $<\,0.7$\\
\tablenotemark{a} & 1 & $< 1.6$ & 0 & $<\,1.6$& 0 & $<\,0.4$& 6 & $<\,0.7$\\
HESS~J1427$-$608 & 5 & $<1.9$ & 0 & $<\,1.2$& 2 & $<\,0.8$& 4 & $<\,0.6$\\
HESS~J1458$-$608 & 13 & $<2.6$ & 13 & $1.7 \pm 0.5 \pm 0.7$ & 0 & $<\,0.5$& 0 & $<\,0.3$\\
\tablenotemark{a} & 13 & $< 2.6$ & 13 & $1.7 \pm 0.5 \pm 0.7$ & 0 & $<\,0.5$& 0 & $<\,0.3$\\
HESS~J1503$-$582 & 10 & $<3.9$ & 1 & $<\,2.2$& 4 & $<\,1.5$& 9 & $<\,1.0$\\
HESS~J1554$-$550 & 0 & $<0.5$ & 0 & $<\,0.6$& 0 & $<\,0.3$& 0 & $<\,0.3$\\
HESS~J1626$-$490 & 1 & $<2.3$ & 1 & $<\,2.0$& 1 & $<\,1.0$& 0 & $<\,0.6$\\
HESS~J1646$-$458A & 0 & $<2.8$ & 0 & $<\,1.7$& 1 & $<\,1.9$& 1 & $<\,1.0$\\
\tablenotemark{a} & 0 & $< 2.7$ & 0 & $<\,1.7$& 1 & $<\,1.9$& 1 & $<\,1.0$\\
HESS~J1646$-$458B & 6 & $<4.7$ & 0 & $<\,2.6$& 4 & $<\,1.9$& 5 & $<\,1.2$\\
\tablenotemark{a} & 4 & $< 4.3$ & 0 & $<\,2.2$& 4 & $<\,1.9$& 5 & $<\,1.3$\\
HESS~J1702$-$420 & 6 & $<4.7$ & 0 & $<\,2.6$& 8 & $<\,2.4$& 1 & $<\,0.8$\\
\tablenotemark{a} & 6 & $< 4.5$ & 0 & $<\,2.5$& 8 & $<\,2.4$& 1 & $<\,0.8$\\
HESS~J1718$-$385 & 3 & $<2.1$ & 0 & $<\,1.5$& 0 & $<\,0.8$& 6 & $<\,0.9$\\
\tablenotemark{a} & 3 & $< 2.1$ & 0 & $<\,1.5$& 0 & $<\,0.8$& 6 & $<\,0.9$\\
HESS~J1729$-$345 & 0 & $<1.4$ & 0 & $<\,1.4$& 0 & $<\,0.8$& 0 & $<\,0.5$\\
HESS~J1809$-$193 & 15 & $<8.5$ & 11 & $4.7 \pm 1.5 \pm 2.1$ & 3 & $<\,2.3$& 1 & $<\,1.0$\\
HESS~J1813$-$178 & 3 & $<2.5$ & 0 & $<\,1.4$& 5 & $<\,1.6$& 1 & $<\,0.7$\\
HESS~J1818$-$154 & 0 & $<1.6$ & 0 & $<\,1.6$& 0 & $<\,0.8$& 0 & $<\,0.4$\\
HESS~J1831$-$098 & 0 & $<1.9$ & 0 & $<\,1.5$& 0 & $<\,1.0$& 0 & $<\,0.6$\\
HESS~J1833$-$105 & 4 & $<2.1$ & 3 & $<\,1.9$& 1 & $<\,0.7$& 0 & $<\,0.7$\\
\tablenotemark{a} & 4 & $< 2.1$ & 3 & $<\,1.9$& 1 & $<\,0.7$& 0 & $<\,0.7$\\
HESS~J1843$-$033 & 0 & $<0.9$ & 0 & $<\,1.0$& 0 & $<\,0.5$& 0 & $<\,0.5$\\
MGRO~J1844$-$035 & 0 & $<1.4$ & 0 & $<\,1.2$& 0 & $<\,0.9$& 0 & $<\,0.5$\\
HESS~J1846$-$029 & 2 & $<2.0$ & 5 & $<\,2.4$& 0 & $<\,0.5$& 0 & $<\,0.4$\\
HESS~J1849$-$000 & 0 & $<1.3$ & 1 & $<\,1.5$& 0 & $<\,0.5$& 0 & $<\,0.5$\\
HESS~J1858$+$020 & 0 & $<1.1$ & 0 & $<\,1.2$ & 0 & $<\,0.5$ & 0 & $<\,0.4$\\
MGRO~J1900$+$039 & 0 & $<1.2$ & 0 & $<\,1.3$& 0 & $<\,0.6$& 0 & $<\,0.4$\\
HESS~J1912$+$101 & 10 & $<4.6$ & 2 & $<\,2.7$& 8 & $<\,2.0$& 4 & $<\,1.1$\\
VER~J1930$+$188 & 0 & $<1.0$ & 1 & $<\,1.1$& 0 & $<\,0.4$& 0 & $<\,0.3$\\
VER~J1959$+$208 & 0 & $<0.3$ & 0 & $<\,0.3$& 0 & $<\,0.3$& 0 & $<\,0.4$\\
MGRO~J2031$+$41A & 14 & $<29.6$ & 9 & $<\,22.6$& 3 & $<\,8.4$& 4 & $<\,2.3$\\
W49A & 3 & $<2.4$ & 0 & $<\,1.6$& 3 & $<\,1.0$& 3 & $<\,0.8$\\
\enddata
\tablecomments{Results of the maximum likelihood spectral fits for sources not detected in our LAT analysis. These results are obtained assuming the sources have the same morphology as was measured by VHE. ``a" in the first column corresponds to the results with the contribution of pulsars associated in Table~\ref{tab:pulsars} subtracted from the SED for the source just above. Columns 2 and 3 respectively give the TS and a 99\% c.l. upper limit on the integrated flux in the 10 to 316\,GeV energy range. Columns 4 to 9 give the TS and the integrated flux in three logarithmically-spaced energy ranges: 10-31\,GeV, 31-100\,GeV and 100-316\,GeV. When the TS in the energy bin is $<10$ a 99\% c.l. upper limit on the flux is given. The two uncertainties respectively correspond to the statistical and systematic uncertainties.}
\end{deluxetable}
\normalsize
\noindent
\end{landscape}
\thispagestyle{empty}
\clearpage

The LAT spectra of the sources have been derived assuming the published VHE morphology. In addition to performing a spectral fit over the entire energy range, we computed a SED by fitting the flux of the source independently in 3 energy bins spaced uniformly in log from 10$\,$GeV to 316$\,$GeV. During this fit, we fixed the spectral index of the source at 2 as well as the model of background sources to the best-fit obtained in the whole energy range. We define a detection in the energy bin when $\text{TS}\geq10$ and otherwise compute a flux upper limit using the same method as for the fit in the full energy range. All spectral results are presented in Tables~\ref{tab:Det_results} and \ref{tab:Not_Det_Res}.

\subsection{Systematics}
\label{systext}

Two main systematic uncertainties can affect the extension fit: uncertainties in our model of the Galactic diffuse emission and uncertainties in our knowledge of the LAT PSF. We used the procedure described in \cite{2012arXiv1207.0027L} and obtained the total systematic error on the source extension by adding the two errors in quadrature.

Three main systematic uncertainties can affect the LAT flux estimate for an extended source: uncertainties on the Galactic diffuse background, on the effective area and on the shape of the source. We combined these errors in quadrature to obtain an estimate of the total systematic uncertainty on spectral parameters.  

The dominant uncertainty comes from the Galactic diffuse emission and was estimated by using the alternative model for Galactic diffuse emission described in \cite{2012arXiv1207.0027L}. The systematic due to the effective area was determined by using modified instrument response functions as explained in \cite{2012ApJS..203....4A}.

The imperfect knowledge of the true $\gamma$-ray morphology introduces a last source of error. We derived an estimate of the uncertainty on the spectral parameters due to the uncertainty on the $\gamma$-ray morphology by computing the difference between the values obtained assuming the published VHE spatial model and those obtained assuming the best-fit GeV extension obtained in this analysis.

As discussed in Section 3.2, for the sources near LAT-detected pulsars, a fourth source of systematic uncertainties can affect the LAT flux estimate : the pulsar model. For these sources, we studied any potential contamination of the putative LAT $\gamma$-ray sources by pulsars by performing a second fit of the regions including the pulsars in our background models. The results are also included in Tables~\ref{tab:Spat_results}, \ref{tab:Det_results} and \ref{tab:Not_Det_Res} and are flagged with an ``a". The effects of pulsar contamination are discussed for individual sources in Section 4.

We caution that some LAT-detected pulsars may have spectra at LAT energies that deviate from the simple exponential cutoff power law model assumed in the 2FGL catalog, for energies above 10$\,$GeV. Aliu et al. (2011) show this to be the case for the Crab pulsar. In such cases, our fit could still be contaminated by the pulsar, especially in the lowest energy bin between 10$\,$GeV and 31.6$\,$GeV. For instance, in the case of Geminga for which this analysis detected no significant PWN like emission, the comparison of the energy flux above 10GeV between the 2FGL and the 1FHL catalogs shows a factor $\sim$2 smaller flux for 2FGL.

Table~\ref{tab:Det_results} shows that HESS~J1825$-$137 and MGRO~J0632+17 have large systematic uncertainties. The uncertainty on HESS~J1825$-$137 mainly comes from the term corresponding to the source morphology. As can be seen in Tables~\ref{tab:TeV_sources} and \ref{tab:Spat_results}, the LAT best-fit Gaussian has $\sigma=0 \fdg 44$ while the VHE symmetric Gaussian has $\sigma=0 \fdg 125$. This difference of morphology yields a difference in maximum likelihood flux of $+155\%$.

The large systematic uncertainty on MGRO~J0632+17 is not surprising since the VHE source has an extension of more than 1$\degr$, while the LAT morphology is best-fit as a point source located at the pulsar's position.

\section{ANALYSIS RESULTS}
\label{res}

We detected 30 statistically-significant LAT $\gamma$-ray sources among the 58 VHE sources. The results of the spatial and spectral analyses are shown in Tables~\ref{tab:Spat_results} and \ref{tab:Det_results}. In addition to describing the LAT data analysis, we attempt to classify the origin of the GeV emission using the spatial and spectral information from LAT data as well as multi-wavelength information. We labeled each source as ``PWN" when there is a clear PWN identification, ``PWNc" when the source is a PWN candidate, ``PSR" when the emission is likely coming from the pulsar only and ``O" for other emission.

\begin{deluxetable}{llccccc}
\tabletypesize{\scriptsize}
\tablecaption{Pulsar and PWN characteristics.
\tabletypesize{\scriptsize}
\label{tab:table_luminosity}}
\tablewidth{0pt}
\tablehead{\colhead{Name} & \colhead{PSR} & \colhead{$\dot{E}$} & \colhead{$\tau_C$} & \colhead{Distance} & \colhead{Refs} & \colhead{$\gamma$-ray pulsar}\\
 \colhead{} & \colhead{} & \colhead{$\left(\text{erg}\,\text{s}^{-1}\right)$} & \colhead{(kyr)} & \colhead{(kpc)} & \colhead{} & \colhead{}}
\startdata
VER~J0006+727 & PSR~J0007+7303 & 4.5e+35 & 13.9 & 1.4$ \pm 0.3$ & (1) & Y \\
Crab & PSR J0534$+$2200 & 4.6e+38 & 1.2 & 2.0$\pm 0.5$ & (2) & Y \\
MGRO~J0631+105 & PSR~J0631+1036 & 1.7e+35 & 43.6 & 1.0$\pm 0.2$ & (3) & Y\\
MGRO~J0632+17 & PSR~J0633+1746 & 3.2e+34 & 342 & 0.2$^{+0.2}_{-0.1}$ & (4) & Y\\
Vela$-$X & PSR~J0835$-$4510 & 6.9e+36 & 11.3 & 0.29$ \pm 0.02$ & (5) & Y\\
HESS~J1018$-$589 & PSR~J1016$-$5857 & 2.6e+36 & 21 & $2.9^{+0.6}_{-1.9}$\tablenotemark{a} & (6,7) & Y\\
HESS~J1023$-$575 & PSR~J1023$-$5746 & 1.1e+37 & 4.6 & 2.8\tablenotemark{b}  & (8) & Y\\
HESS~J1026$-$582 & PSR~J1028$-$5819 & 8.4e+35 & 90 & 2.3$\pm 0.3$ & (9) & Y\\
HESS~J1119$-$614 & PSR~J1119$-$6127 & 2.3e+36 & 1.6 & 8.4$\pm 0.4$ & (10) & Y\\
HESS~J1303$-$631 & PSR~J1301$-$6305 & 1.7e+36 & 11 & 6.7$^{+1.1}_{-1.2}$ & (11, 12) & N \\
HESS~J1356$-$645 & PSR~J1357$-$6429  & 3.1e+36 & 7.3 & 2.5$^{+0.5}_{-0.4}$ & (13) & Y\\
HESS~J1418$-$609 & PSR~J1418$-$6058 & 4.9e+36 & 1.0 & 1.6$\pm 0.7$ & (14) & Y  \\
HESS~J1420$-$607 & PSR~J1420$-$6048  & 1.0e+37 & 13 & 5.6$\pm 0.9$ & (11,15) & Y \\
HESS~J1458$-$608 & PSR~J1459$-$6053 & 9.1e+35 & 64.7 & 4\tablenotemark{d} & (16) & Y\\
HESS~J1514$-$591 & PSR~J1513$-$5908 & 1.7e+37 & 1.56 & 4.2$\pm 0.6$& (12) & Y \\
HESS~J1554$-$550 & \nodata & \nodata & 18\tablenotemark{c} & 7.8$\pm 1.3$\tablenotemark{e} & (17, 18, 19) & N\\
HESS~J1616$-$508 & PSR~J1617$-$5055 & 1.6e+37 & 8.13 & 6.8$\pm 0.7$\tablenotemark{f}& (12) & N\\
HESS~J1632$-$478 & \nodata & 3.0e+36 \tablenotemark{g}& 20 & 3.0\tablenotemark{d} & (20) & N\\
HESS~J1640$-$465 & \nodata & 4.0e+36 \tablenotemark{g} & \nodata & \nodata & (22, 23) & N\\
HESS~J1646$-$458B & PSR~J1648$-$4611 & 2.1e+35 & 110 & $5.0\pm0.7$ & (21) & Y\\
HESS~J1702$-$420 & PSR~J1702$-$4128 & 3.4e+35 & 55 & 4.8$\pm 0.6$ & (24) & Y\\
HESS~J1708$-$443 & PSR~J1709$-$4429 & 3.4e+36 & 17.5 & 2.3$\pm 0.3$ & (25) & Y\\
HESS~J1718$-$385 & PSR~J1718$-$3825 & 1.3e+36 & 89.5 & 3.6$\pm 0.4$ & (26, 12) & Y\\
HESS~J1804$-$216 & PSR~J1803$-$2137 & 2.2e+36 & 16 & 3.8$^{+0.4}_{-0.5}$ & (12) & N \\
HESS~J1809$-$193 & PSR~J1809$-$1917 & 1.8e+36 & 51.3 & 3.5$\pm 0.4$\tablenotemark{f}& (12) & N\\
HESS~J1813$-$178 & PSR~J1813$-$1749 & 6.8e+37 & 5.4 & 4.7\tablenotemark{h} & (12, 27) & N\\
HESS~J1818$-$154 & PSR~J1818-1541 & 2.3e+33 & 9 \tablenotemark{i} & 7.8$^{+1.6}_{-1.4}$\tablenotemark{f} & (28) & N\\
HESS~J1825$-$137 & PSR~J1826$-$1334 & 2.8e+36 & 21 & 3.9$\pm 0.4$ & (12, 29) & N \\
HESS~J1831$-$098 & PSR~J1831$-$0952 & 1.1e+36 & 128 & 4.0$\pm$0.4\tablenotemark{f} & (12, 30) & N\\
HESS~J1833$-$105 & PSR~J1833$-$1034 & 3.4e+37 & 4.85 & 4.7$\pm 0.4$ & (31) & Y\\
HESS~J1837$-$069 & PSR~J1838$-$0655 & 5.5e+36 & 2.23 & $6.6\pm0.9$ & (32) &N\\
HESS~J1841$-$055 & PSR~J1838$-$0537 & 5.9e+36 & 4.97 & 1.3\tablenotemark{b}  & (33) & Y\\
HESS~J1846$-$029 & PSR~J1846$-$0258 & 8.1e+36 & 0.73 & 5.1\tablenotemark{b} & (12) & N\\
HESS~J1848$-$018 & \nodata & \nodata & \nodata & 6\tablenotemark{d} & (34) & N\\
HESS~J1849$-$000 & PSR~J1849$-$0001 & 9.8e+36 & 42.9 & 7\tablenotemark{d} & (35) & N\\
HESS~J1857+026 & PSR~J1856+0245 & 4.6e+36 & 20.6 & 9.0$\pm$1.2\tablenotemark{f} & (12) & N\\
MGRO~J1908+06 & PSR~J1907+0602 & 2.8e+36 & 19.5 & $3.2\pm0.3$ & (36) & Y\\
HESS~J1912+101 & PSR~J1913+1011 & 2.9e+36 & 169 & 4.8$^{+0.5}_{-0.7}$\tablenotemark{f} & (12) & N\\
VER~J1930+188 & PSR~J1930+1852 & 1.2e+37 & 2.89 & 9$^{+7}_{-2}$1\tablenotemark{f} & (37) & N\\
MGRO~J1958+2848 & PSR~J1958+2846 & 3.4e+35 & 21.7 & \nodata & (12) & Y\\
VER~J1959+208 & PSR~J1959+2048 & 1.6e+35 & \nodata & $2.5\pm 0.1$ & (38, 39) & Y\\
MGRO~J2019+37 & PSR~J2021+3651 & 3.4e+36 & 17.2 &  10$^{+2}_{-4}$  & (40, 41) & Y\\
MGRO~J2228+61 & PSR~J2229+6114 & 2.2e+37 & 10.5 &  0.8$\pm 0.2$  & (42, 43) & Y\\
\enddata\\
\begin{flushleft}
a- SNR distance.\\
b- Pseudo distance based on the observed correlation between the $0.1-100\,$GeV
$\gamma$-ray luminosity and $\dot{E}$ \citep{2010ApJ...725..571S}. This relation suffers various caveats which translates to large uncertainties in the derived distance value and is therefore replaced by a conservative upper limit in 2PC.\\
c- Sedov model, see (31).\\
d- No available uncertainty given in the reference.\\
e- Relation between the hydrogen column density ($N_H$) and E(B-V) and between E(B-V) and the distance.\\
f- Distance estimated using the pulsar dispersion measure and the NE2001 model \citep{2002astro.ph..7156C} which is available as off-line code as done in 2PC. To estimate the distance errors we apply a 20\% Dispersion Measure uncertainty as is used in the online DM-distance estimator tool.\\
g- X-ray vs spin-down luminosity correlation, see (41).\\
h- Distance derived from HI absorption.\\
i- Sedov age, see (16).
\end{flushleft}
\tablecomments{This table describes the pulsar/PWN properties used in this analysis for the population studies in Figures \ref{fig:rapportTeV} and \ref{fig:dotelpwn}. Column 2 gives the name of the associated pulsar. A ``\nodata" means that no pulsar has been detected but the parameters of an assumed pulsar can be estimated. Columns 3, 4, and 5 show the pulsar's spin down power, characteristic age and distance respectively. Footnotes indicate cases when the distance is not estimated from the pulsar dispersion measure or cases when the age estimate is not the pulsar's characteristic age. Column 7 says if the pulsar has been detected in $\gamma$-rays: ``Y" means that it has and ``N" means that it has not according to 2PC.\\References:  (1) \cite{1993AJ....105.1060P}, (2) \cite{1973PASP...85..579T}, (3) \cite{2010ApJ...708.1426W}, (4) \cite{2012ApJ...755...39V}, (5) \cite{2003ApJ...596.1137D}, (6) \cite{2001ApJ...557L..51C}, (7) \cite{Ruiz86}, (8) \cite{2010ApJ...725..571S}, (9) \cite{2008MNRAS.389.1881K}, (10) \cite{2004MNRAS.352.1405C}, (11) \cite{dalton1303}, (12) \cite{2005yCat.7245....0M}, (13) \cite{2006MNRAS.372..777L}, (14) \cite{1997ApJ...476..347Y}, (15) \cite{2012ApJ...750..162K}, (16) \cite{2012arXiv1205.0719D}, (17) \cite{1996ApJ...471..887S}, (18) \cite{1999ApJ...511..274S}, (19) \cite{2009ApJ...691..895T}, (20) \cite{2010AA...520A.111B}, (21) \cite{2005AJ....129.1993M}, (22) \cite{2007ApJ...662..517F}, (23) \cite{2009ApJ...706.1269L}, (24) \cite{2003MNRAS.342.1299K}, (25) \cite{1995AA...293..795J}, (26) \cite{2001MNRAS.328...17M}, (27) \cite{2007AA...470..249F}, (28) \cite{2011ICRC....7..247H}, (29) \cite{2011ApJ...738...42G},  (30) \cite{2011ICRC....7..243S},  (31) \cite{2006ApJ...637..456C}, (32) \cite{2008ApJ...681..515G}, (33) \cite{2012ApJ...755L..20P}, (34) \cite{2008AIPC.1085..372C}, (35) \cite{2011ApJ...729L..16G}, (36) \cite{2010ApJ...711...64A}, (37) \cite{2008AJ....136.1477L}, (38) \cite{2012ApJ...744...33G}, (39) \cite{2012ApJ...760...92H}, (40) \cite{1997ApJ...476..347Y}, (41) \cite{2004ApJ...612..389H}, (42) \cite{2009ApJ...706.1331A}, (43) \cite{2001ApJ...560..236K}.}
\end{deluxetable}
\normalsize
\noindent

Table~\ref{tab:table_luminosity} summarizes the PSR/PWN systems' characteristics such as the age or the distance. The characteristic age of the pulsar is given by $\tau_C= P / 2 \dot{P}$ where P is the pulsar's rotational period. This formula is obtained assuming that the pulsar is a dipole and that the initial period of the pulsar is negligible compared to its current value. The putative pulsars powering HESS~J1554$-$550, HESS~J1632$-$478, HESS~J1640$-$465, HESS~J1818$-$154 and HESS~J1848$-$018 are not yet detected. However, Table~\ref{tab:table_luminosity} gives the system age or distance estimated using another method than the pulsar dispersion measure, e.g. distance measure derived by $\text{H}_{\text{I}}$ absorption or using a statistical relation between the hydrogen column density ($N_H$) and E(B$-$V) \citep{1975ApJ...198..103R} and a relation between E(B$-$V) and the distance \citep{1978AA....64..367L}. 

\subsection{Spatial Results}
\label{morph_res}

Assigning an association between a LAT source and an VHE source depends on two considerations: the spatial and spectral consistency with the counterpart. Spatial consistency has two aspects: centroid coincidence and extension compatibility. For assessing the degree of spatial coincidence, we considered a LAT emission to be spatially coincident with the VHE source if the distance between the LAT and VHE experiments best-fit postions was inside a circle whose radius is the quadratic sum of: 1) the 68\% c.l. uncertainty on the LAT best-fit position, 2) the 68\% c.l. position uncertainty of the VHE source, and 3) the 99\% containment radius of the source based on the VHE fitted angular extent. These data are summarized in Table~\ref{tab:CompTeVGeV}. Only one source is not spatially coincident using this criterion: HESS~J1018$-$589. We removed from this table the sources likely to be associated to pulsar emission. These sources will be discussed in Section \ref{pulsarsect}.

\tabletypesize{\scriptsize}
\begin{deluxetable}{lllll}

\tablewidth{0pt}
\tablecaption{Comparison between the LAT and VHE positions.
\label{tab:CompTeVGeV}}
\tablehead{ \colhead{Name} & \colhead{$\text{Unc}_{\text{GeV}}$} & \colhead{$\text{Unc}_{\text{TeV}}$} & \colhead{$r_{99\%}$ TeV} & \colhead{Distance}\\
\colhead{} & \colhead{(deg)} & \colhead{(deg)} & \colhead{(deg)} & \colhead{(deg)}}
\startdata
HESS~J1018$-$589 &  0.04  & 0.02 & 0.00 & 0.12\\
HESS~J1023$-$575 & 0.04 &  0.09 & 0.55 & 0.10\\
HESS~J1119$-$614  & 0.05  & \nodata & 0.15 & 0.09\\
HESS~J1303$-$631  & 0.05  & 0.01 & 0.49 & 0.41\\
HESS~J1356$-$645 &  0.05  & 0.03 & 0.61 & 0.19\\
HESS~J1420$-$607 &  0.04 &  0.02 & 0.18 & 0.01\\
HESS~J1507$-$622 &  0.06  & 0.04 & 0.46 & 0.21\\
HESS~J1514$-$591  & 0.03 &  0.07 & 0.23 & 0.06\\
HESS~J1614$-$518 &  0.05 &  0.03 & 0.58 & 0.19\\
HESS~J1616$-$508 &  0.04 &  0.01 & 0.42 & 0.16\\
HESS~J1632$-$478 &  0.04 & 0.04 & 0.41 & 0.15\\
HESS~J1634$-$472 &  0.03 & 0.05 & 0.33 & 0.18\\
HESS~J1640$-$465 &  0.05 &  0.01 & 0.12 & 0.07\\
HESS~J1804$-$216 &  0.04 & 0.02  & 0.65 & 0.06\\
HESS~J1825$-$137 & 0.06 &  0.03 & 0.38 & 0.28\\
HESS~J1834$-$087 & 0.05 & 0.02 & 0.27 & 0.06\\
HESS~J1837$-$069 & 0.06 &  0.02 & 0.26 & 0.12\\
HESS~J1841$-$055 &  0.06 &  0.05 & 1.00 & 0.22\\
HESS~J1848$-$018 &  0.04 &  \nodata & 0.97 & 0.11\\
HESS~J1857$+$026 &  0.04 &  0.05 & 0.29 & 0.14\\
VER~J2016$+$372 & 0.05 & \nodata & \nodata & 0.11\\
\enddata
\tablecomments{Comparison of the localization and localization uncertainty for the sources observed at LAT and VHE energies and classified as ``PWN", ``PWNc" or ``O". Columns 2, 3, 4 and 5 respectively give the averaged uncertainty on the position obtained using \emph{Fermi}-LAT data and assuming a point source if $\text{TS}_{\text{ext}}<16$ or a Gaussian if $\text{TS}_{\text{ext}}>16$, the quadratic sum of the statistical (68\% confidence radius) and systematic uncertainties on the VHE position given in Table 1, the 99$\%$ containment radius of the VHE emission assuming the extension listed in Table 1 and the distance between the position of the source fit at LAT and VHE energies.}
\end{deluxetable}
\normalsize
\noindent

Although VER~J2016+372 is spatially extended as announced in \cite{2011arXiv1110.4656A}, neither the extension nor the uncertainties of the spatial parameters are yet available. There is little doubt that these parameters will show that the VERITAS and LAT emission are spatially coincident. This case will be discussed in Section \ref{Osources}.

A finite measured extension of a $\gamma$-ray source is a direct way to discriminate between a pulsar and a PWN. Therefore, we searched for significant angular extensions of the LAT sources. Of the 21 sources not labeled as PSR in Section \ref{pulsarsect}, eight are significantly extended: HESS~J1303$-$631, HESS~J1614$-$518, HESS~J1616$-$508, HESS~J1632$-$478, HESS~J1804$-$216, HESS~J1825$-$137, HESS~J1837$-$069 and HESS~J1841$-$055. Six of them were already detected by \cite{2012arXiv1207.0027L}. Only HESS~J1632$-$478 has an extension inconsistent with that work. For this source, \cite{2012arXiv1207.0027L} measured an extension of ($0 \fdg 35\,\pm\,0 \fdg 06)$ assuming a disk, while we obtained ($0 \fdg 30\,\pm\,0 \fdg 04)$ for a Gaussian. Comparison between 68\% containment radius of a disk and a Gaussian yields $\sigma_{Disk} \sim 1.84 \sigma_{Gaussian}$ \citep{2012arXiv1207.0027L}, where $\sigma_{Disk}$ and $\sigma_{Gaussian}$ respectively represent the disk radius and the Gaussian standard deviation. Therefore, for HESS J1632$-$478, our analysis led to an extension bigger than that obtained in \cite{2012arXiv1207.0027L}. This difference certainly comes from the three additional 2FGL sources (2FGL~J1631.7$-$4720c, 2FGL~J1630.2$-$4752 and 2FGL~J1632.4$-$4820c) included in the model used by \cite{2012arXiv1207.0027L}. These sources are not included in the 1FHL catalog and are below the pre-defined $\text{TS}>$25 threshold for additional sources. Since they are unidentified, we chose to not add them.

\begin{figure}[h!]
\centering
\includegraphics[width=0.5\textwidth]{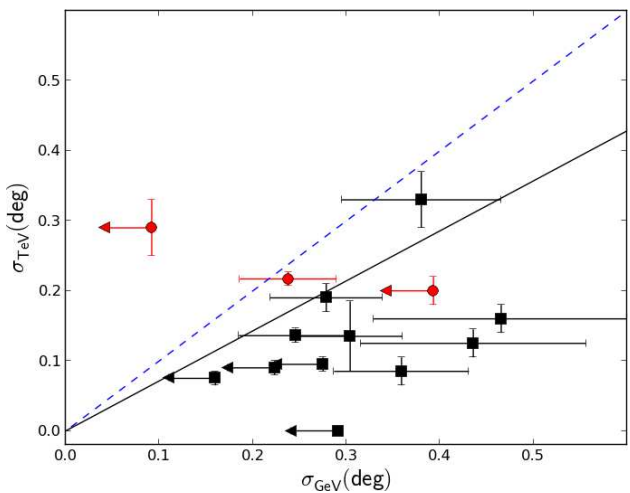}
\caption{\label{extplot}Comparison of the sizes of the sources observed at LAT and VHE energies. Contributions of pulsars listed in Table 2 were taken into account by modeling them as point sources with spectral parameters fixed at their values in the 2FGL catalog. For sources not significantly extended at LAT energies, a 99\% c.l. upper limit on the extension is shown. The red circles represent the sources within $0 \fdg 5$ of a $\gamma$-ray pulsar and the black squares represent the other sources. This plot shows a zoom on $\sigma_{TeV}$ and $\sigma_{GeV} \in [0\degr ,0 \fdg 6]$. The dashed line shows $\sigma_{TeV} = \sigma_{GeV}$. The solid line shows the $\sigma_{GeV} = 1.4 \times \sigma_{TeV}$ fit result (see Section \ref{morph_res}).
\label{fig:sigmavssigma}}
\end{figure}

Figure~\ref{extplot} compares the sizes measured by the LAT and VHE. The only source with $\sigma_{\text{GeV}} < \sigma_{\text{TeV}}$ is HESS~J1708$-$443 whose LAT emission is likely coming from the pulsar. As we discussed above and in Section \ref{PWNc}, we suspect that the extensions of HESS~J1303$-$631 and HESS~J1632$-$478 are overestimated. We assumed a linear relation between the VHE experiment and the LAT extension ($\sigma_{\text{GeV}} =\alpha \times \sigma_{\text{TeV}}$) and fit the data points, obtaining $\alpha = 1.4 \pm 0.2$ with a $\chi^2/\text{d.o.f.} = 15/7$. In the case in which we fixed $\alpha=1$ we obtained a $\chi^2/\text{d.o.f.}=24/8$. This implies that, with the current statistics, the LAT extension is larger than the one measured by VHE experiments at the 2.8$\sigma$ level. Additional exposure and a more precise analysis of crowded regions (e.g. the regions of HESS~J1303$-$631, HESS~J1632$-$478, HESS~J1837$-$069 or HESS~J1841$-$055) are needed to draw any conclusion. Indeed, in these regions the $\gamma$-ray emission could be due to more than one source, as discussed below in the case of HESS~J1841$-$055 or as discussed in \cite{2008ApJ...681..515G} for the case of HESS~J1837$-$069.

\subsection{Pulsars Detected Above 10$\,$GeV}
\label{pulsarsect}

\begin{figure}[h!]
\centering
\subfigure{
\includegraphics[width=0.47\textwidth]{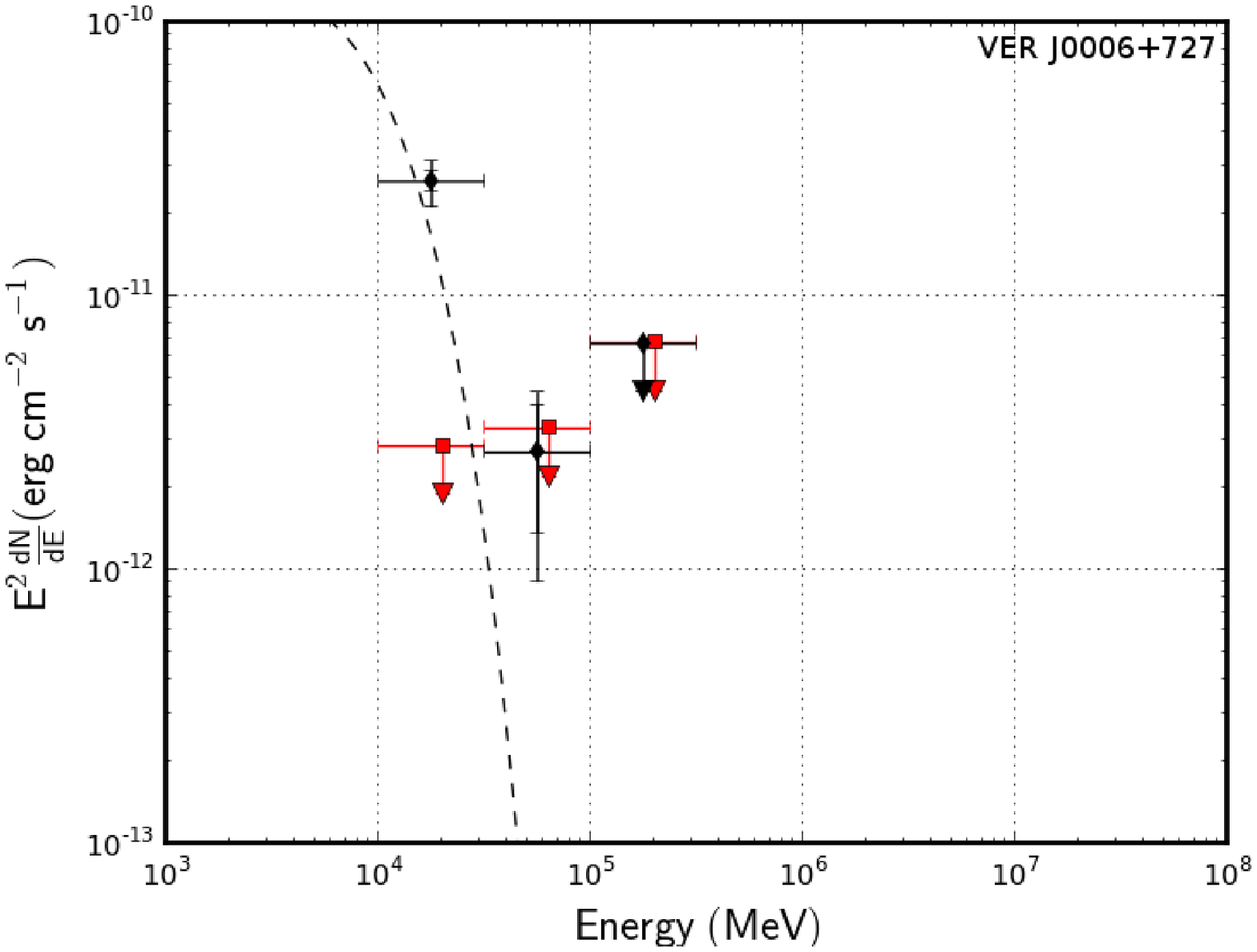}
}
\subfigure{
\includegraphics[width=0.47\textwidth]{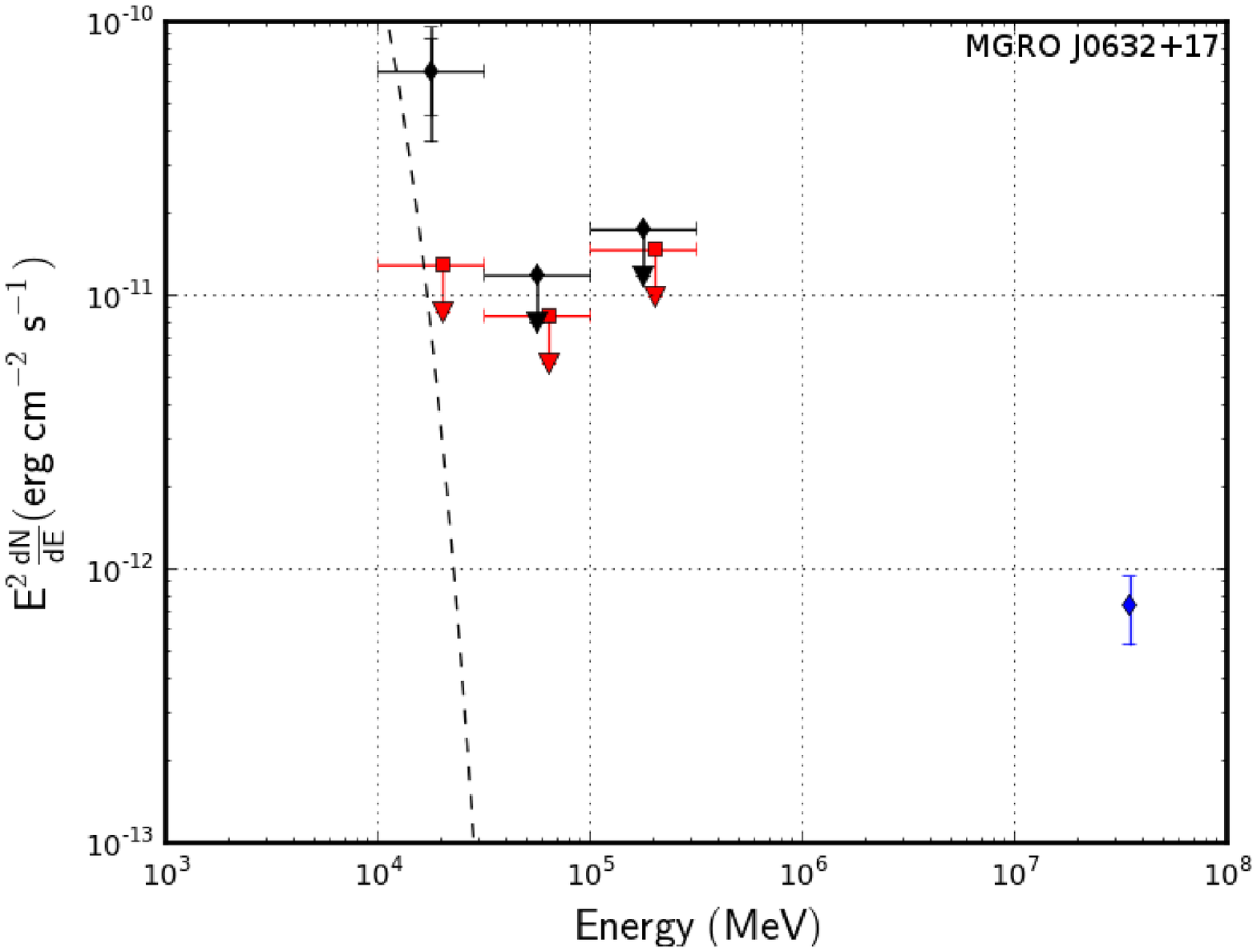}
}
\caption{\label{fig:mgroj0632}\label{fig:verj0006}\label{fig:sedsourcespuls} Multi-wavelength SEDs of VER~J0006$+$727 and MGRO~J0632$+$17. The black diamonds represent the SED of the source obtained at LAT energies. For sources with an associated LAT-detected pulsar, the red squares represent the SED with any potential emission from the LAT-detected pulsar subtracted using the spectral parameters from the 2FGL catalog. To enhance readability, we offset the square markers in energy by a factor of 13\%.  Finally, the blue diamond markers represent the SED obtained by VHE experiments. Table \ref{tab:pulsars} lists the references for the VHE SED of each source. For sources with an associated LAT-detected pulsar, the best-fit spectrum of the pulsar from 2FGL is included as a dashed line. }
\end{figure}

\begin{figure}[h!]
\centering
\subfigure{
\includegraphics[width=0.47\textwidth]{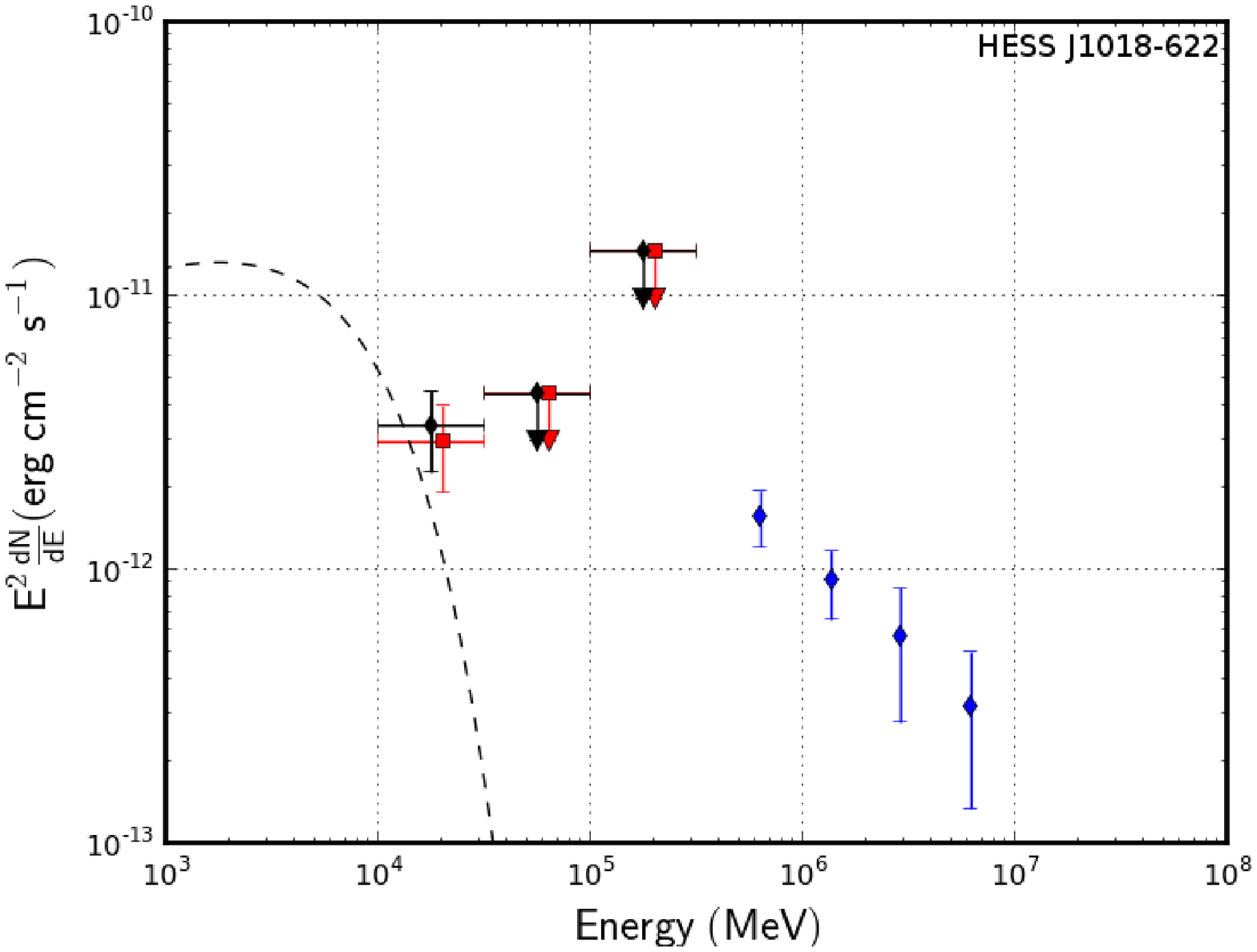}
}
\subfigure{
\includegraphics[width=0.47\textwidth]{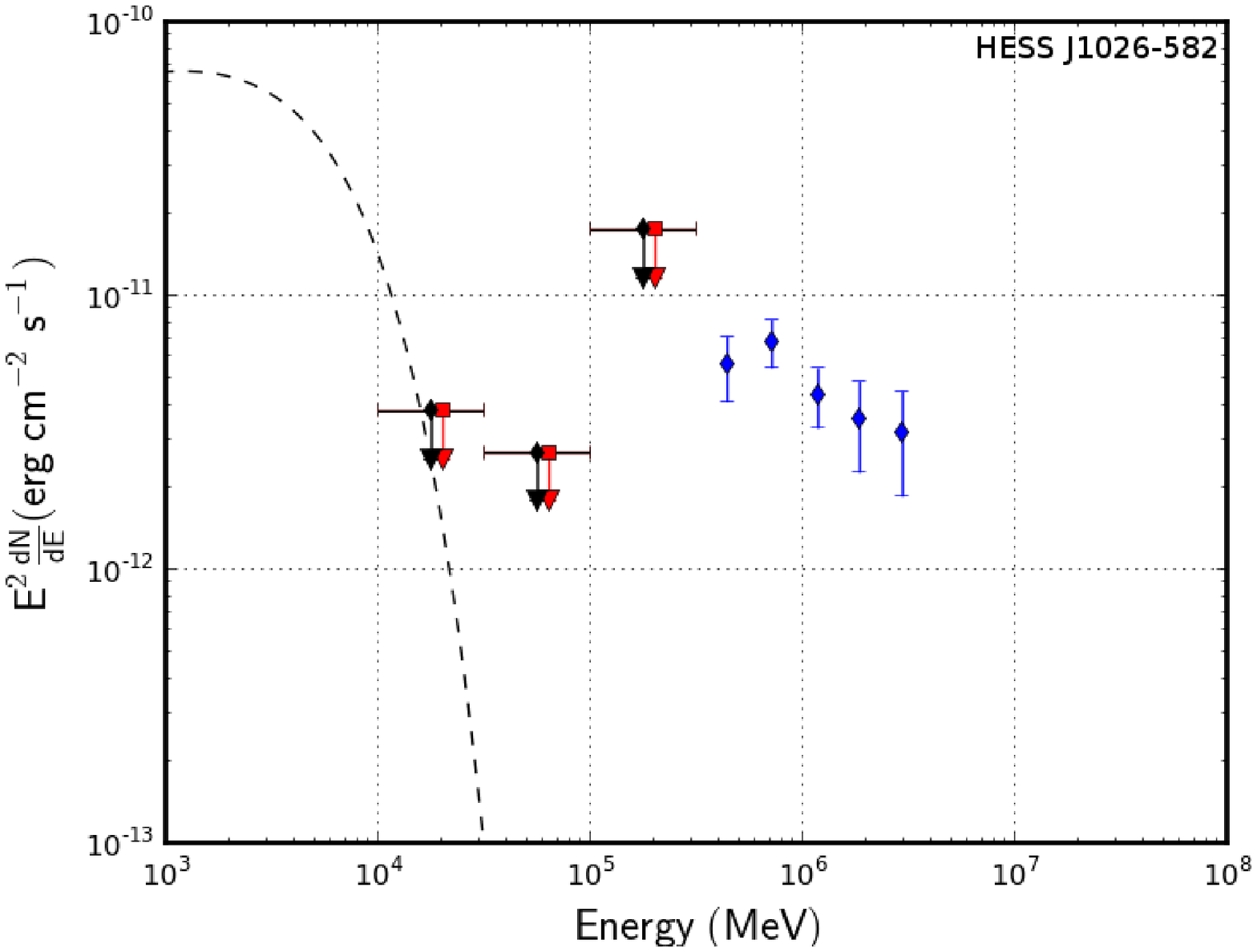}
}
\subfigure{
\includegraphics[width=0.47\textwidth]{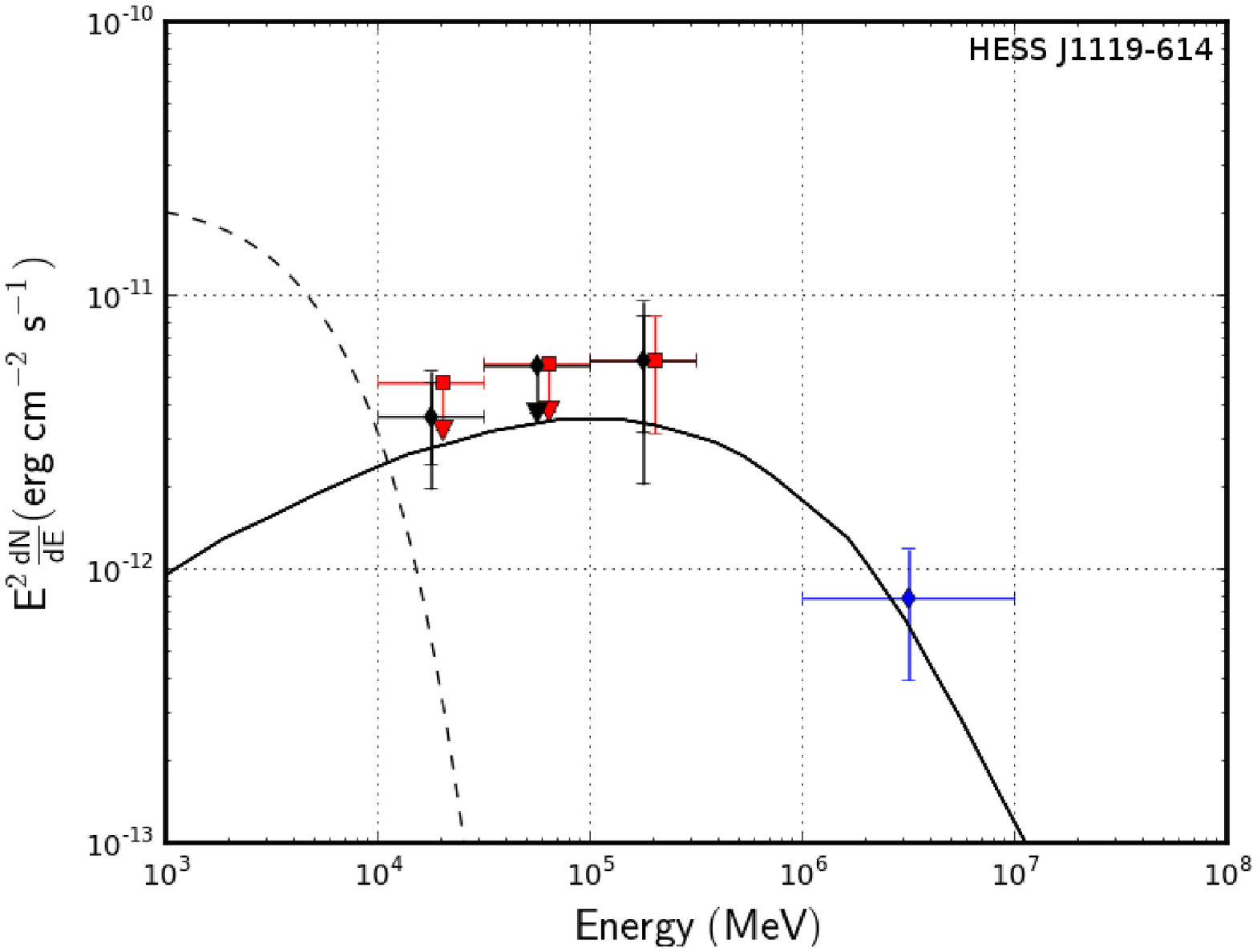}
}
\subfigure{
\includegraphics[width=0.47\textwidth]{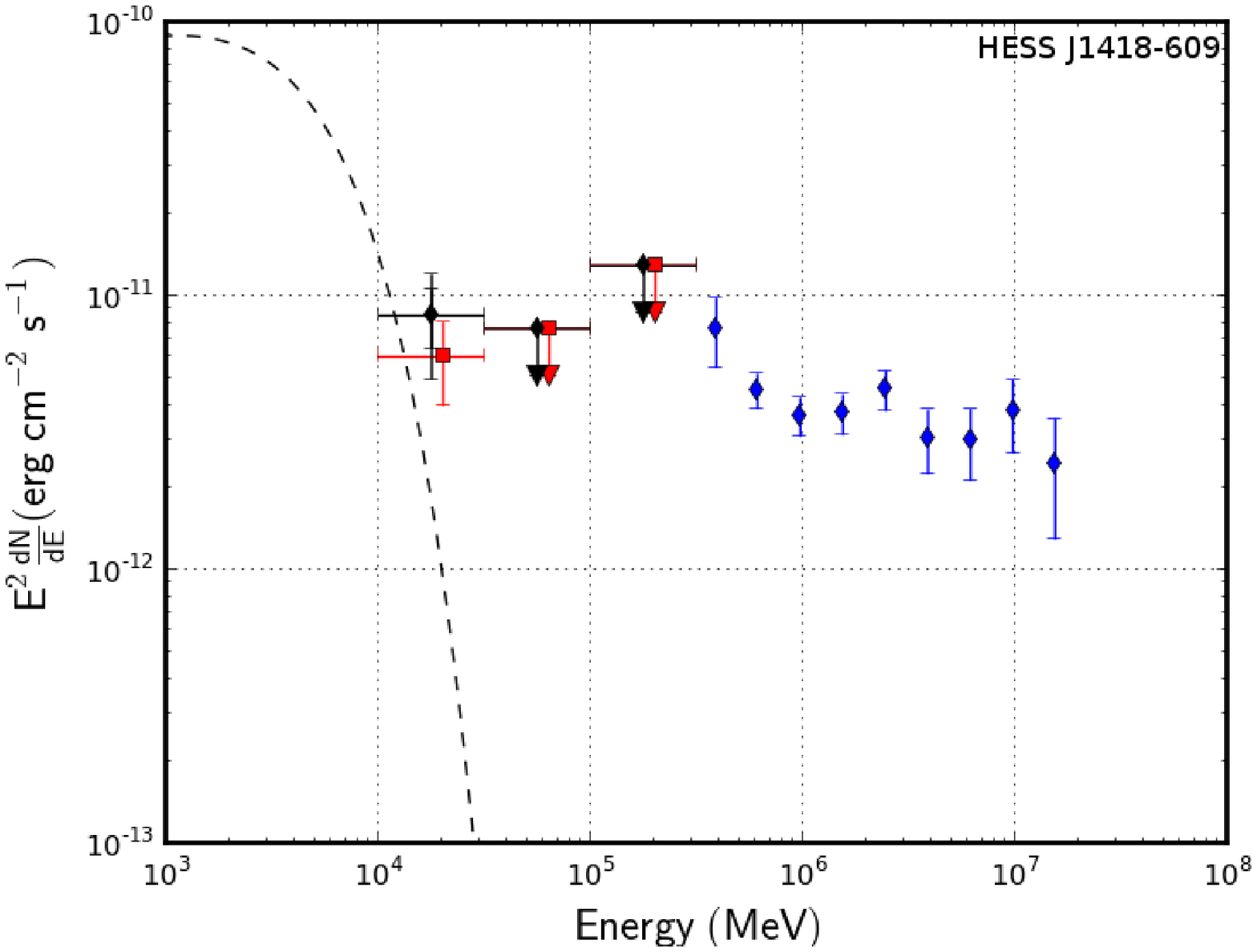}
}
\subfigure{
\includegraphics[width=0.47\textwidth]{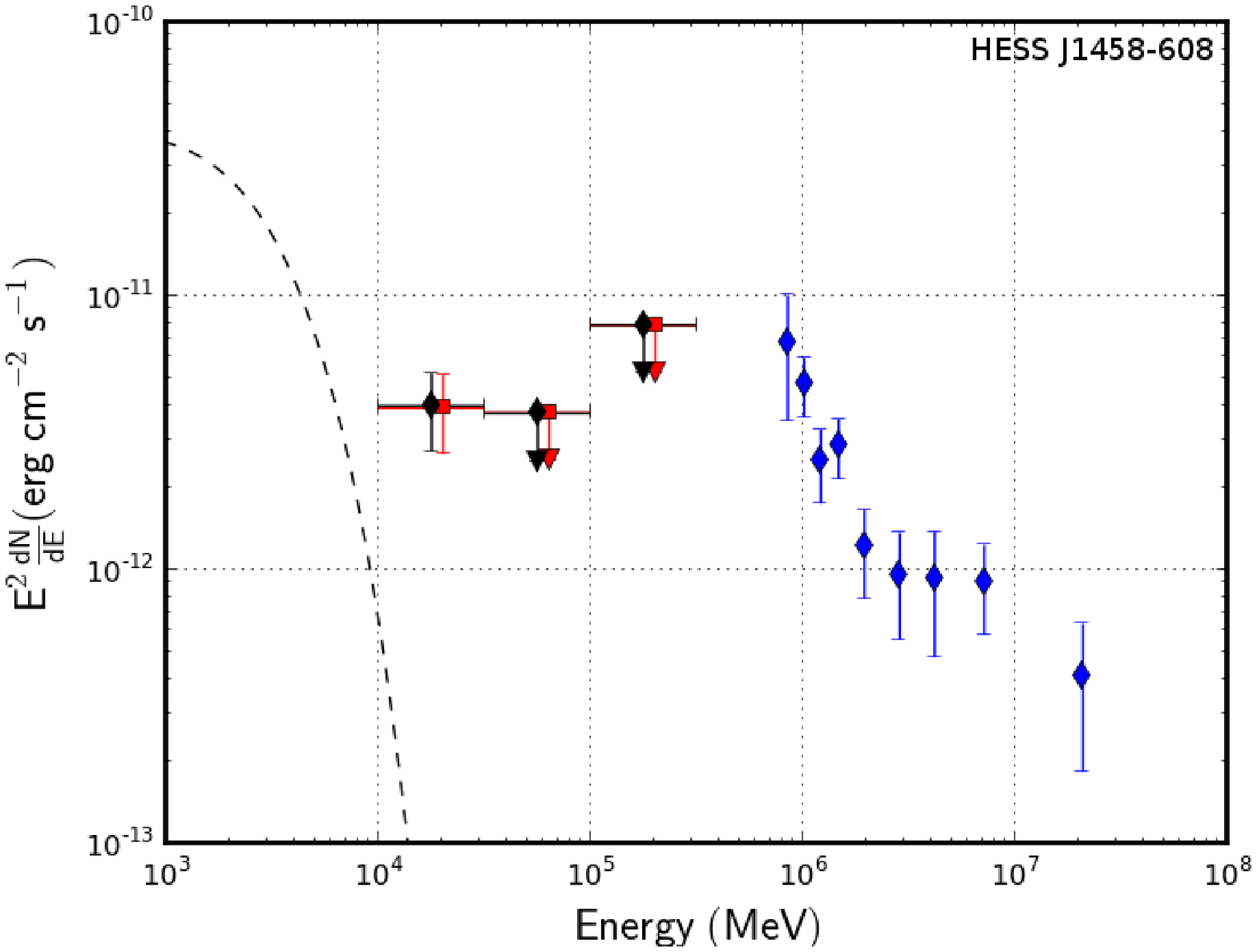}
}
\subfigure{
\includegraphics[width=0.47\textwidth]{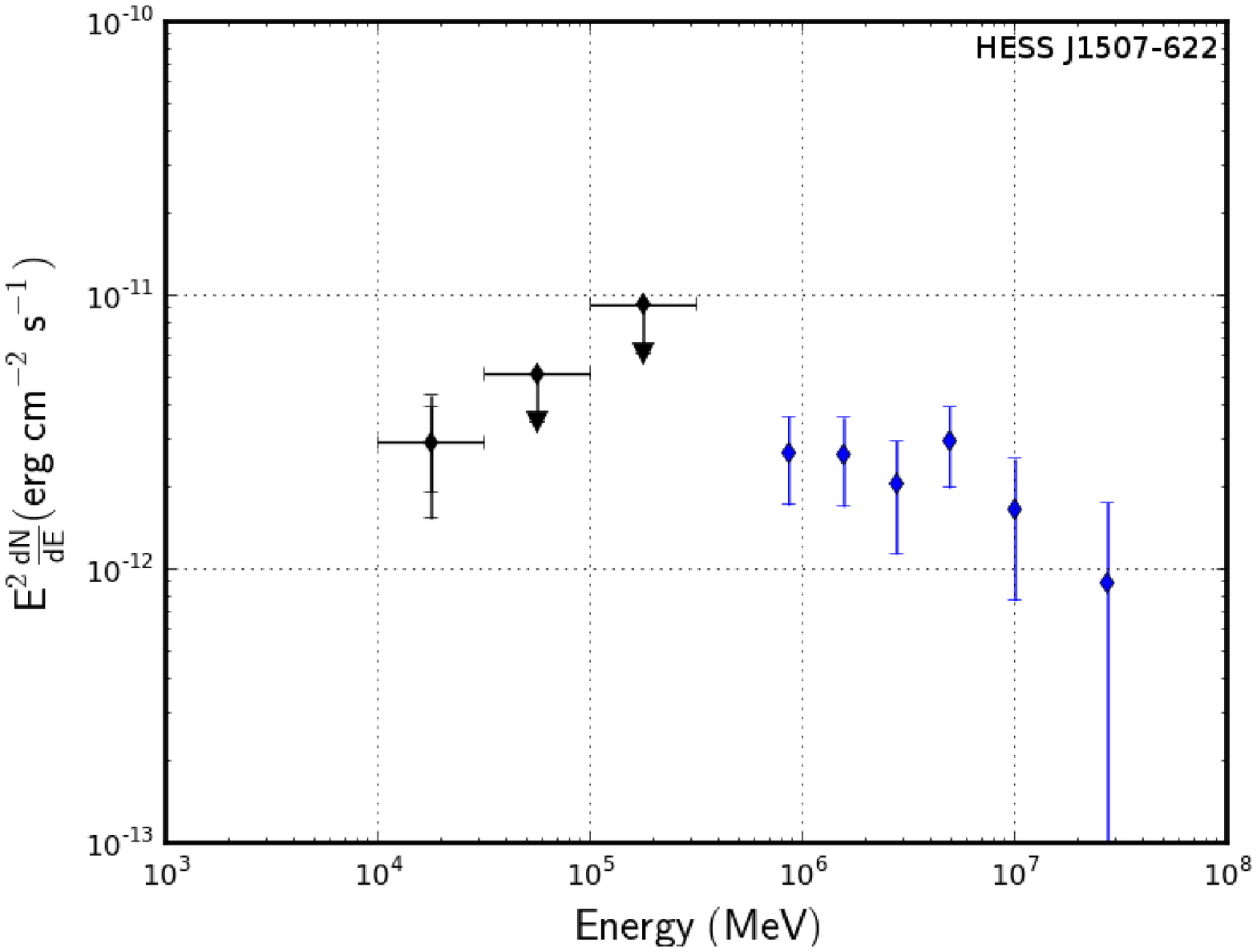}
}
\caption{\label{fig:hessj1119}\label{fig:1026}\label{fig:1458}The LAT and VHE SEDs of HESS~J1018$-$589, HESS~J1119$-$614, HESS~J1418$-$609, HESS~J1458$-$608 and HESS~J1507$-$622 following conventions of Figure~\ref{fig:mgroj0632}. The solid line corresponds to the model proposed by \cite{Mayerdiploma} for HESS~J1119$-$614. In the case of HESS~J1119$-$614, the spectral point has been computed using the integral flux given in \citep[][see Section~\ref{PWNc} for details]{2010AIPC.1248...25K}. The horizontal error bars show the energy range of the used integral flux.}
\end{figure}

\begin{figure}[h!]
\centering
\subfigure{
\includegraphics[width=0.47\textwidth]{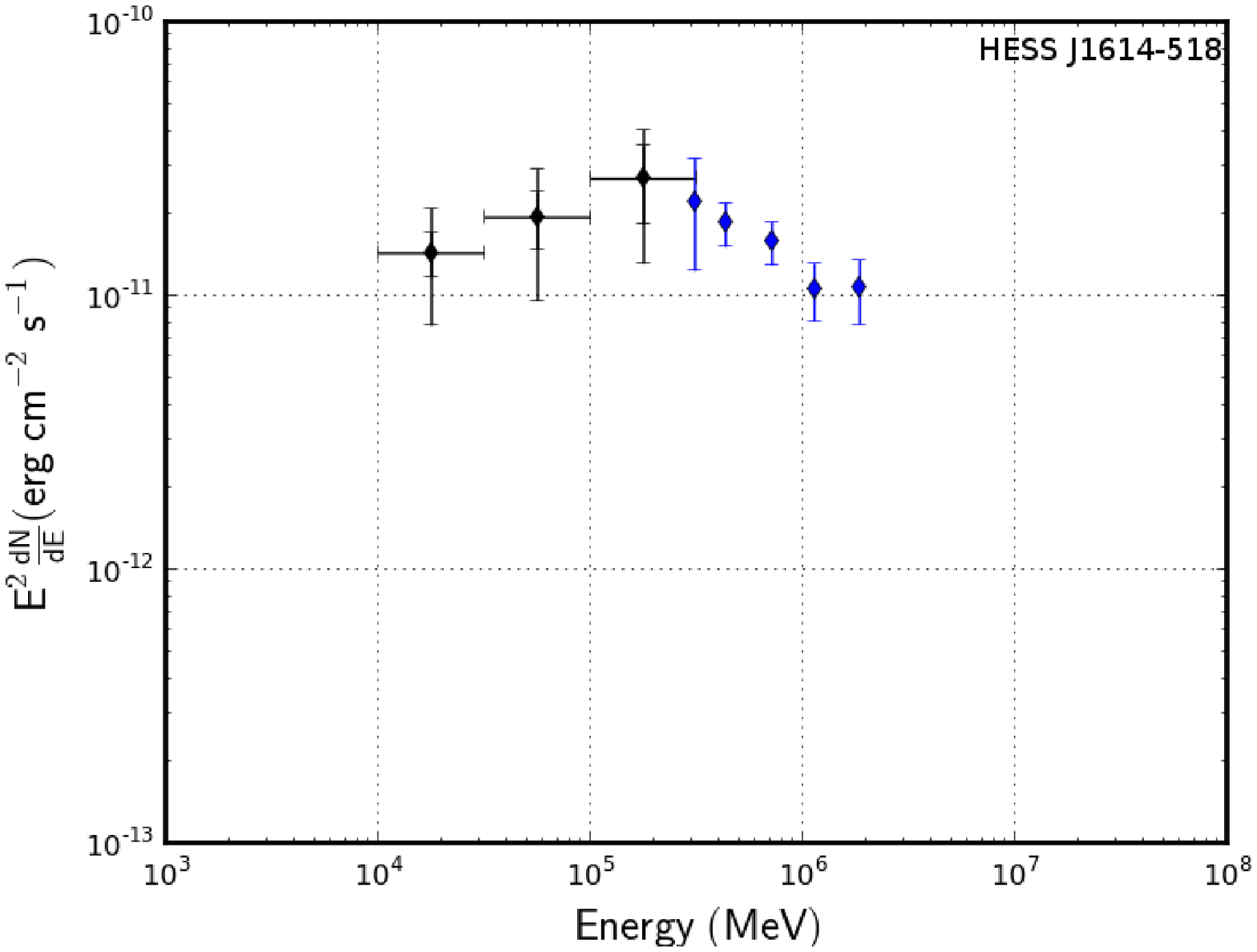}
}
\subfigure{
\includegraphics[width=0.47\textwidth]{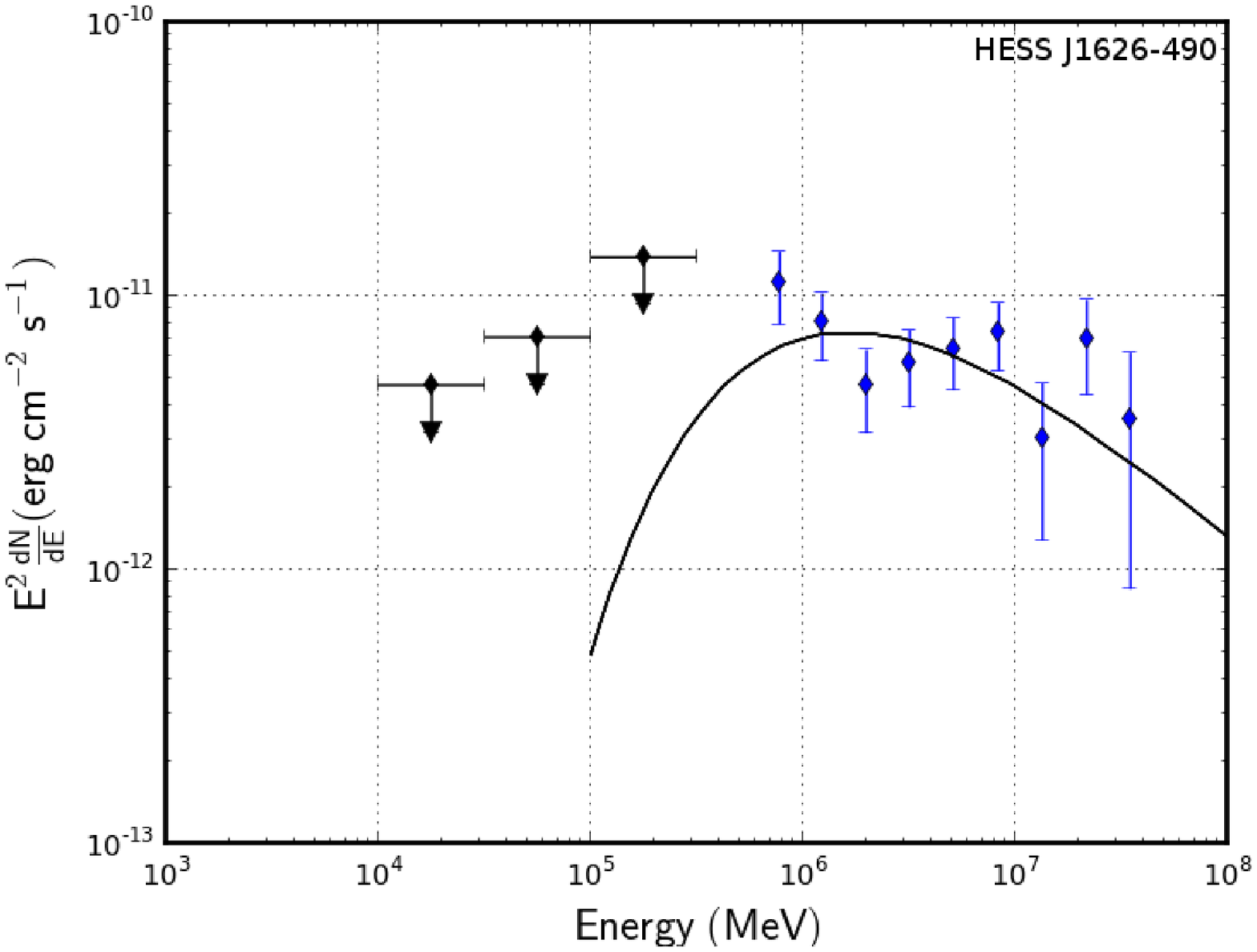}
}
\subfigure{
\includegraphics[width=0.47\textwidth]{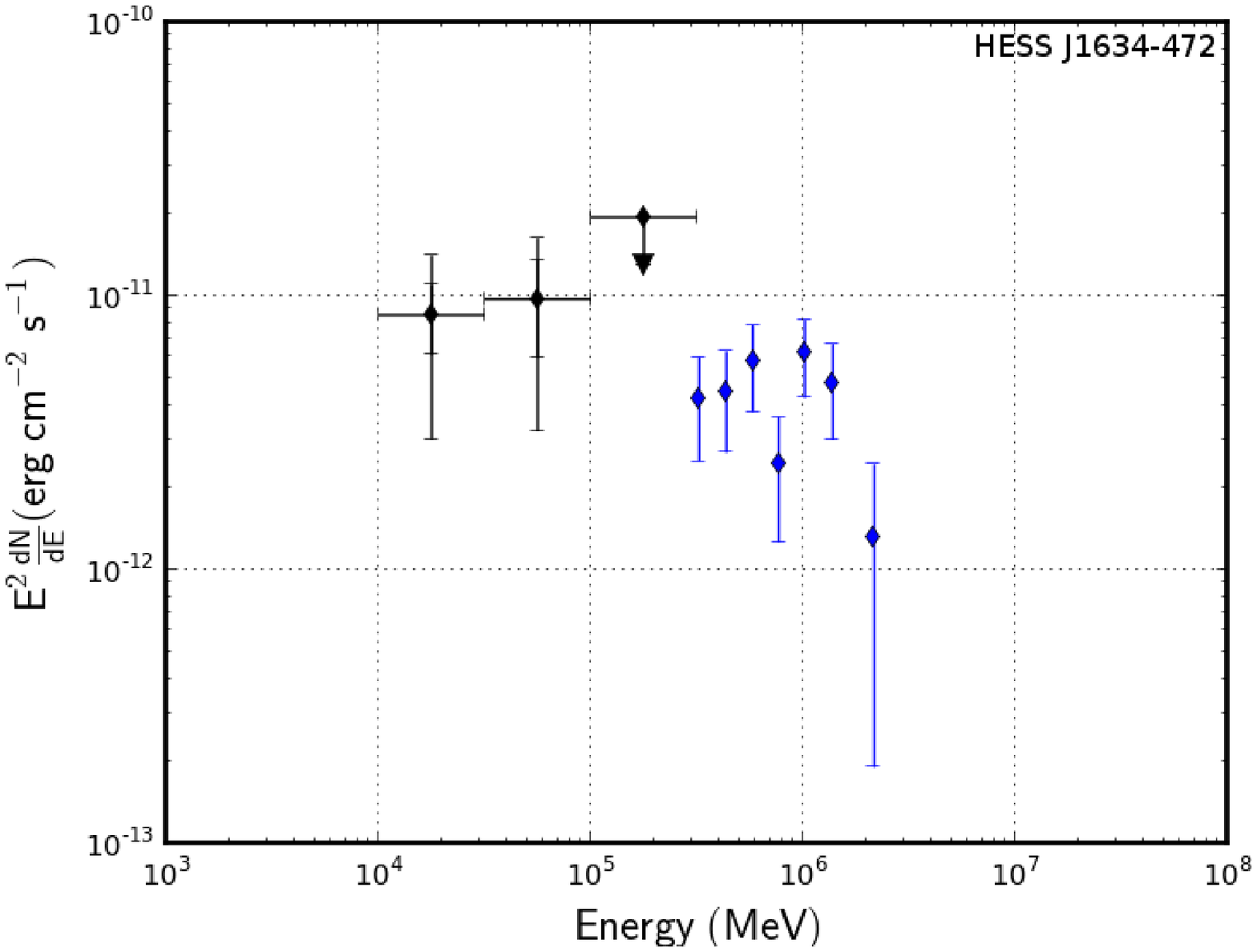}
}
\subfigure{
\includegraphics[width=0.47\textwidth]{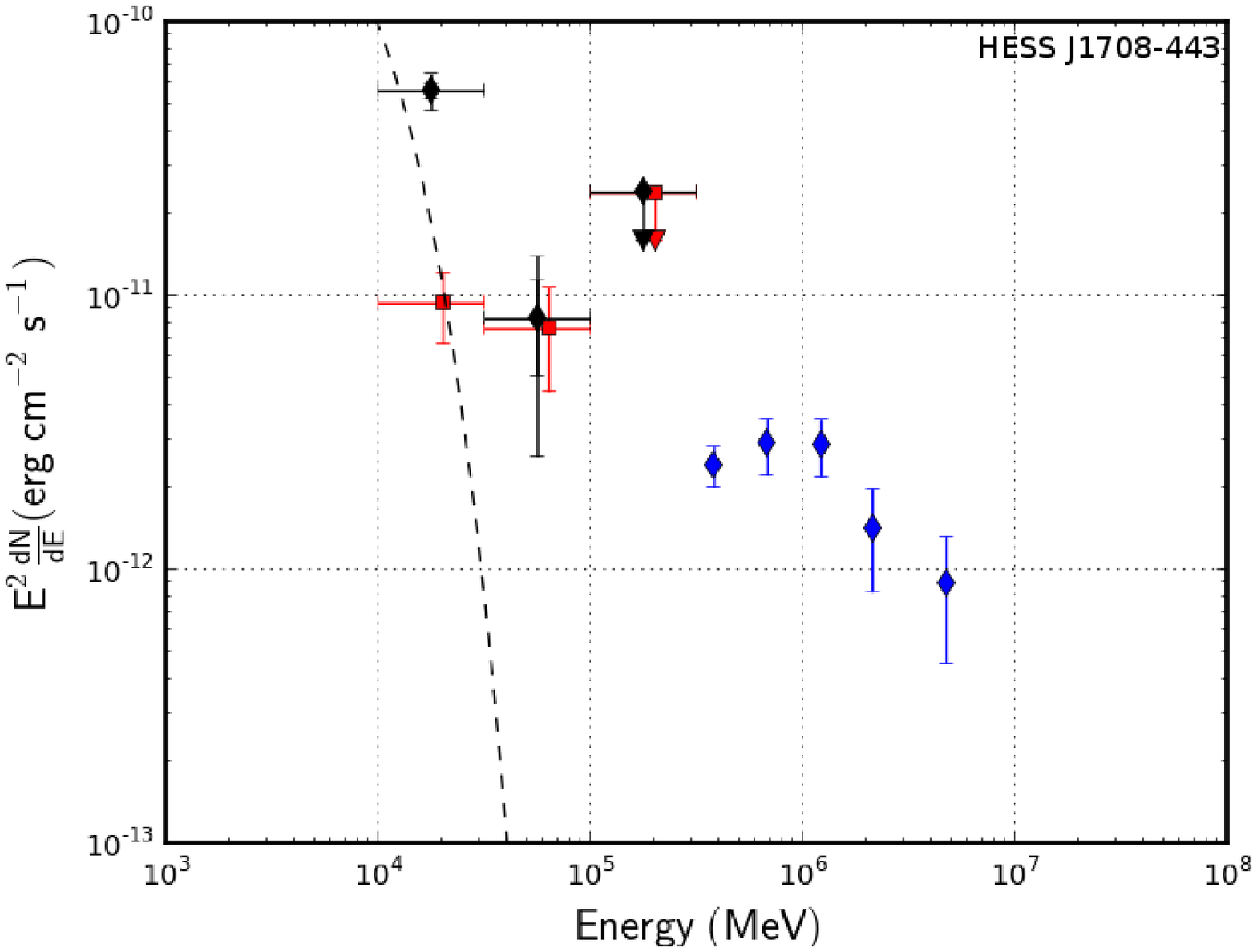}
}
\subfigure{
\includegraphics[width=0.47\textwidth]{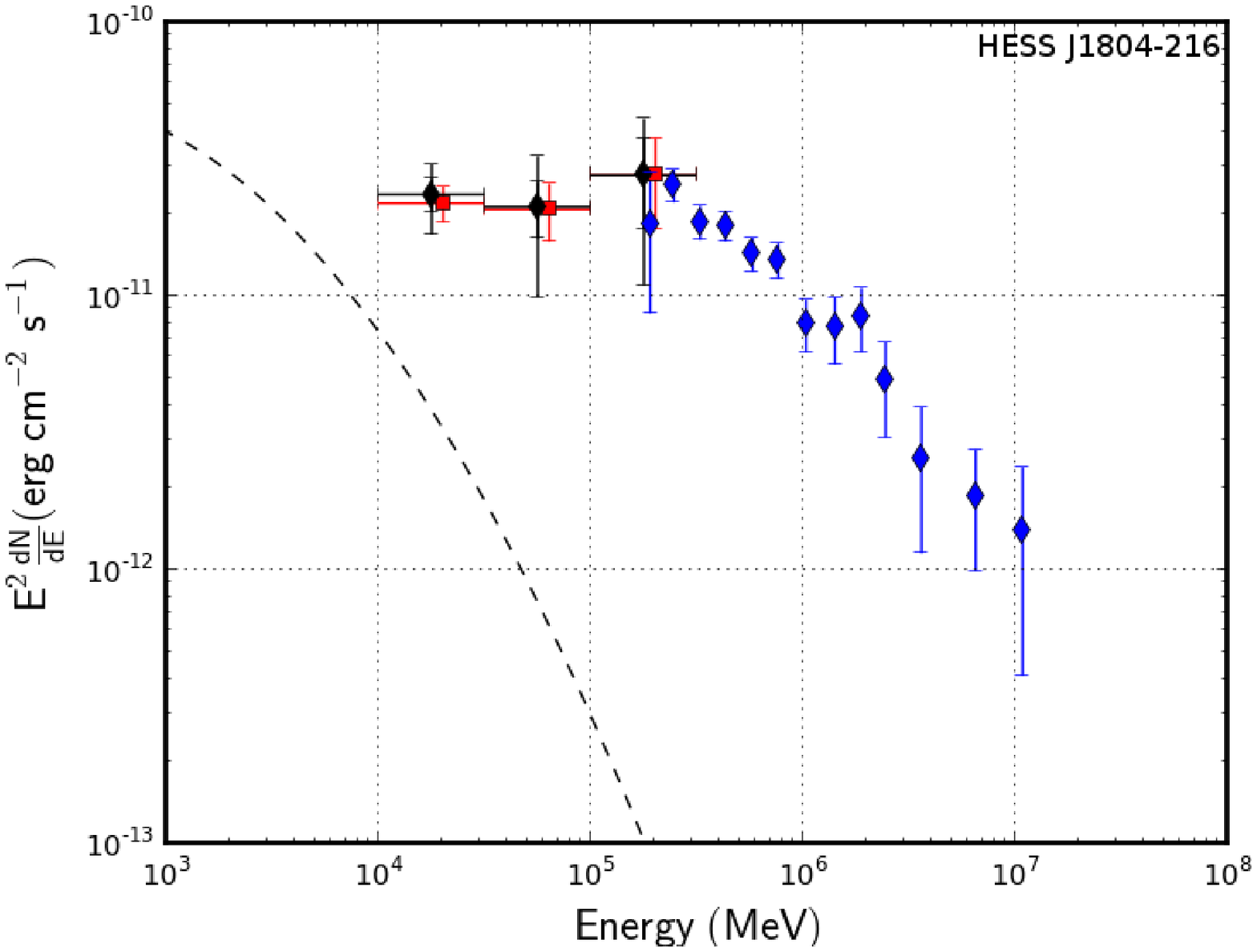}
}
\subfigure{
\includegraphics[width=0.47\textwidth]{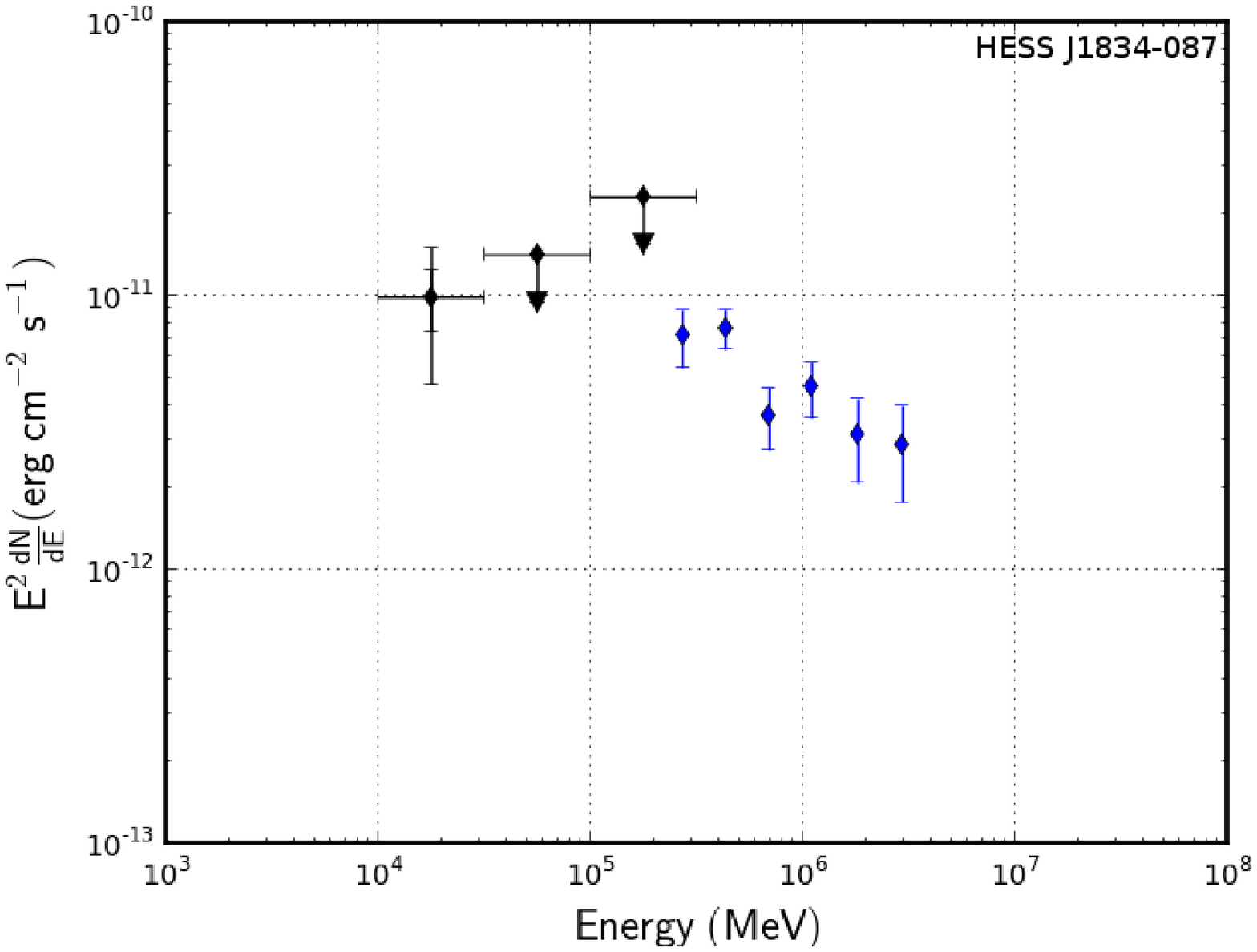}
}
\caption{\label{fig:hessj1718}\label{fig:1626}\label{fig:hess1708}The LAT and VHE SEDs of HESS~J1614$-$518, HESS~J1626$-$490, HESS~J1634$-$472, HESS~J1708$-$443, HESS~J1804$-$216 and HESS~~J1834$-$087 following the conventions of Figure~\ref{fig:mgroj0632}. The solid line corresponds to the model proposed by \cite{2011ICRC....7...44E} for HESS~J1626$-$490.}
\end{figure}

\begin{figure}[h!]
\centering
\subfigure{
\includegraphics[width=0.47\textwidth]{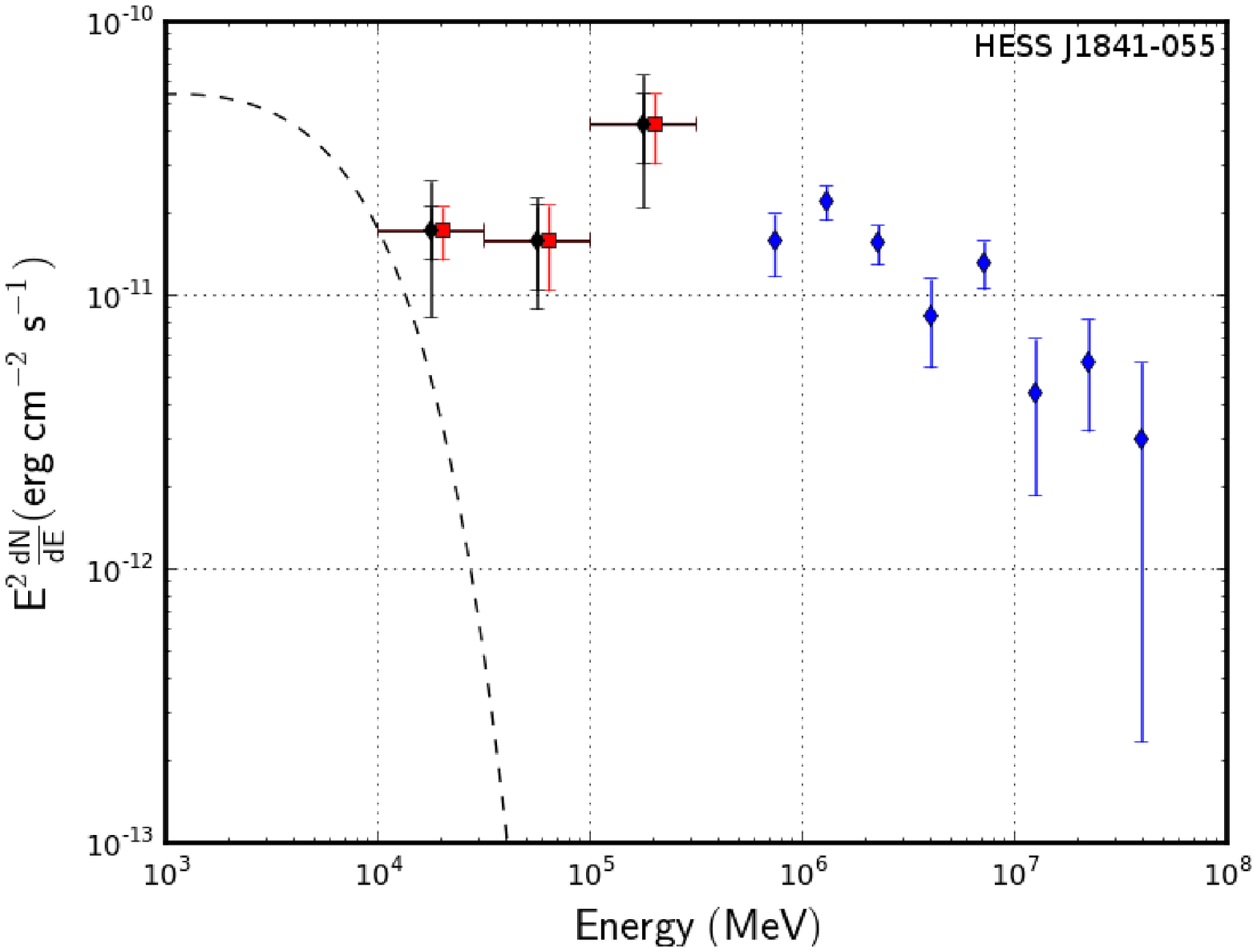}
}
\subfigure{
\includegraphics[width=0.47\textwidth]{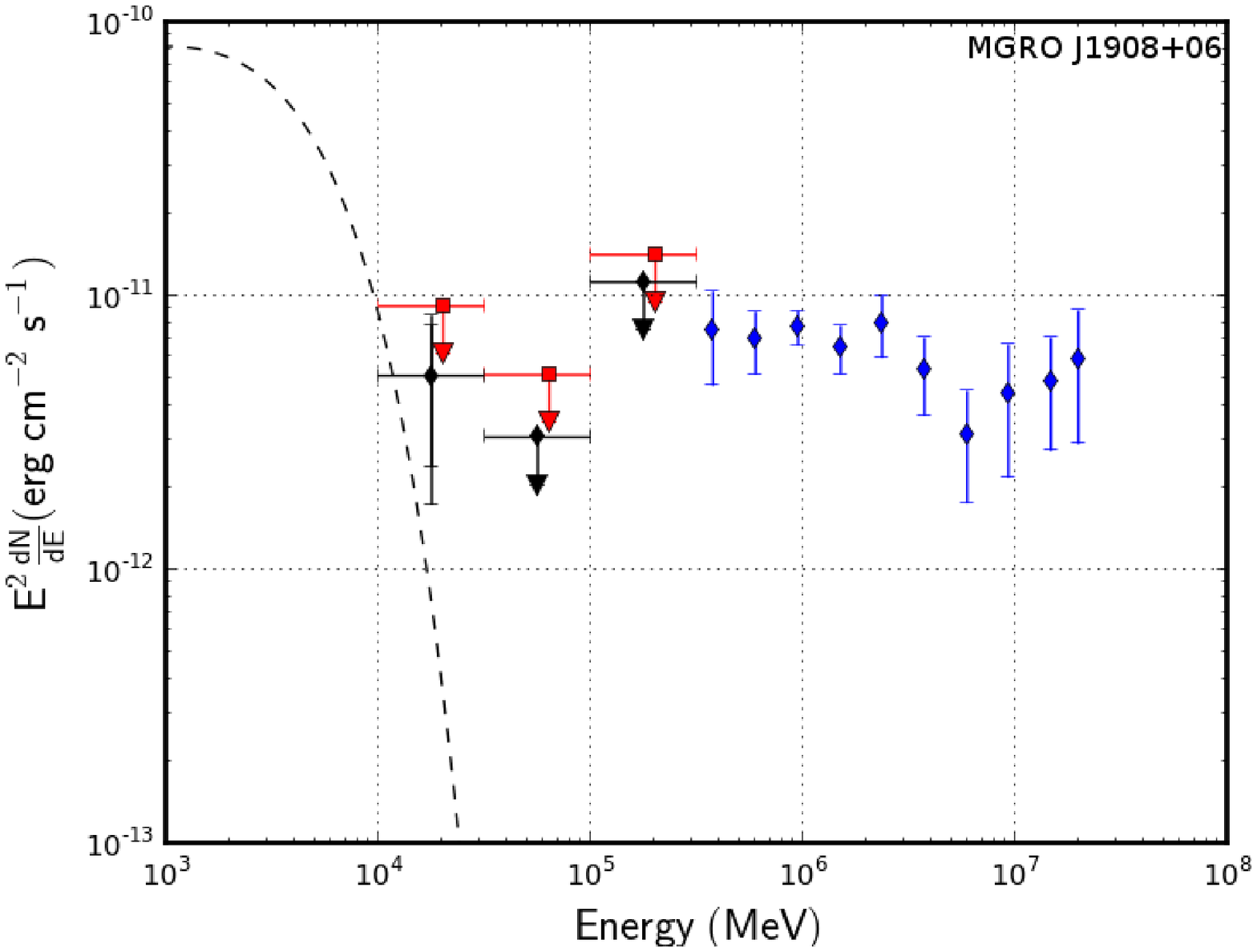}
}
\subfigure{
\includegraphics[width=0.47\textwidth]{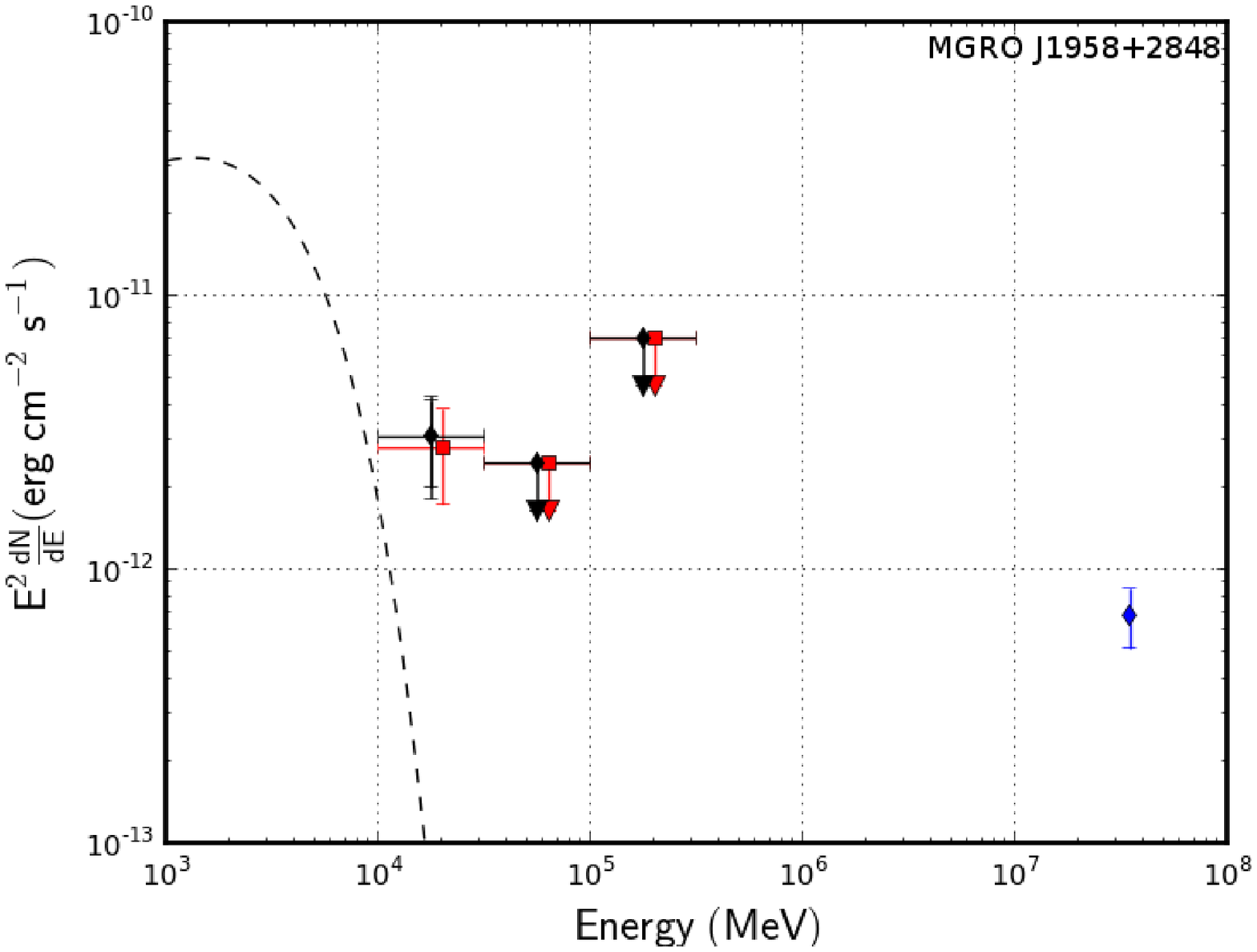}
}
\subfigure{
\includegraphics[width=0.47\textwidth]{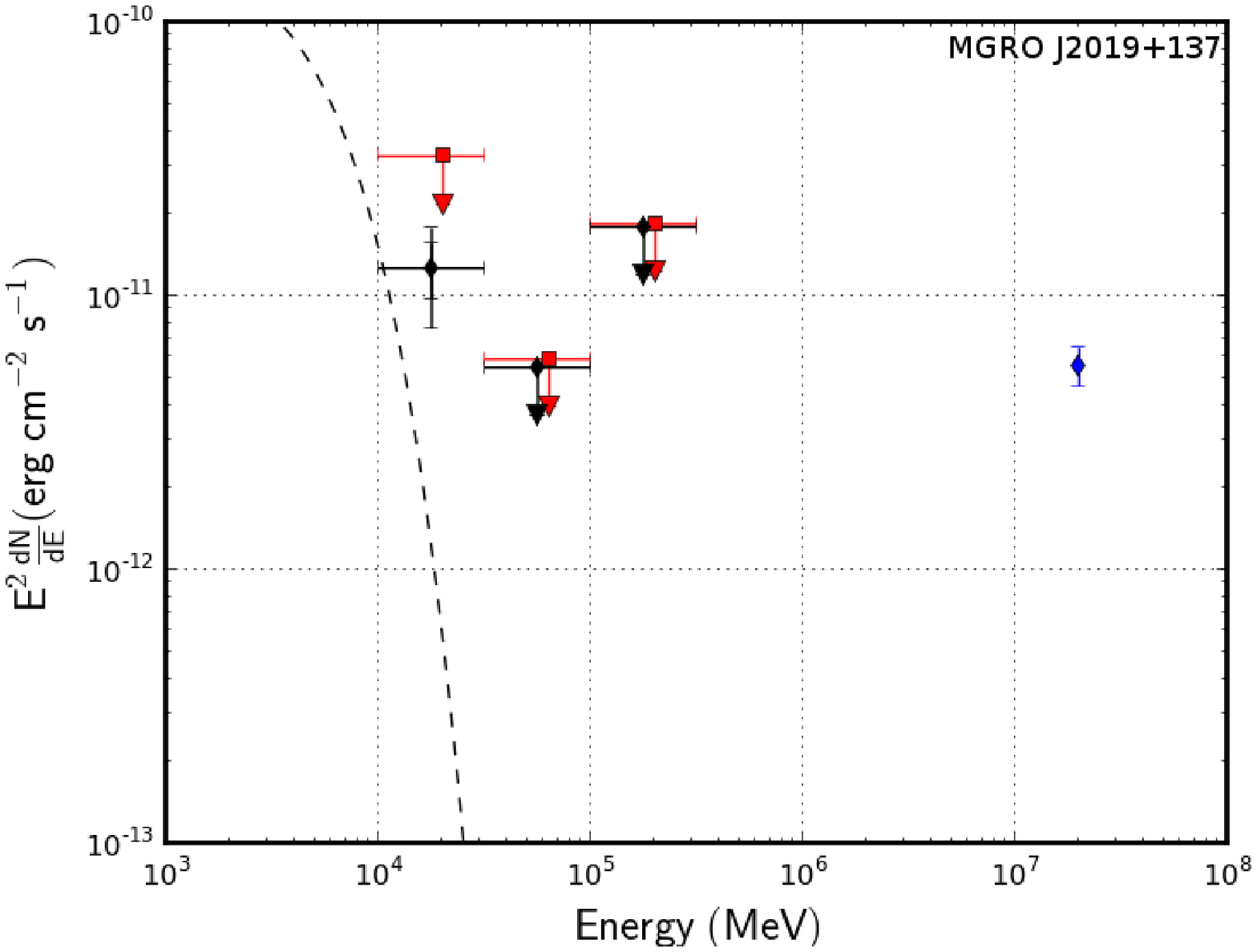}
}
\subfigure{
\includegraphics[width=0.47\textwidth]{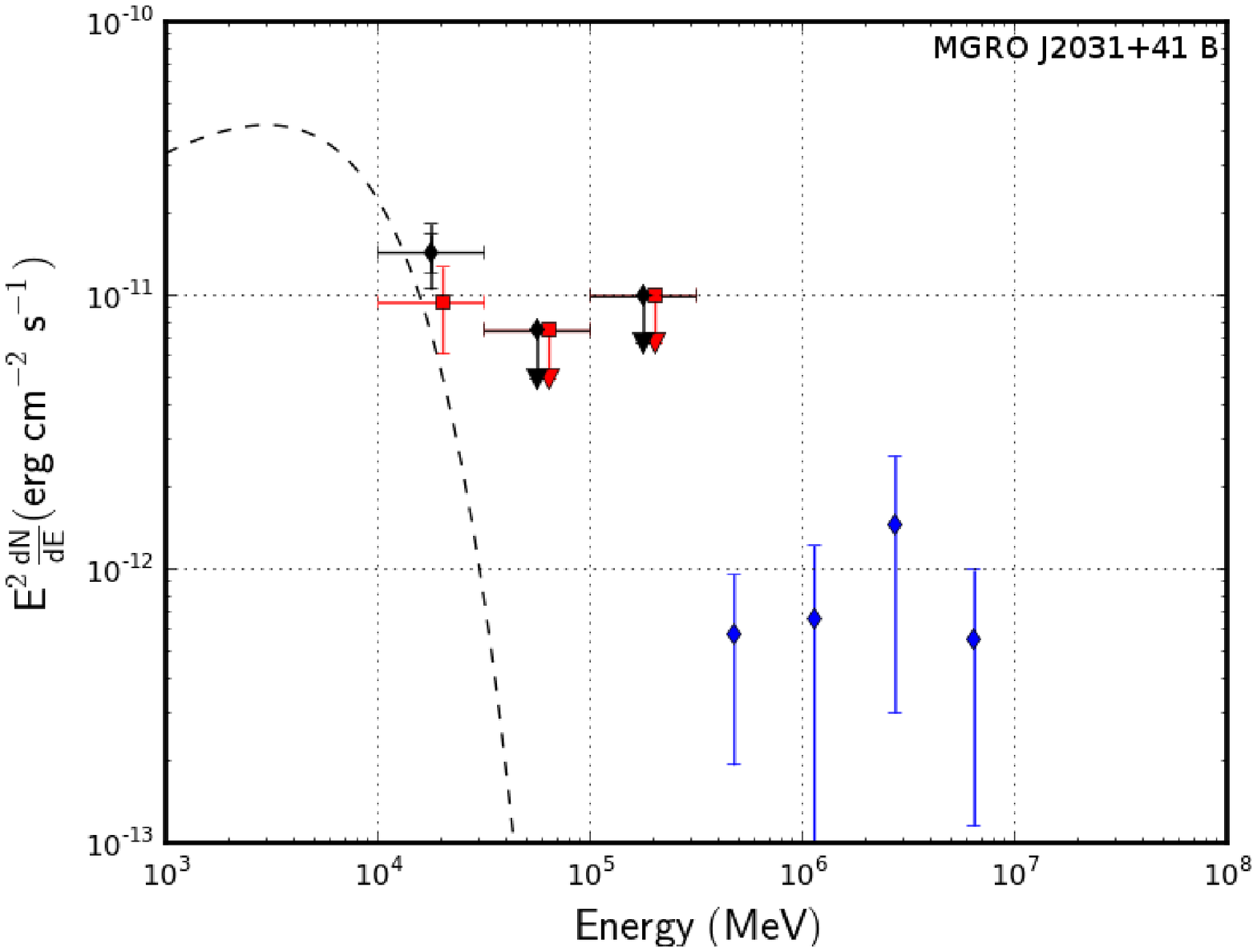}
}
\subfigure{
\includegraphics[width=0.47\textwidth]{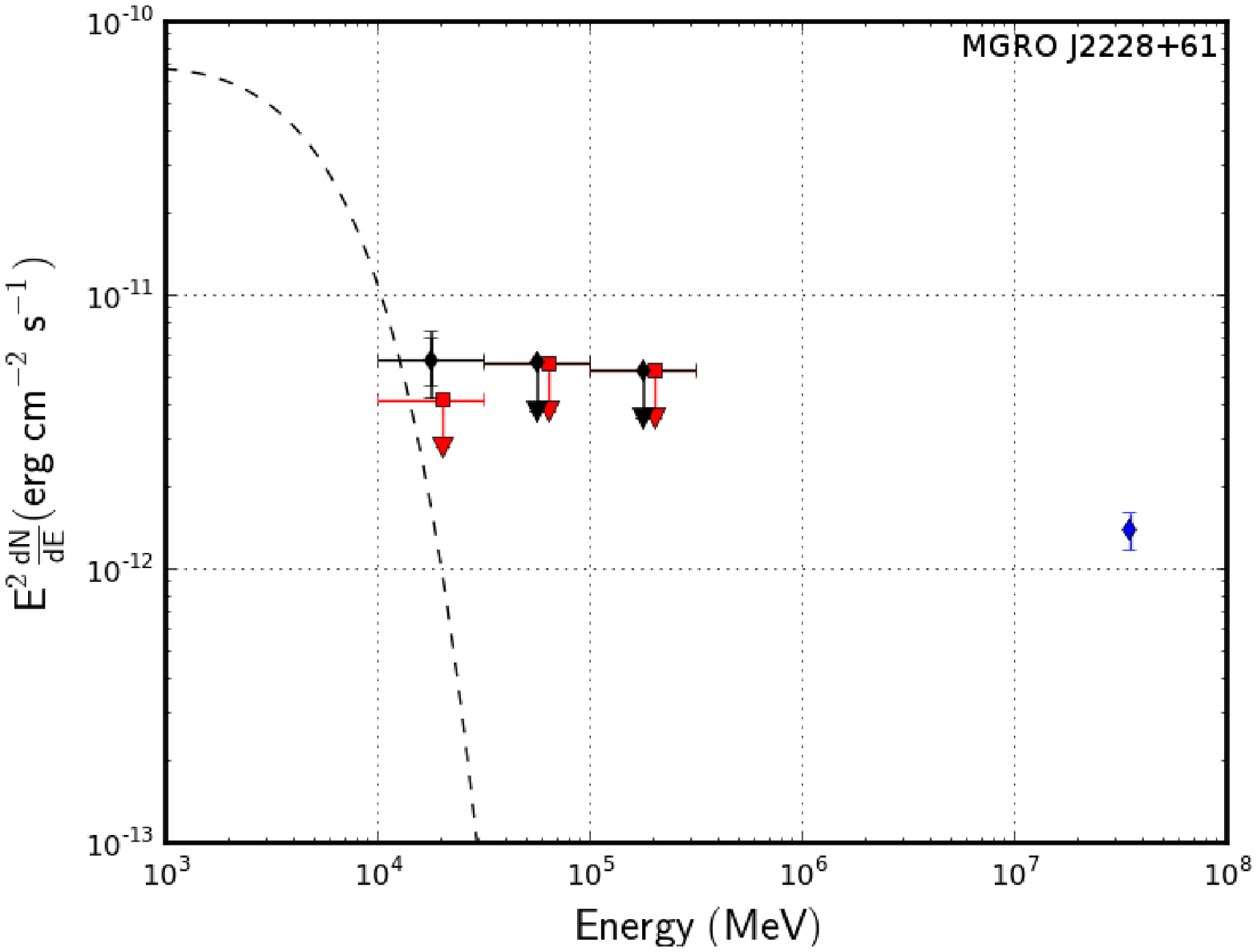}
}
\caption{\label{fig:1841}\label{fig:mgro2019}The LAT and VHE SEDs of HESS~J1841$-$055, MGRO~J1908$+$06, MGRO~J1958$+$2848, MGRO~J2019$+$137, MGRO~J2031$+$41B and MGRO J2228$+$61 following the conventions of Figure~\ref{fig:mgroj0632}.}
\end{figure}

In the LAT energy range, 22 sources are found to be point-like, i.e. not significantly extended. Table~\ref{tab:Det_results} demonstrates that among these 22 point-like sources, nine have a soft spectral index ($\Gamma>3$):  HESS~J1418$-$609, HESS~J1708$-$443, MGRO~J0632+17, MGRO~J1908$+$06, MGRO~J1958$+$2848, MGRO~J2019$+$37, MGRO~J2228+61, MGRO~J2031+41B and VER~J0006+727. Tables~\ref{tab:pulsars} and \ref{tab:CompGeVpuls} demonstrate that all of them are located close to a LAT-detected pulsar. Table~\ref{tab:Det_results} shows that their $\gamma$-ray emission significantly decreases when the contribution of the associated LAT-detected pulsar is included. Furthermore, the spectra obtained above 10$\,$GeV for these sources agree with the nearby pulsars' spectra given in the 2FGL catalog as shown in Figures~\ref{fig:sedsourcespuls}, \ref{fig:1026}, \ref{fig:hessj1718} and \ref{fig:1841}. These figures only show SEDs for sources with no X-ray and radio information available. The spectra for the other sources will be discussed in Section~\ref{PWNc}.


\clearpage

\tabletypesize{\scriptsize}
\begin{deluxetable}{llll}
\tablewidth{0pt}
\tablecaption{Comparison between the LAT emission and pulsar positions.
\label{tab:CompGeVpuls}}
\tablehead{ \colhead{Name} & \colhead{$\text{Unc}_{\text{GeV}}$} & \colhead{$\text{Unc}_{\text{PSR}}$} & \colhead{distance}\\
\colhead{} & \colhead{(arcmin)} & \colhead{(arcmin)} & \colhead{(arcmin)}}
\startdata
VER~J0006$+$727 & 1.2 & 0.4 & 1.2\\
MGRO~J0632$+$17 & 0.6 & 0.2 & 0.6\\
HESS~J1418$-$609 & 1.8 & 0.8 & 2.0\\
HESS~J1708$-$443 & 0.6 & 0.3 & 0.6\\
MGRO~J1908$+$06 & 2.4 & 0.8 & 2.5\\
MGRO~J1958$+$2848 & 2.4 & 1.1 & 1.2\\
MGRO~J2019$+$37 & 1.8 & 0.5 & 1.2 \\
MGRO~J2031$+$41B & 3.6 & 0.8 & 1.2\\
MGRO~J2228$+$61 & 1.8 & 0.6 & 0.6\\
\enddata\\
\tablecomments{For the nine sources classified as "PSR", a comparison of the localization of the source in the 10 GeV to 316 GeV energy range with the localization in 2FGL from 100 MeV to 100 GeV. Columns 2, 3 and 4 respectively give the 68\% uncertainty on the point source position obtained in this work, the averaged 68\% uncertainty on the 2FGL position and the distance between the 2FGL position and the position obtained in this work. The average on the 2FGL position uncertainty is obtained as $\sqrt{a \times b}$ where $a$ and $b$ are the length of the semi-minor and semi-major axes of the 68\% confidence ellipse defined in 2FGL. }
\end{deluxetable}

\normalsize
\noindent

To study the contamination of the LAT data for the VHE sources by these pulsars, we accounted for the pulsar contribution to photons in the region as explained in Section~\ref{Model}. Tables~\ref{tab:Spat_results} and \ref{tab:Det_results} and Figures~\ref{fig:sedsourcespuls}, \ref{fig:1026}, \ref{fig:hessj1718} and \ref{fig:1841} show the results of this new fit using the 2FGL models of the pulsars. For these nine sources, in Figures~\ref{fig:sedsourcespuls}, \ref{fig:1026}, \ref{fig:hessj1718} and \ref{fig:1841} the low-energy part of the emission ($<$30\,GeV) tends to disappear, suggesting that we primarily detected pulsar emission. For these reasons, we infer that the observed LAT emission coming from these sources is likely due to the pulsars themselves. We labeled these nine sources as ``PSR" in Table~\ref{tab:Spat_results}.

\subsection{PWNe and PWNe Candidates}
\label{PWNeC}
The hardness of the $\gamma$-ray spectrum and the presence of a pulsar energetic enough to power a potential PWN are important criteria for source classification as a PWN. In addition to these criteria, to label a source as a ``PWN" or a ``PWNc" for PWN candidate, we required that its LAT spectrum connects with the VHE spectrum, i.e. that the differential fluxes at the highest LAT energies and in the VHE low energy range be approximately equivalent, such that they can be interpreted as a continuous multi-wavelength spectrum. An exception has been made for HESS~J1848$-$018, which is classified as ``PWNc" as explained in Section \ref{hj1848}. Based on multi-wavelength analyses, we labeled 3 sources as ``PWN": HESS~J1356$-$645, HESS~J1514$-$591 and HESS~J1825$-$137. HESS~J1356$-$645 is considered as ``PWN" based upon the morphology of the source observed in the radio and X-ray domain \citep{2011AA...533A.103H}. HESS~J1514$-$591 is considered as a ``PWN" because of the correlation observed between the H.E.S.S. and X-ray emissions and the observation of pulsar jets in the X-ray domain \citep{2005AA...435L..17A}. And finally,  HESS~J1825$-$137 is considered to be a ``PWN" since \cite{2006AA...460..365A} have shown that it has an energy-dependent morphology in the H.E.S.S. energy range.

Detections of 7 of the PWNe candidates have been previously published. For HESS~J1023$-$575 \citep{2011ApJ...726...35A}, HESS~J1640$-$465 \citep{2010ApJ...720..266S}, HESS~J1857+026 \citep{Rousseau1857}, HESS~J1616$-$508, HESS~J1632$-$478 and HESS~J1837$-$069 \citep{2012arXiv1207.0027L} and HESS~J1848$-$018 \citep{2010AA...518A...8T}, the LAT emission has been discussed in previous work and proposed to be of PWN origin.

We detected 4 new PWNe candidates (HESS~J1119$-$614, HESS~J1303$-$631, HESS~J1420$-$607 and HESS~J1841$-$055) and 1 new PWN (HESS~J1356$-$645). These sources are spatially consistent with pulsars able to power them (respectively PSR~J1119$-$6127, PSR~J1301$-$6305, PSR~J1420$-$6048, PSR~J1838$-$0537 and PSR~J1357$-$6429). Moreover, their LAT emissions are best modeled by hard spectra and are compatible with the VHE SEDs. The new LAT sources and PWNe candidates are described in Section~\ref{PWNc}.

\subsection{``O"-Type Sources}
\label{Osources}

The 7 sources that cannot be associated to a PWN or a pulsar are labeled "O" for Other. These sources cannot be associated because either no sufficiently energetic pulsar is known at that location or the spectral connection from the LAT to the VHE measurements does not support a PWN interpretation.

3 sources have relatively hard spectra at LAT energies that connect spectrally to the associated VHE source: HESS~J1614$-$518, HESS~J1634$-$472 and HESS~J1804$-$216. These sources are not classified as PWNe due to previous multi-wavelength analyses. The former is spatially coincident with 5 detected pulsars, but none is luminous enough to power a PWN that would explain the $\gamma$-ray emission \citep{2008AIPC.1085..241R}. \cite{2011PASJ...63S.879S} found two X-ray counterparts to the H.E.S.S. source and proposed an SNR identification. As discussed in \cite{2012arXiv1207.0027L}, the nature of the source remains unclear. HESS~J1634$-$472 does not have any counterpart pulsars energetic enough to power it. Finally, \cite{2012ApJ...744...80A} studied the link between the H.E.S.S. and the LAT emission coming from the region of HESS~J1804$-$216 and concluded that the emission is more likely due to the energy-dependent scattering of particles accelerated in a SNR than to a PWN.

The remaining 4 sources are HESS~J1018$-$589, HESS~J1507$-$622, HESS~J1834$-$087 and VER~J2016+372. In the LAT energy range, they are point-like with relatively soft spectra ($3 > \Gamma > 2$). Therefore, the association of the LAT emission with their VHE counterparts is uncertain. HESS~J1018$-$589 is close to both the $\gamma$-ray binary 1FGL~J1018.6$-$5856 \citep{2012arXiv1202.3164T} and the nearby SNR~G284.3$-$1.8. The LAT source appears to be spatially coincident with SNR~G284.3$-$1.8. SNR~G284.3$-$1.8 is not included in our list of candidates and will be analyzed in the SNR catalog. \cite{2012AA...545A..94D} were not able to conclusively determine the origin of the $\gamma$-ray emission of HESS~J1507$-$622 and confirm the result of \cite{2011AA...525A..45H}: a leptonic scenario seems favored, while a hadronic one seems unlikely. HESS~J1834$-$087 and VER~J2016$+$372 lack a pulsar energetic enough to power their emission. But \cite{2006ApJ...643L..53A} have suggested that the interaction of SNR~W41 with a nearby molecular cloud could explain HESS~J1834$-$087. Another analysis of LAT and H.E.S.S. observations of HESS~J1834$-$087 will be presented elsewhere. 

\subsection{Descriptions of New LAT PWNe Candidates}
\label{PWNc}

In this section, we compare source spectra with previously published models (the references to those models are provided below for each case). We do not refit those models incorporating the LAT data. Therefore, disagreements between the published models and the LAT data do not necessarily imply that the models are ruled out as it may be possible to accommodate our results under these models with a different set of parameters.

\subsubsection{HESS~J1119$-$614}

During the Parkes multibeam pulsar survey, \cite{2000ApJ...541..367C} discovered PSR~J1119$-$6127, a young ($\tau_C = 1.6\,$kyr) pulsar with a high spin-down power $\dot{E} = 2.3 \times 10^{36}\,$erg$\,$s$^{-1}$ within the composite SNR G292.2$-$0.5. This pulsar was also detected by the LAT \citep{2011ApJ...743..170P}. Using \emph{Chandra} observations, \cite{2003ApJ...591L.143G} and \cite{2008ApJ...684..532S} reported the presence of a faint and compact PWN close to this pulsar. More recently, a H.E.S.S. source coincident with PSR~J1119$-$6127 and G292.2$-$0.5 was announced\footnote{\url{http://cxc.harvard.edu/cdo/snr09/pres/DjannatiAtai\_Arache\_v2.pdf}}.

Using the method described above, a signal is detected at the position of HESS~J1119$-$614 with a TS of 27 (4.9$\,\sigma$ with two d.o.f.). Nevertheless, as can be seen on Figure~\ref{fig:hessj1119} and in Table~\ref{tab:Det_results}, the spectrum is contaminated by a low-energy component associated with PSR~J1119$-$6127. We assess the contamination from the pulsar by subtracting the emission from 2FGL J1118.8$-$6128 associated with PSR~J1119$-$6127. This decreases the significance of the detection to TS=16 (3.6$\,\sigma$ with 2$\,$d.o.f.), which is just above the threshold from Section 3.3. Subtracting the pulsar emission affects the spectrum of our source by making it harder and changing the lowest energy point to become an upper limit (see Figure~\ref{fig:hessj1119}). 

Figure~\ref{fig:hessj1119} shows the SED of HESS~J1119$-$614. Because only an integral flux between 1 and 10$\,$TeV is available for this source \citep{2010AIPC.1248...25K}, we computed a spectral point at 3.16$\,$TeV assuming a fiducial spectral index of 2.4 in the H.E.S.S. energy range. We represented this point on the figure. The leptonic model proposed by \cite{Mayerdiploma} is a one-zone model in which accelerated electrons cool radiatively by IC scattering on CMB photons, stellar photons from the vicinity and photons radiated by dust, and by synchrotron losses. It implies an initial period of the pulsar $P_0 =21.4\,$ms, an initial magnetic field inside the PWN $B_0\,\sim 400\,\mu$G (leading to a current magnetic field of B$\,\sim 32\,\mu$G), a braking index of $n=2.91$ \citep{2000ApJ...541..367C} and a conversion efficiency of rotational energy into relativistic particles of $\eta = 0.3$. The leptonic model matches the new LAT data points.

The energetics of PSR~J1119$-$6127, the detection of a compact PWN in X-rays and the leptonic model proposed by \cite{Mayerdiploma} support the hypothesis in which the LAT-H.E.S.S. emission originates in the PWN inside G292.2$-$0.5. Furthermore, the parameters derived in \cite{Mayerdiploma} and the jet-like morphology in the  X-ray data are reminiscent of the case of MSH 15$-$52 \citep{2010ApJ...714..927A,1996PASJ...48L..33T}.

\subsubsection{HESS~J1303$-$631}

HESS~J1303$-$631 was serendipitously discovered in 2004 \citep{2005AA...439.1013A} during an observation campaign of the binary system PSR~B1259$-$63. It is the first H.E.S.S. source classified as a UNID due to the lack of detected counterparts in radio and X-rays with \emph{Chandra} \citep{2005ApJ...629.1017M}. \cite{dalton1303} found only one plausible counterpart in the vicinity of HESS~J1303$-$631: PSR~J1301$-$6305  with a spin down power of $\dot{E} = 1.70 \times 10^{36}\,$erg$\,$s$^{-1}$. The authors also presented the detection of a very weak X-ray PWN using \emph{XMM-Newton} observations. This, with the energy dependent morphology observed by H.E.S.S. led to the conclusion that HESS~J1303$-$631 is an old PWN offset from the pulsar powering it.

\begin{figure}[h!]
\centering
\subfigure{
\includegraphics[width=0.49\textwidth]{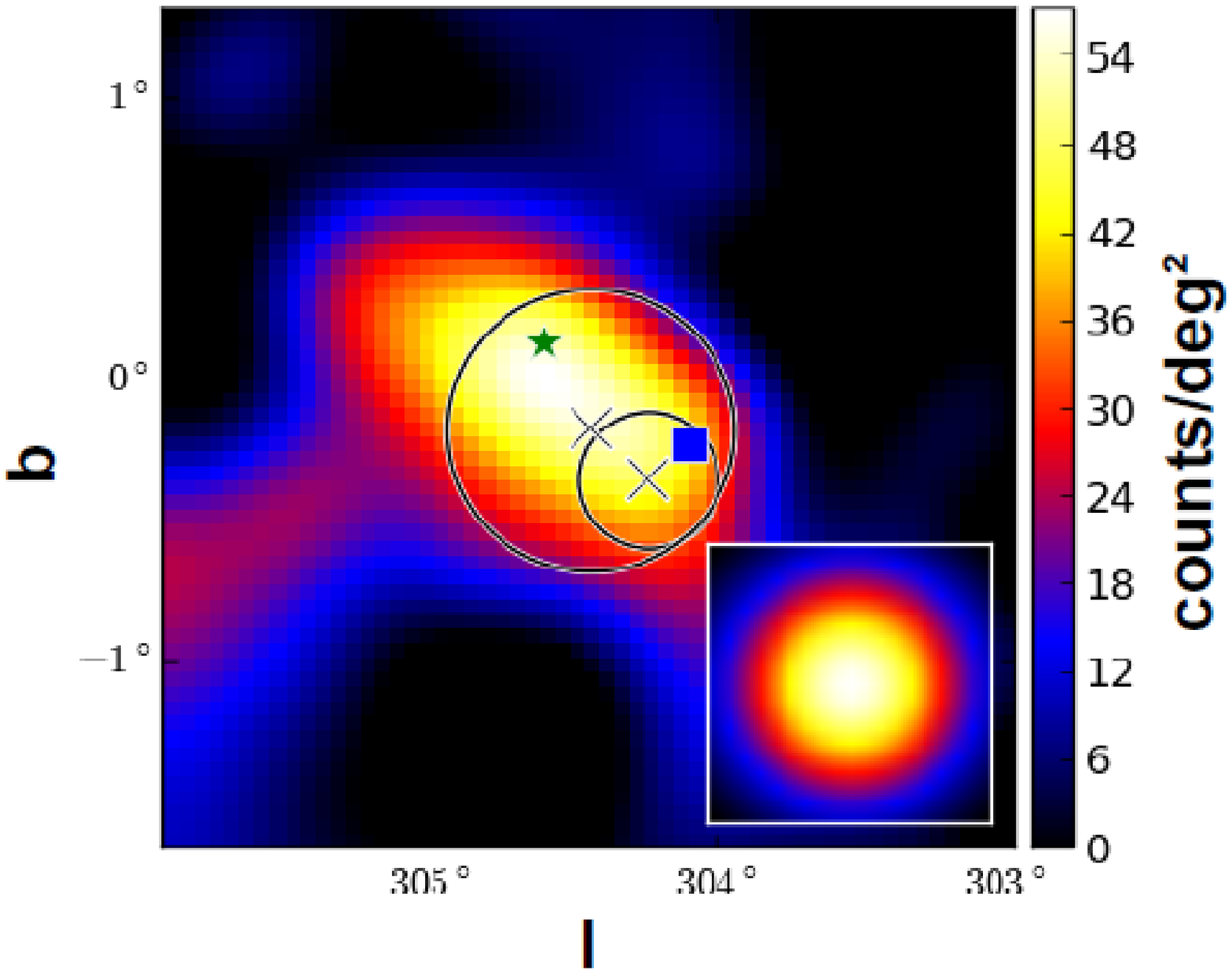}}
\subfigure{
\includegraphics[width=0.49\textwidth]{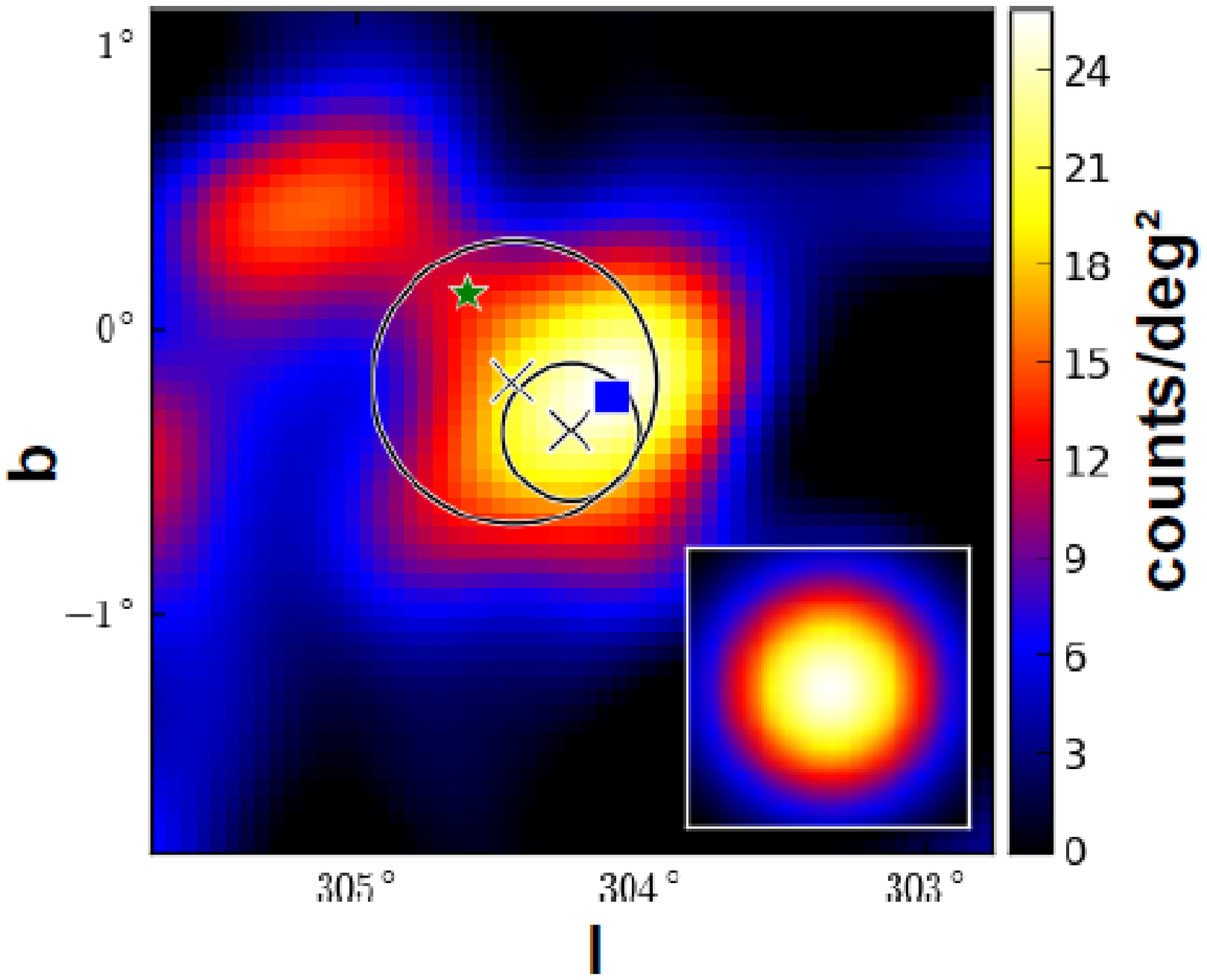}}
\caption{Count map of the region of HESS~J1303$-$631 above 10 GeV (top) and above 31 GeV (bottom). We subtracted the Galactic and isotropic diffuse emission. The count map is smoothed by a Gaussian of $0 \fdg 27$ corresponding to the PSF above 10 GeV. The green star indicates the position of the SNR Kes 17 and the blue square represents the position of PSR~J1301$-$6305. The small and big circles respectively show the extension of the H.E.S.S. Gaussian proposed by \cite{2005AA...439.1013A} and the extension of the Gaussian derived in this work. The lower right inset is the model predicted emission from a point-like source with the same spectrum as HESS~J1303$-$631 smoothed by the same kernel.
\label{1303}}
\end{figure}

\cite{2011ApJ...740L..12W} found no significant emission from HESS~J1303$-$631 using 30 months of \emph{Fermi}-LAT data between 1 and 20$\,$GeV. With 15 months of additional data and a higher energy threshold, our analysis now provides a first detection of LAT emission coincident with the HESS source. Nevertheless, Figure~\ref{1303} shows that the detected emission might be contaminated by $\gamma$-ray emission from the nearby SNR Kes~17. Since we cannot separate these two sources using our strategy with the current statistics, we decided to include the effect of source confusion in our estimate of systematic uncertainties for HESS~J1303$-$631. Therefore, we ran the analysis again, adding a source at the position of Kes~17. We quantify the systematic error as the differences in parameter values resulting from fitting with the two background models. The maximal variation is in the lowest energy bin of the SED.

\begin{figure}[h!]
\begin{center}
\begin{tabular}{cc}
\includegraphics[width=0.47\textwidth]{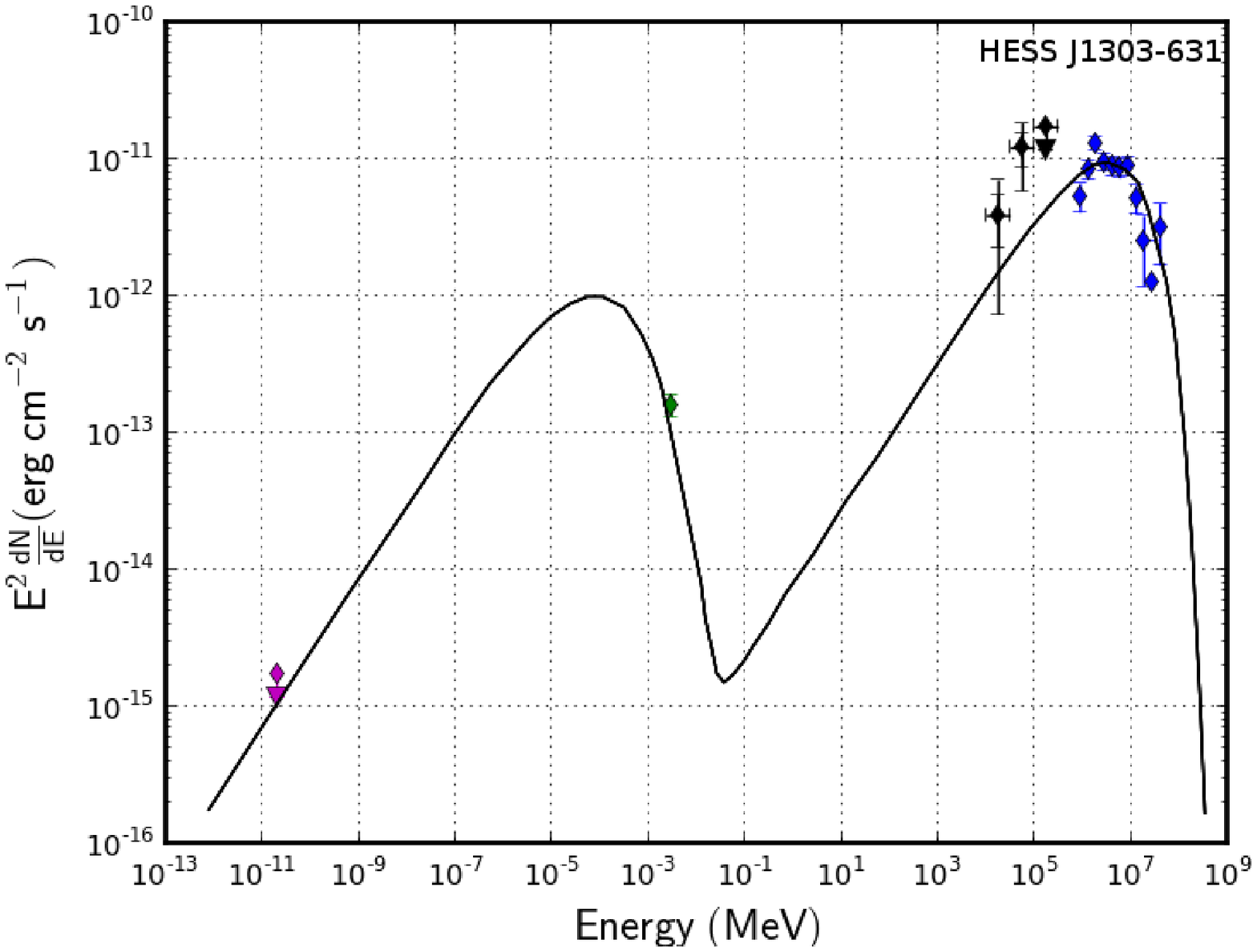} & \includegraphics[width=0.47\textwidth]{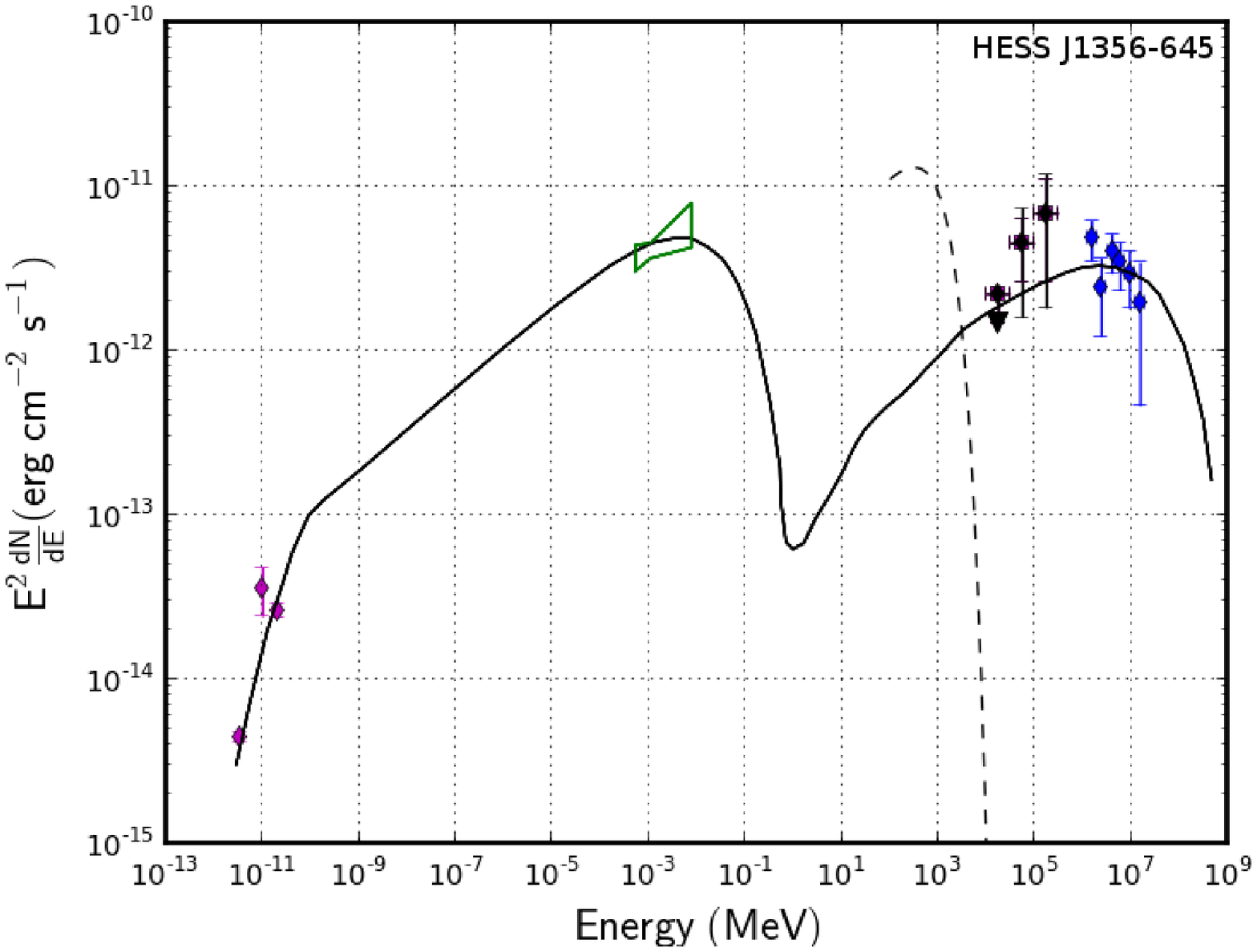}\\
\includegraphics[width=0.47\textwidth]{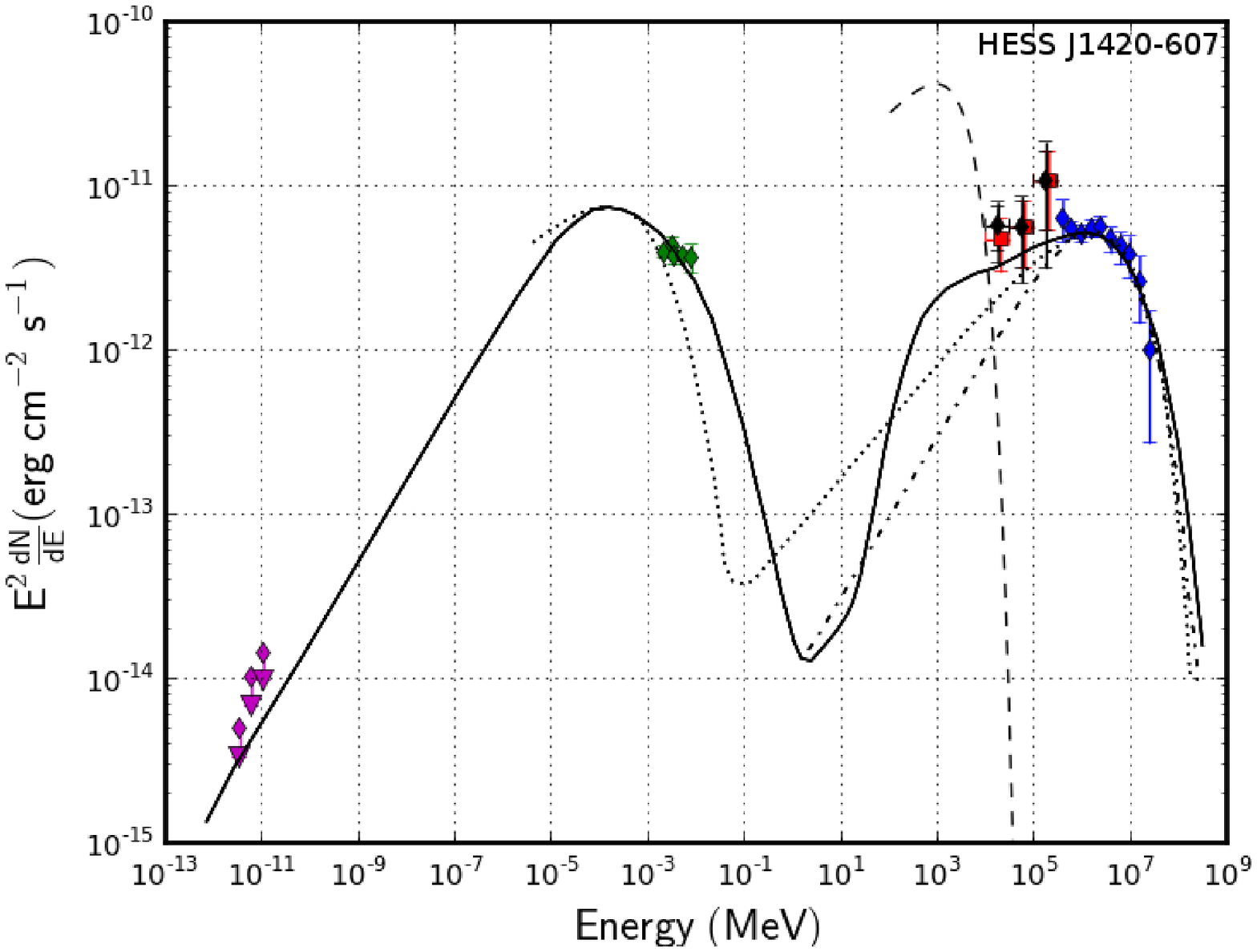} & \includegraphics[width=0.47\textwidth]{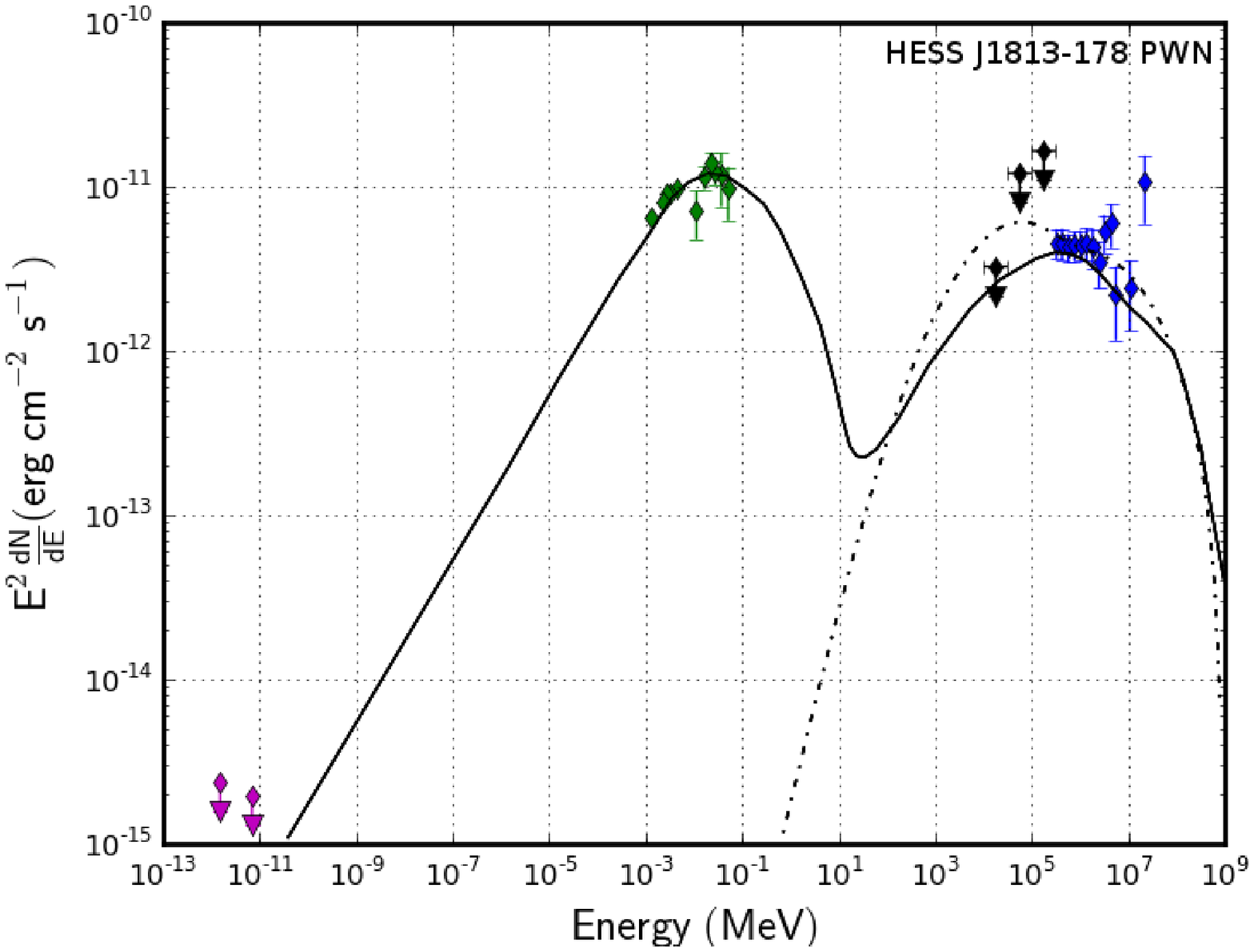}\\
\includegraphics[width=0.47\textwidth]{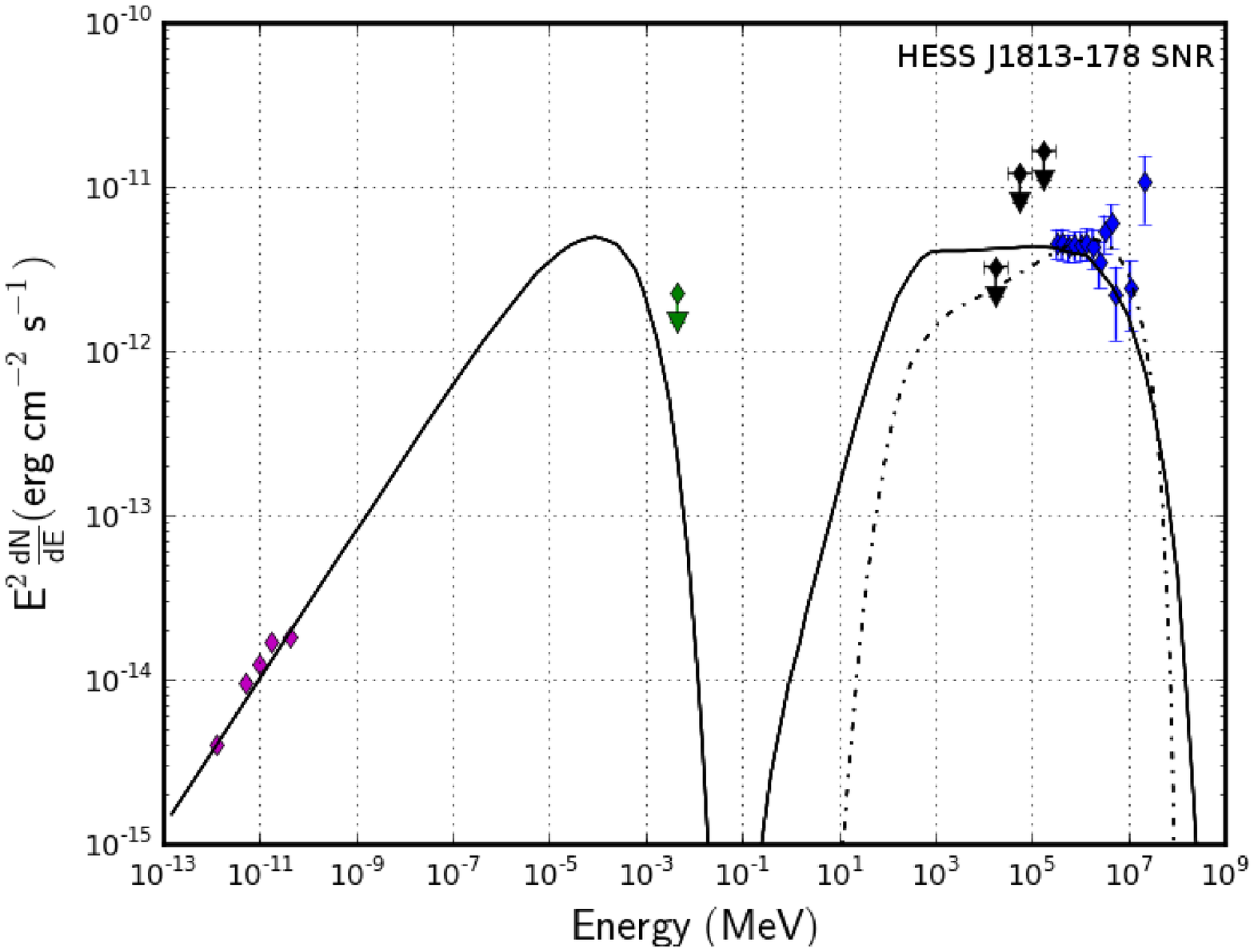} & \includegraphics[width=0.47\textwidth]{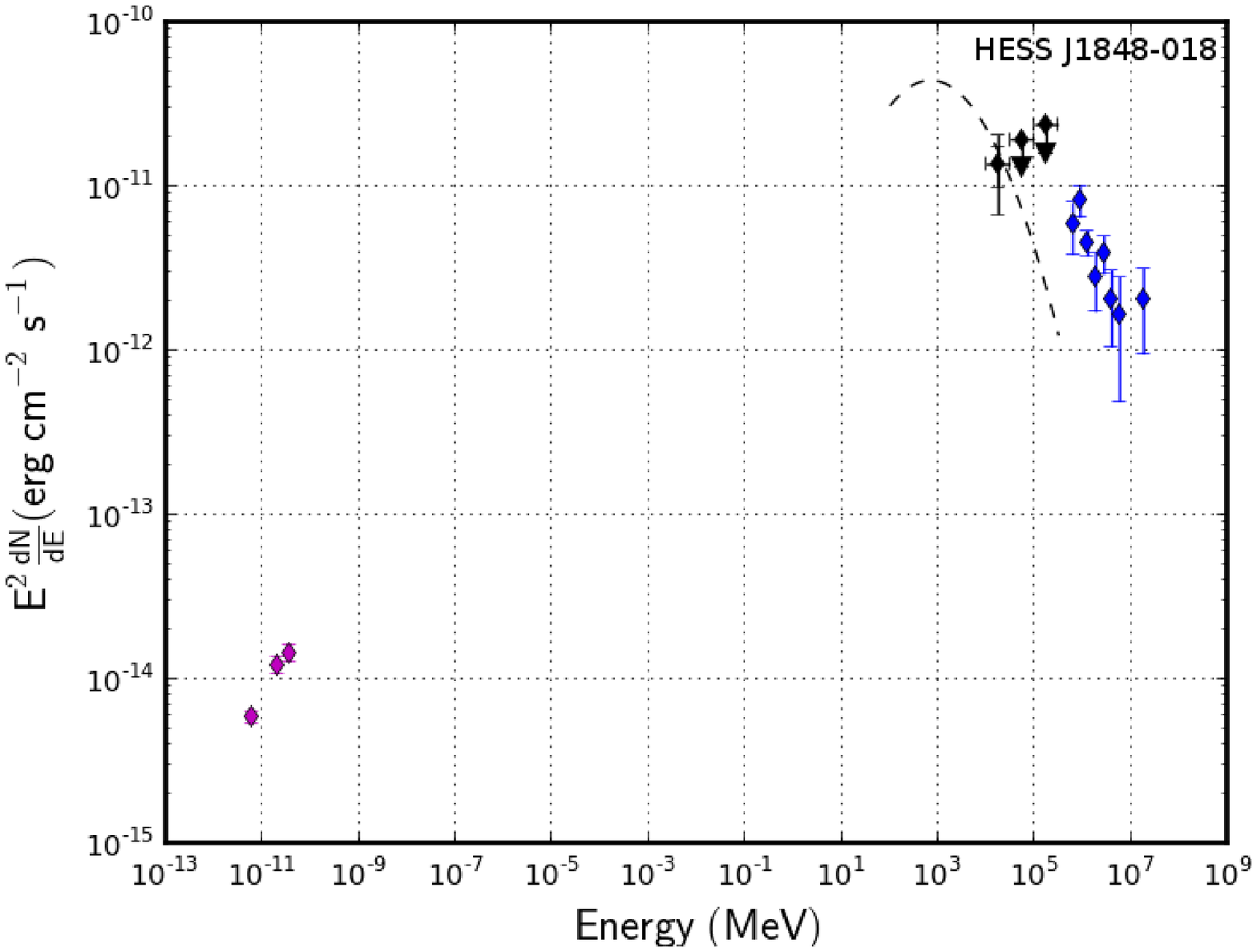}
\end{tabular}
\end{center}
\caption{\label{fig:1848}\label{fig:hessj1813}\label{fig:hessj1420}\label{fig:hess1356}\label{fig:hessj1303}
\scriptsize{Multi-wavelength SEDs of HESS~J1303$-$631, HESS~J1356$-$645, HESS~J1420$-$607, HESS~J1813$-$178 and HESS~J1848$-$018. The blue, green and magenta points represent the H.E.S.S., X-ray and radio spectra respectively. The conventions used for the LAT spectral points are the same as in Figure~\ref{fig:hessj1718}.  The dashed line corresponds to the model of associated pulsars. Details on the models are given in the text. The solid, dotted and dashed-dotted lines represent multi-wavelength models from other publications. Models and references: HESS~J1303$-$631: leptonic model proposed by \cite{dalton1303}. HESS~J1356$-$645: leptonic model proposed by \cite{2011AA...533A.103H}. HESS~J1420$-$607: the solid and dashed lines show the hadronic+leptonic and leptonic models proposed by \cite{2010ApJ...711.1168V} and the dotted line represents the leptonic model proposed by \cite{2012ApJ...750..162K}. HESS~J1813$-$178 PWN: the solid and dashed lines respectively show the leptonic models proposed by \cite{2007AA...470..249F} and \cite{2010ApJ...718..467F}. HESS~J1813$-$178 SNR: the solid and dashed lines respectively show the hadronic models proposed by \cite{2007AA...470..249F} and \cite{2010ApJ...718..467F}.}} 
\end{figure}

For our best LAT morphology of a Gaussian of dispersion $0 \fdg 45$ (see Table~\ref{tab:Spat_results}), we obtained an index of $\Gamma = 1.71 \pm 0.26 \pm 0.37$ (we obtained $\Gamma=1.53 \pm  0.23 \pm 0.37$ assuming the H.E.S.S. best fit Gaussian, Table~\ref{tab:Det_results}). This hard index is in the range of values obtained for LAT detected PWNe and is inconsistent with the spectral index of $\sim 2.4$ derived by \cite{2011ApJ...740L..12W} for Kes~17. This is an evidence that the $\gamma$-ray emission above 10$\,$GeV is dominated by the PWNe candidate. As can be seen in Figure~\ref{fig:hessj1303}, even though the connection between the LAT and the H.E.S.S. energy range is not perfect,  LAT and H.E.S.S. spectra are not inconsistent, suggesting a potential physical relationship.

Figure~\ref{fig:hessj1303} shows the SED of HESS~J1303$-$631 together with the one-zone leptonic model proposed by \cite{dalton1303}. In this model, VHE $\gamma$-rays are created via IC scattering of electrons on the CMB photons. Infrared and optical target photons are neglected. The model reproduces the radio, X-ray and H.E.S.S. data with an electron spectral index of  $1.8 \pm 0.1$, a cut-off energy of $31^{+5}_{-4}\,$TeV, and an average magnetic field of $1.4 \pm 0.2\,\mu$G. However, the flux predicted in the LAT energy range is well below the flux detected by the \emph{Fermi}-LAT. This may be due to the absence of infrared and optical photon fields in the model described above or to the contamination produced by Kes 17. A specific analysis is needed to make conclusions about the constraints that the \emph{Fermi}-LAT could yield on the $\gamma$-ray emission of this source. 

\subsubsection{HESS~J1356$-$645}

HESS~J1356$-$645 is an extended source detected by H.E.S.S. during the Galactic Plane Survey~\citep{2011AA...533A.103H}. It lies close to the pulsar PSR~J1357$-$6429 discovered during the Parkes multibeam survey of the Galactic Plane \citep{2004ApJ...611L..25C}. Its high spin-down power of $\dot{E} = 3.1 \times 10^{36}\,$erg\,s$^{-1}$ makes it a good candidate to power a PWN. Analysis of archival radio and X-ray data from \emph{ROSAT/PSPC} and \emph{XMM/Newton} have revealed a faint extended structure coincident with the VHE emission~\citep{2011AA...533A.103H} providing another argument in favor of the PWN scenario. In parallel, \cite{2011AA...533A.102L} announced the detection of a pulsed signal from PSR~J1357$-$6429 in the $\gamma$-ray and X-ray energy ranges using \emph{Fermi}-LAT and \emph{XMM-Newton} data. However, using 29 months of LAT data between 0.1 and 100$\,$GeV, no counterpart to the H.E.S.S. emission was found in the off-pulse window of the pulsar.

The 16 additional months of observations by \emph{Fermi}-LAT and the higher maximum energy used (316$\,$GeV instead of 100$\,$GeV) in our dataset now enables the detection of a faint counterpart to the H.E.S.S. emission with a TS = 24 (4.7$\,\sigma$ assuming two d.o.f.). 

With its spectral cutoff at low energy, $\sim$800$\,$MeV \citep{2011AA...533A.102L}, PSR~J1357$-$6429 is not significant in the 10$\,$GeV to 316$\,$GeV energy range. Therefore, we do not expect to see any changes in the spectral parameters when adding PSR~J1357$-$6429 to the model of the region. This is verified in Table~\ref{tab:Det_results} as well as in Figure~\ref{fig:hess1356}. 

The combined LAT-H.E.S.S. data in Figure~\ref{fig:hess1356} provide new information concerning the spectral shape of the $\gamma$-ray emission. The spatial and spectral consistency between the LAT and H.E.S.S. emission suggests a physical relationship, leading to the assumption that the LAT and the H.E.S.S. emission are due to the same object. Assuming that the $\gamma$-ray signal comes from the PWN powered by PSR~J1357$-$6429, \cite{2011AA...533A.103H} proposed a leptonic scenario (black curve) which provides an excellent fit of the new multi-wavelength data. This one-zone model is based on the evolution of an electron population injected with an exponentially cutoff power-law spectrum of index 2.5 and cut-off energy of 350$\,$TeV. These electrons cool radiatively through IC scattering on the CMB, Galactic infrared (T$\sim\,$35$\,$K and 350$\,$K) and optical (T$\sim\,$4600$\,$K) photons and through synchrotron emission in a magnetic field $\sim\,$3.5$\,\mu$G.

The similarities between PSR~J1357$-$6429 and the Vela pulsar and between their PWNe led \cite{2011AA...533A.103H} to discuss two leptonic emission components. In the case of Vela$-$X, the ``halo" is seen in the LAT and radio energy ranges and the ``cocoon" in the H.E.S.S. and X-ray energy ranges, while in the case of HESS~J1356$-$645, a single lepton population explains the broad-band spectrum with reasonable parameters. Unlike Vela$-$X , PSR~J1357$-$6429 is very faint in radio and X-ray, and observations in these bands do not provide morphological constraints. Future multi-wavelength data are greatly needed to better describe this source.

\subsubsection{HESS~J1420$-$607}

The complex of compact and extended radio/X-ray sources, called Kookaburra~\citep{1999ApJ...515..712R}, spans over one square degree along the Galactic Plane. It has been extensively studied to explain the EGRET source 3EG~J1420$-$6038/GEV~J1417$-$6100 \citep{1999ApJS..123...79H,1997ApJ...488..872L}. Within its North-East excess, designated `K3', the young and energetic pulsar PSR~J1420$-$6048 with period 68$\,$ms, characteristic age $\tau_C = 13\,$kyr, and spin down power $10^{37}\,$erg$\,$s$^{-1}$ was discovered \citep{2001ApJ...552L..45D}. X-ray observations by \emph{ASCA} and later by \emph{Chandra} and \emph{XMM-Newton} revealed extended X-ray emission surrounding this pulsar and identified as a potential PWN \citep{2001ApJ...561L.187R,2005ApJ...627..904N}. On the South-West side of the Kookaburra complex lies a bright nebula exhibiting extended hard X-ray emission, G313.1+0.1 \citep[aka the ``Rabbit",][]{1999ApJ...515..712R}. This X-ray excess was also proposed as a PWN contributing to the $\gamma$-ray emission detected by EGRET.

The H.E.S.S. Galactic Plane survey revealed two VHE sources in this region: HESS~J1420$-$607 and HESS~J1418$-$609 \citep{2006AA...456..245A}. HESS~J1420$-$607 is centered north of PSR~J1420$-$6048 (near K3), while HESS~J1418$-$609 is coincident with the Rabbit nebula. More recently, \emph{Fermi}-LAT detected pulsed $\gamma$-ray emission from PSR~J1420$-$6048 and PSR~J1418$-$6058, the latter being a new $\gamma$-ray pulsar found through blind frequency searches \citep{2010ApJS..187..460A, 2009Sci...325..840A}. PSR~J1418$-$6058 is coincident with an X-ray source in the Rabbit PWN and has a spin-down power high enough to power the PWN candidate HESS~J1418$-$609. 

\begin{figure}[h!]
\centering
\subfigure{
\includegraphics[width=0.49\textwidth]{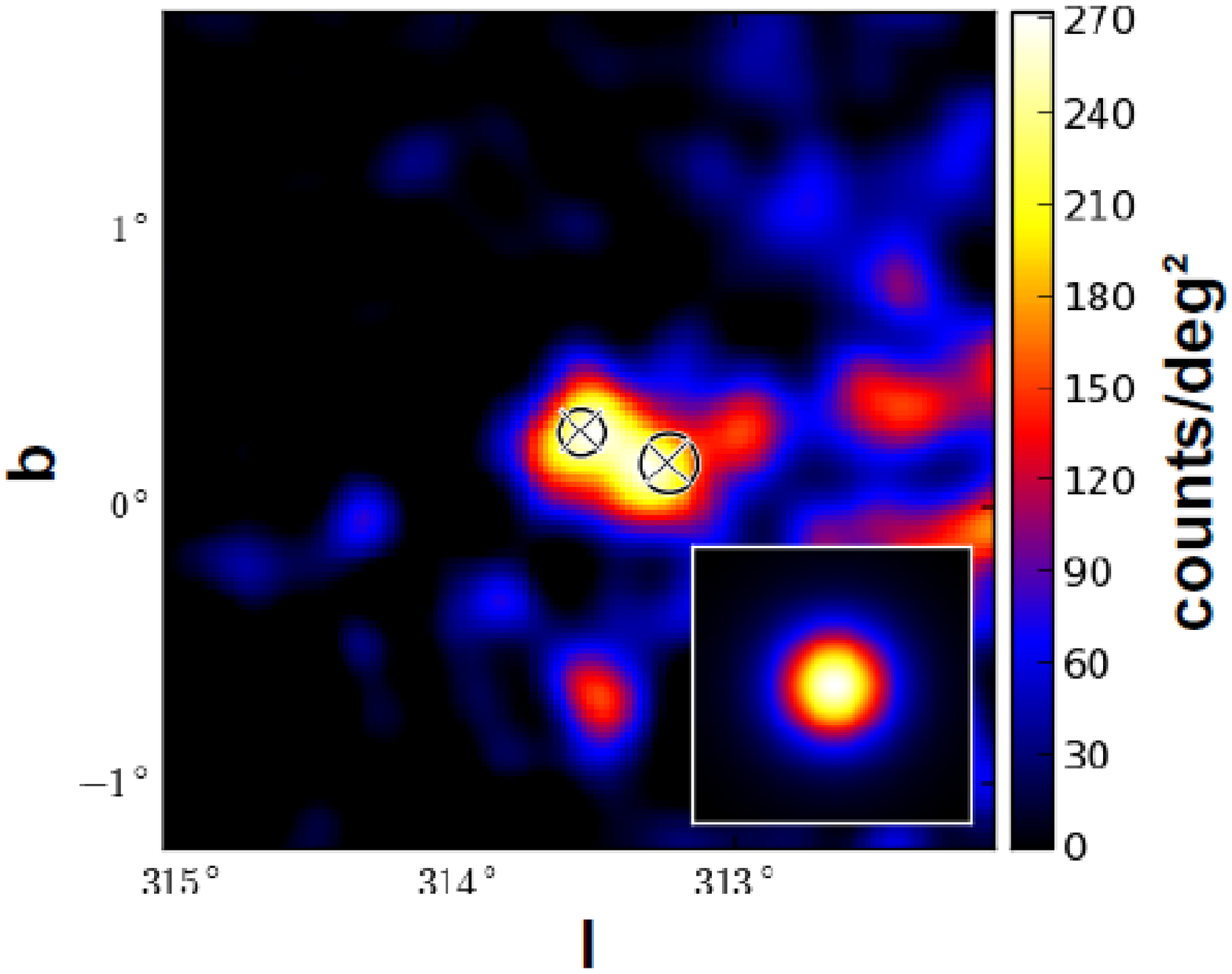}}
\subfigure{
\includegraphics[width=0.49\textwidth]{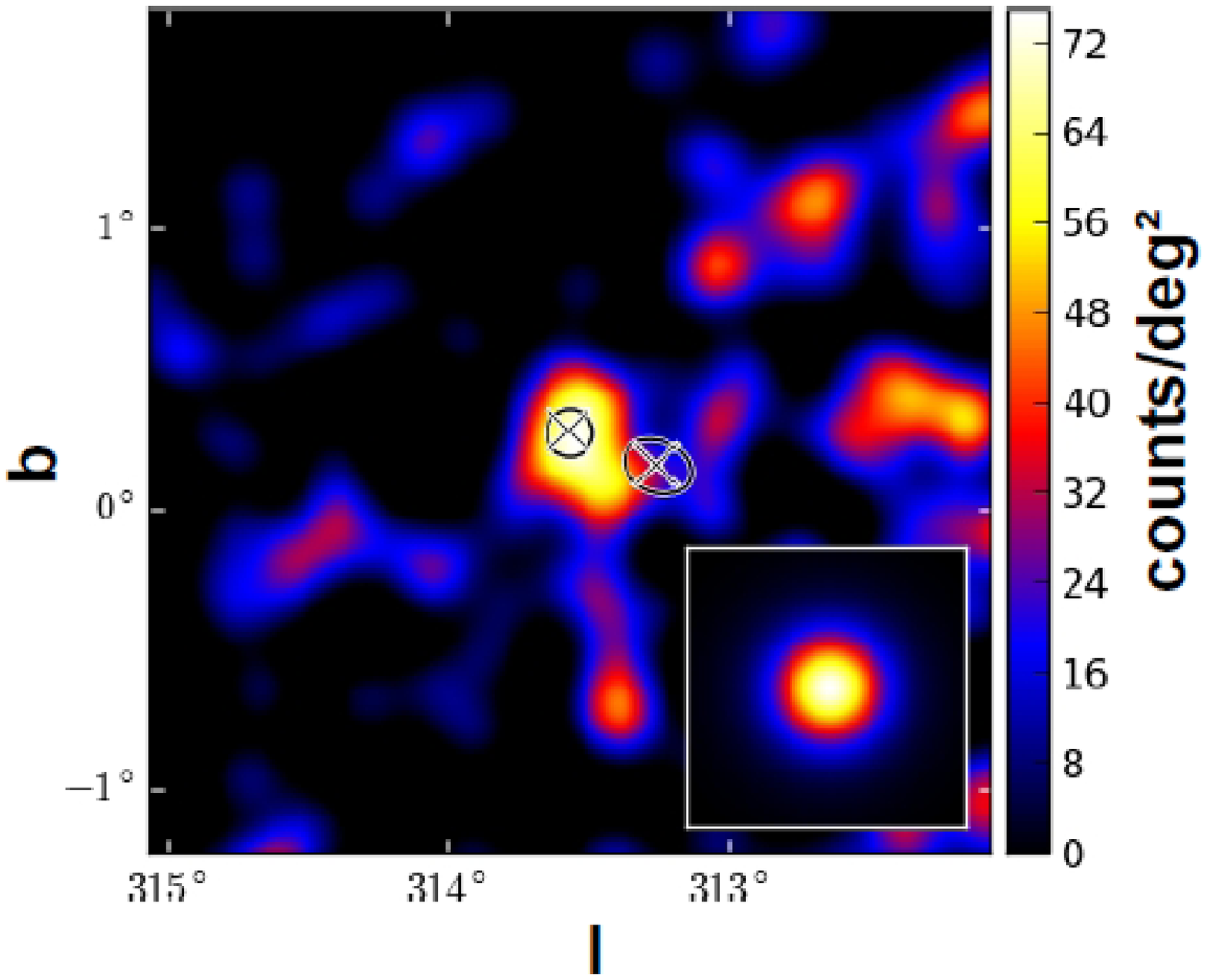}}
\caption{Smoothed count map of the region of the Kookaburra complex observed by \emph{Fermi} above 10 GeV (top) and 31 GeV (bottom). The Galactic and isotropic diffuse emissions are subtracted. The left and right circles show the best fit obtained by VHE experiments respectively for the K3 nebula and the Rabbit nebula. The lower right inset is the model predicted emission from a point-like source with the same spectrum as HESS~J1420$-$607 smoothed by the same kernel. There is significant emission from both regions for energies above 10 GeV but there is only significant emission from HESS~J1420$-$607 for energies above 30 GeV. 
\label{fig:K3countsmap}}
\end{figure}

Figure~\ref{fig:K3countsmap} shows two smoothed count maps centered on the location of the K3 nebula. The Galactic and the isotropic diffuse emission were subtracted to show the excesses coming from HESS~J1420$-$607 and HESS~J1418$-$609. Above 10$\,$GeV, HESS~J1418$-$609 and HESS~J1420$-$607 are confused, due to the size of the LAT PSF and to the limited statistics. However, on the count map above 31$\,$GeV, the emission from HESS~J1418$-$609 disappears, confirming the soft spectrum and the pulsar-like emission coming from HESS~J1418$-$609 and the harder spectrum of HESS~J1420$-$607.

\cite{2010ApJ...711.1168V} used a two-zone time dependent numerical model with constant injection luminosity to investigate the physical properties of HESS~J1420$-$607. The authors injected relativistic particles following a power-law spectrum into the inner nebula zone; they evolved this spectrum over time and injected the resultant spectrum into the outer nebula zone. Figure~\ref{fig:hessj1420} shows the results obtained for a hadronic + leptonic model on the assumption of a low density environment ($n\,\sim$1$\,$cm$^{-3}$) and magnetic fields of 12$\,\mu$G and 9$\,\mu$G respectively in the inner and outer nebula. The same figure presents a leptonic scenario assuming magnetic fields of 12$\,\mu$G and 8$\,\mu$G respectively in the inner and outer nebula. The strength of this magnetic field implies a lepton spectral break at $\sim$100 TeV after evolution in the inner nebula. More recently, \cite{2012ApJ...750..162K} proposed a one-zone leptonic model assuming a power-law injection spectrum with an index of 2.3 with a cut-off at $\sim$40$\,$TeV and a magnetic field of $\sim$3$\,\mu$G; the model SED is shown in the same figure.

As explained in Section 3, the model used for PSR~J1420$-$6048 was fixed to the 2FGL spectrum. Therefore, if this model does not perfectly reproduce the data with 21 additional months, the LAT spectral points of HESS~J1420$-$608 might still be contaminated by the pulsar's emission. With the current statistics, all models reproduce the LAT and H.E.S.S. data reasonably well. A future off-pulse analysis of LAT data for this pulsar performed with more statistics could help discriminate between the models.

\subsubsection{HESS~J1841$-$055}

HESS~J1841$-$055 was discovered during the H.E.S.S. Galactic Plane Survey \citep{2008AA...477..353A} and remained unidentified. The emission is highly extended and shows possibly three peaks, and more than one possible counterparts, suggesting that the H.E.S.S. emission is composed of more than one source. Using \emph{INTEGRAL} data, \cite{2009ApJ...697.1194S} proposed the high-mass X-ray binary system AX J1841.0$-$0536 as a potential counterpart, at least for part of the emission. \cite{2012ApJS..199...31N} and \cite{2012PhRvD..85h3008N} detected three sources coincident with HESS~J1841$-$055. \cite{2011ICRC....6..197T} proposed the association of HESS~J1841$-$055 to an ancient PWN powered by PSR~J1841$-$0524, PSR~J1838$-$0549 or both as each pulsar taken independently would need an efficiency greater than 100\% to solely power a potential PWN associated with the H.E.S.S. source. More recently, the blind search detection of the new $\gamma$-ray pulsar PSR~J1838$-$0537 with \emph{Fermi}-LAT provided another potential counterpart of the H.E.S.S. source. Indeed, assuming a distance of 2$\,$kpc, \cite{2012ApJ...755L..20P} estimated that PSR~J1838$-$0537 is sufficiently energetic to power the whole H.E.S.S. source with a conversion efficiency of 0.5\%, similar to other suggested pulsar/PWN associations \citep{2008ApJ...682L..41H}.

In this work, HESS~J1841$-$055 is detected as a significantly extended source (TS$_{ext}$=32) at a position consistent with the H.E.S.S. source. The LAT extension of $0 \fdg 38 \pm 0 \fdg 09$ comparable with the $0 \fdg 33 \pm 0 \fdg 04$ of the H.E.S.S. source. 

As can be seen in Figure~\ref{fig:1841}, the \emph{Fermi}-LAT spectral points connect to the H.E.S.S. ones. The spatial consistency and spectral connection between H.E.S.S. and LAT emissions suggest a physical relationship. The hard \emph{Fermi}-LAT spectrum detected implies that a curvature must arise between the H.E.S.S. energy range and the LAT energy range. This is typical of most PWNe detected by \emph{Fermi}-LAT and H.E.S.S. that present IC emission peaking at a few hundreds of GeV and would favor the PWN scenario. However, this source is very extended at both wavelengths and could be composed of several $\gamma$-ray sources. Follow-up observations with VHE experiments and continued observation with \emph{Fermi}-LAT are needed to unveil the real nature of HESS~J1841$-$055.

\subsubsection{HESS~J1848$-$018}
\label{hj1848}

HESS~J1848$-$018 was discovered during the extended H.E.S.S. Galactic Plane Survey \citep{2008AIPC.1085..372C}, in the direction of, but slightly offset from, the star-forming region W43 (aka G30.8$-$0.2). The H.E.S.S. emission is characterized by significant extension ($0 \fdg 32 \pm 0 \fdg 02$), a power-law spectrum of index $\sim 2.8$ and an integrated flux above 1$\,$TeV $\sim 2 \times 10^{-12}\,$cm$^{-2}$\,s$^{-1}$. The absence of an energetic pulsar or SNR within $0 \fdg 5$ from HESS~J1848$-$018 favors an association with the star-forming region W43. The only potential counterpart for this source found in radio and X-rays is the Wolf-Rayet star WR 121a.

Located $0 \fdg 2$ from the centroid of HESS~J1848$-$018, WR 121a is a WN7 subtype star, in a binary system \citep{2011AA...532A..92L}, associated with W43 and characterized by extreme mass loss rates. \cite{2008AIPC.1085..372C} also proposed an association of the H.E.S.S. emission with the molecular clouds contained in W43. These molecular clouds could lead to the production of high-energy $\gamma$-rays from the neutral pion decays following \emph{p-p} collisions in the ambient gas.

In the LAT energy range, \cite{2010AA...518A...8T} proposed an association with a spatially coincident source 0FGL~J1848.6$-$0138. \cite{2011MmSAI..82..739L} analyzed the \emph{Fermi}-LAT data around HESS~J1848$-$018 and detected a source with a 3.7$\,\sigma$ evidence for an extension ($\sigma \sim 0 \fdg 3$). This disfavors models in which the LAT emission would be produced by a pulsar alone. However, statistics were not large enough to discriminate between one extended source and several point sources. Moreover, the spectrum was well described by a log normal representation (Eq. \ref{logp}) and the SED was very similar to those obtained for most pulsars detected by \emph{Fermi}-LAT. Therefore, the emission could be a composite of a radio-faint pulsar and an additional source.

In our analysis, HESS~J1848$-$018 is detected as a faint point-like source but consistent with the extension reported by \cite{2011MmSAI..82..739L}. Figure~\ref{fig:1848} includes the H.E.S.S. spectral points from \cite{2008AIPC.1085..372C} and the radio points corresponding to the W43 central cluster from \cite{2011AA...532A..92L}. We note that the point obtained in our analysis is consistent with the dashed curve, which represents the spectrum derived by \cite{2011MmSAI..82..739L}. It is not absolutely clear from this figure whether the LAT and H.E.S.S. spectra have a common or a distinct origin, and future multi-wavelength data would be needed to discriminate between the pulsar/PWN and the massive star formation region hypotheses. 

\subsection{Constraints Obtained from Non-Detections}
\label{nondet}
This section presents the sources for which no $\gamma$-ray emission is detected but the corresponding upper limits constrain the models.

\subsubsection{HESS~J1026$-$582}

HESS~J1026$-$582 was discovered by H.E.S.S. during an improved analysis of the region of HESS~J1023$-$575 \citep{2011AA...525A..46H}. While HESS~J1023$-$575 is close to the energetic pulsar PSR~J1023$-$5746 ($\dot{E}=10^{37}\,$erg\,s$^{-1}$) discovered by the LAT using blind search algorithms \citep{2010ApJ...725..571S}, the authors proposed an association between HESS~J1026$-$582 and PSR~J1028$-$5819 discovered by the Parkes 64-m telescope \citep{2008MNRAS.389.1881K} and also detected by the LAT \citep{2009ApJ...695L..72A}. The proximity of the pulsar suggested a PWN scenario to explain the VHE emission. This hypothesis is supported by the spin-down power of PSR~J1028$-$5819, $\dot{E} = 8.43 \times 10^{35}$\,erg$\,$s$^{-1}$ \citep{2009ApJ...695L..72A}, typical of observed PWNe. \cite{2012AA...543A.130M} observed marginal X-ray emission but could not definitely classify the emission as being characteristic of a PWN. Follow-up observations with \emph{XMM-Newton} and \emph{Chandra} are needed to confirm this detection. 

No significant LAT emission coming from the location of the H.E.S.S. excess is detected in our analysis. The very low TS value of 1.0 with an integrated flux less than $1.6\times10^{-10}\,$ph$\,$cm$^{-2}\,$s$^{-1}$ (see Table~\ref{tab:Not_Det_Res}) gives little hope for a future detection by the LAT. The upper limits in Figure~\ref{fig:1026} show that a rising spectrum is needed in the LAT energy range. This suggests an IC peak at energies higher than 100 GeV consistent with \emph{Fermi}-LAT observations of other PWNe. However, the lack of multi-wavelength data (especially in radio and X-rays) prevents clear identification of this source.

\subsubsection{HESS~J1458$-$608}

PSR~J1459$-$6053, discovered in $\gamma$-rays, is an energetic and older pulsar with a spin-down power of $\dot{E} = 9.2 \times 10^{35}$\,erg\,s$^{-1}$, and a characteristic age of $\tau_C=64$\,kyr \citep{2010ApJS..187..460A}. An X-ray counterpart to PSR~J1459$-$6053 was discovered by \emph{Swift} \citep{2011ApJS..194...17R} and \emph{Suzaku} \citep{Kanaiphd}. HESS~J1458$-$608 was discovered 9.6$^{\prime}$ away from PSR~J1459$-$6053 after a dedicated observation \citep{2012arXiv1205.0719D}. The proximity to the pulsar and the extension of HESS~J1458$-$608 suggested that both objects could be related in a PSR/PWN scenario. In this case, the lack of or the faint X-ray emission could be explained by the system's age.

In our work HESS~J1458$-$608 was not significantly detected above 10\,GeV. Table~\ref{tab:Not_Det_Res} shows that the observed marginal emission comes from the energy bin between 10 and 31\,GeV. Figure~\ref{fig:1458} shows that subtracting the pulsar's contribution does not change the SED. This comes from the fact that the spectrum of the pulsar in the 2FGL catalog above 10\,GeV is negligible compared to the SED. The upper limits computed in the energy bins between 31 and 316\,GeV, where no pulsar emission is expected, show that a change in the slope of the spectrum is needed between the H.E.S.S. and the LAT component. This is consistent with an IC peak above 100\,GeV in the range observed for the PWNe detected with the \emph{Fermi}-LAT. However, the H.E.S.S. spectrum of HESS~J1458$-$608 differs from other PWNe, since it seems to show a hardening at high energy.

\subsubsection{HESS~J1626$-$490}

HESS~J1626$-$490 is another unidentified source detected during the H.E.S.S. Galactic Plane Survey \citep{2008AA...477..353A}. According to a leptonic model, \cite{2011ICRC....7...44E} found no X-ray source bright enough to be consistent with the H.E.S.S. emission using \emph{XMM-Newton} observations. However, the authors suggested that a hadronic scenario based on the interaction of SNR~G335.2+00.1 with a molecular cloud could explain the H.E.S.S. emission. This hypothesis is supported by a density depression in $\text{H}_\text{I}$ that could be explained by a recent event such as a supernova.

With a TS of 1.5, HESS~J1626$-$490 is not detected in our analysis. The model presented in \cite{2011ICRC....7...44E} and shown in Figure~\ref{fig:1626} reproduces the H.E.S.S. SED and predicts emission below the LAT upper limits. A radio or/and X-ray detection of synchrotron emission from a PWN or the detection of a pulsar could call this model into question.

\subsubsection{HESS~J1813$-$178}

HESS~J1813$-$178 was discovered during the H.E.S.S. survey of the Inner Galaxy \citep{2005Sci...307.1938A} and also detected by MAGIC \citep{2006ApJ...637L..41A}. The source was identified as being SNR G12.8$-$0.0 following its radio detection by \cite{2005ApJ...629L.105B}. Using \emph{XMM-Newton} observations, \cite{2007AA...470..249F} detected a complex morphology composed of a point-like source and an extended nebula. The morphological and spectral similarities of the central object with a PWN led \cite{2007AA...470..249F} to propose a PWN/SNR scenario to describe the X-ray sources. This hypothesis was strengthened by the discovery of PSR~J1813$-$1749 \citep{2009ApJ...700L.158G}. This pulsar is one of the most energetic pulsars in our Galaxy with a spin-down power of $\dot{E} = 5.6 \times 10^{37}\,$erg\,s$^{-1}$ but has not yet been detected by the LAT. However, the nature of the H.E.S.S. emission remains unclear as either the SNR or the PWN could produce emission at these energies.

Our analysis yielded a TS of 2.5 with an upper limit on the integrated flux of $2.4 \times 10^{-10}$\,ph\,cm$^{-2}$\,s$^{-1}$ assuming the H.E.S.S. morphology as spatial shape. Figure~\ref{fig:hessj1813} shows the multi-wavelength SED of HESS~J1813$-$178. The upper limits derived using the procedure described in Section~3 show that the spectrum of HESS~J1813$-$178 cannot be flat between the H.E.S.S. and the LAT energy ranges and suggest a peak with an energy cutoff located between the two energy ranges. 

\cite{2007AA...470..249F} and \cite{2010ApJ...718..467F} investigated a leptonic model in which the X-ray core and VHE $\gamma$-ray emission are associated. Both take into account IC scattering on CMB, infrared and near infrared photon fields and synchrotron emission produced with a rather low magnetic field (B $\sim 7\,\mu$G). The main difference between these two models lies in the injected electron population, which follows a power-law spectrum with an index of 2.0 in \cite{2007AA...470..249F} and a Maxwellian + power-law tail spectrum \citep{2008ApJ...682L...5S} with an index of 2.4 in \cite{2010ApJ...718..467F}. 

Both also investigated the possibility for the H.E.S.S. signal to be created by the SNR shell. We overlaid on Figure~\ref{fig:hessj1813} the models proposed by \cite{2007AA...470..249F} and \cite{2010ApJ...718..467F}. The main differences between these two models lies in the injected proton and electron populations which follow power-law spectra with an index of 2.1 in \cite{2007AA...470..249F} and is computed in a semi-analytical nonlinear model in \cite{2010ApJ...718..467F}.

The upper limits derived in our analysis rejected one of the two models proposed in each hypothesis. Therefore, no conclusion can be reached on the nature of the H.E.S.S. emission being a PWN or a SNR. However, in light of the different models discussed one can exclude a model assuming a Maxwellian + power-law tail injection spectrum with an index of 2.4 with the parameters derived in \cite{2010ApJ...718..467F} in the PWN scenario. These upper limits also constrain the hadronic model by rejecting a power-law injection spectrum of 2.1 with the parameters proposed by \cite{2007AA...470..249F}. Therefore, whatever the origin of the $\gamma$-ray emission, the injected spectrum of the primary electrons and protons needs to be relatively hard in order to stay below the \emph{Fermi}-LAT upper limits ($\Gamma \leq 2.1$).

\section{DISCUSSION}
\label{discussion}

In this section, we investigate the correlations of pulsar age and spin-down power with the flux of the associated PWN in the keV, GeV and TeV bands. Table~\ref{tab:table_luminosity} lists the ages, spin-down luminosities, and distances for the associated pulsars or other distance information, and Table~\ref{tab:table_luminosity2} lists the X-ray, GeV and TeV flux (or flux upper limit) for the associated PWNe. As a first step, we studied the relation between the spectra in the VHE energy range and the LAT energy range. We assumed that the emission measured in the TeV and GeV bands comes from the falling and rising edges, respectively, of the IC peak in the SED produced by the same population of electrons and we studied the spectral shape of this IC peak. \cite{2012arXiv1202.1455M} have shown that a negative correlation between the energy of the peak and the pulsar characteristic age would be expected from the evolution of the cooling time with energy. We attempted to find this relation starting from the LAT and VHE data.

\clearpage

\begin{deluxetable}{llccccc}
\tabletypesize{\scriptsize}
\tablecaption{PWNe properties.
\tabletypesize{\scriptsize}
\label{tab:table_luminosity2}}
\tablewidth{0pt}
\tablehead{ \colhead{Name} & \colhead{$\text{G}_{10\,\text{GeV}}^{316\,\text{GeV}}$} & \colhead{$\text{G}_{1\,\text{TeV}}^{30\,\text{TeV}}$} & \colhead{$\text{G}_{2\,\text{keV}}^{10\,\text{keV}}$} & \colhead{$\text{L}_{10\,\text{GeV}}^{316\,\text{GeV}}$} & \colhead{Refs}\\
 \colhead{} & \colhead{$\left(10^{-12}\,\text{erg}\,\text{cm}^{-2}\,\text{s}^{-1}\right)$} & \colhead{$\left(10^{-12}\,\text{erg}\,\text{cm}^{-2}\,\text{s}^{-1}\right)$} & \colhead{$\left(10^{-12}\,\text{erg}\,\text{cm}^{-2}\,\text{s}^{-1}\right)$} & \colhead{$\left(10^{34}\,\text{erg}\,\text{s}^{-1}\right)$} & \colhead{}}
\startdata
VER~J0006+727 &$<\,6.9 $ & \nodata & \nodata &$<\,0.2 $& \nodata \\
Crab & $486 \pm 188$ & $80 \pm 17$ & $21000 \pm 4200$ & $23\pm 15$ & (1)\\
MGRO~J0631+105 &$<\,6.0 $ & \nodata & \nodata &$<\,0.1 $& \nodata \\
MGRO~J0632+17 &$<\,29 $ & \nodata & \nodata &$<\,0.01 $& \nodata \\
Vela$-$X & $134\pm11$ & $79\pm22$ & $54 \pm 11$ & $0.14 \pm 0.02$ & (1)\\
HESS~J1018$-$589 &$6.8 \pm 6.3$ &$0.9 \pm 0.4$ & \nodata &$ 0.7 \pm 1.0 $ & (2) \\
HESS~J1023$-$575 &$27 \pm 12$ &$4.8 \pm 1.7$ & \nodata &$ 2.5 \pm 1.1 $ & (3) \\
HESS~J1026$-$582 &$<\,9.4 $ &$5.9 \pm 4.4$ & \nodata &$<\,0.6 $& (3) \\
HESS~J1119$-$614 &$9.1 \pm 5.2$ &$2.3 \pm 1.2$\tablenotemark{a} & \nodata &$ 7.7 \pm 4.4 $ & (4) \\
HESS~J1303$-$631 &$16 \pm 11$ &$27 \pm 1$ &$0.16 \pm 0.03$ &$ 8.3 \pm 6.6 $ & (5) \\
HESS~J1356$-$645 &$16 \pm 10$ &$6.7 \pm 3.7$ &$0.06 \pm 0.01$ &$ 1.2 \pm 0.9 $ & (6) \\
HESS~J1418$-$609 &$<\,25 $ &$3.4 \pm 1.8$ &$3.1 \pm 0.2$ &$<\,0.8 $& (7,4,8) \\
HESS~J1420$-$607 &$23 \pm 9$ &$15 \pm 3$ &$1.3 \pm 0.3$ &$ 8.6 \pm 4.5 $ & (1,7,8) \\
HESS~J1458$-$608 &$<\,15 $ &$3.9 \pm 2.4$ & \nodata &$<\,2.8 $& (9) \\
HESS~J1514$-$591 &$46 \pm 13$ &$20 \pm 4$ &$29 \pm 6$ &$ 10 \pm 4 $ & (1) \\
HESS~J1554$-$550 &$<\,3 $ & $1.6 \pm 0.5$ &$3 \pm 1$ &$<\,2 $& (10,11) \\
HESS~J1616$-$508 &$46 \pm 14$ &$21 \pm 5$ &$4.2 \pm 0.8$ &$ 26 \pm 9 $ & (1) \\
HESS~J1632$-$478 &$79 \pm 19$ &$15 \pm 5$ &$0.43 \pm 0.08$ &$ 8.5 \pm 2.1 $ & (12) \\
HESS~J1640$-$465 &$30 \pm 11$ &$5.5 \pm 1.2$ &$0.5 \pm 0.1$ &$ 26 \pm 10 $ & (13) \\
HESS~J1646$-$458B &$<\,24 $ &$5 \pm 2$ & \nodata &$<\,10 $& (14) \\
HESS~J1702$-$420 &$<\,26 $ &$9 \pm 3$ &$0.01 \pm 0.00$ &$<\,7 $& (15) \\
HESS~J1708$-$443 &$29 \pm 13$ &$23 \pm 7$ & \nodata &$ 2 \pm 1 $ & (16) \\
HESS~J1718$-$385 &$<\,12 $ &$4 \pm 2$ &$0.14 \pm 0.03$ &$<\,3 $& (1) \\
HESS~J1804$-$216 &$74 \pm 21$ &$12 \pm 2$ &$0.07 \pm 0.01$ &$ 13 \pm 5 $ & (17) \\
HESS~J1809$-$193 &$<\,48 $ &$19 \pm 6$ &$0.23 \pm 0.05$ &$<\,7 $& (1) \\
HESS~J1813$-$178 &$<\,14$ &$5.0 \pm 0.6$ & \nodata & $<\,3.7$ & (18, 19)\\ 
HESS~J1818$-$154 &$<\,8.9 $ &$1.3 \pm 0.9$ & \nodata &$<\,6.5 $& (20) \\
HESS~J1825$-$137 &$59 \pm 77$ &$61 \pm 14$ &$0.4 \pm 0.1$ &$ 12 \pm 16 $ & (1) \\
HESS~J1831$-$098 &$<\,11 $ &$5.1 \pm 0.6$ & \nodata &$<\,2.2 $& (21) \\
HESS~J1833$-$105 &$<\,12 $ &$2.4 \pm 1.2$ &$40 \pm 0$ &$<\,3.2 $& (1) \\
HESS~J1837$-$069 &$70 \pm 23$ &$22 \pm 9$ &$0.6 \pm 0.2$ &$ 36 \pm 12 $ & (4) \\
HESS~J1841$-$055 &$89 \pm 20$ &$24 \pm 3$ & \nodata &$ 1.8 \pm 0.4 $ & (22) \\
HESS~J1846$-$029 & $<11$ & $9 \pm 2$ & $29 \pm 1$ & $<0.2$ & (4)\\
HESS~J1848$-$018 &$30 \pm 17$ &$4 \pm 1$ & \nodata & $13 \pm 7$ & (23) \\
HESS~J1849$-$000 &$<\,7 $ &$2.1 \pm 0.4$ &$0.9 \pm 0.2$ &$<\,4 $& (24) \\
HESS~J1857+026 &$58 \pm 10$ &$18 \pm 3$ & \nodata &$57 \pm 18$ & (22) \\
MGRO~J1908+06 &$<\,32 $ &$12 \pm 5$ & \nodata &$<\,4 $& (25) \\
HESS~J1912+101 &$<\,27 $ &$7 \pm 4$ & \nodata &$<\,6 $& (26) \\
VER~J1930+188 &$<\,5.5 $ &$2.3 \pm 1.3$ &$5.2 \pm 0.1$ &$<\,5.4 $& (4, 27) \\
VER~J1959+208 &$<\,1.9 $ & \nodata & \nodata &$<\,0.1 $&  (1)\\
MGRO~J2019+37 &$<\,27 $ & \nodata & \nodata &$<\,21 $& \nodata \\
MGRO~J2228+61 &$<\,12 $ & \nodata &$0.88 \pm 0.02$ &$<\,7.4 $& (4) \\
\enddata\\
\begin{flushleft}
a- This flux has been computed from the luminosity given in (15). Since no H.E.S.S. spectral index is available, we assumed a fiducial index of 2.4.
\end{flushleft}
\tablecomments{The flux of the PWNe candidates measured by the LAT (10-316 GeV, column 2), by VHE experiments (1-30 TeV, column 3), and in the 2-10 keV X-ray energy range. Column 5 is the luminosity computed assuming the pulsar distances from Table \ref{tab:table_luminosity}. \\References : (1) \cite{2009ApJ...694...12M} and references therein, (2) \citep{2011arXiv1111.2591M}, (3) \cite{2011AA...525A..46H}, (4) \cite{2010AIPC.1248...25K}, (5) \cite{dalton1303}, (6) \cite{2011AA...533A.103H}, (7) \cite{2012ApJ...750..162K}, (8) \cite{2006AA...456..245A}, (9) \cite{2012arXiv1205.0719D}, (10) \cite{2009ApJ...691..895T}, (11) \cite{2012arXiv1201.0481A},  (12) \cite{2010AA...520A.111B}, (13) \cite{2006ApJ...636..777A}, (14) \citep{2012AA...537A.114A}, (15) \citep{2006ApJ...636..777A}, (16) \cite{2011AA...528A.143H}, (17) \citep{2006ApJ...636..777A}, (18) \cite{2007AA...470..249F}, (19) \cite{2010ApJ...718..467F}, (20) \cite{2011arXiv1112.2901H}, (21) \citep{2011ICRC....7..243S}, (22) \cite{2008AA...477..353A}, (23) \cite{2008AIPC.1085..372C}, (24) \cite{2008AIPC.1085..312T}, (25) \citep{2009AA...499..723A}, (26) \cite{2008AA...484..435A} and (27) \cite{2010ApJ...719L..69A}.}
\end{deluxetable}
\normalsize
\noindent
\clearpage

Among the 58 sources analyzed, 36 sources have an associated pulsar. We selected the 14 sources flagged as PWN or PWNc in Table~\ref{tab:Spat_results} as they are likely due to IC emission. 

We searched for a correlation between the peak energy and the characteristic age of a pulsar by estimating the peak energy of the assumed IC component. To characterize the IC peak we used the log-normal representation,

\begin{equation}
\label{logp}
\frac{dN}{dE}=N_0 \times \left(\frac{E}{E_0}\right)^{-\left[ \alpha + \beta \times \log_{10}\left(\frac{E}{E_0}\right) \right]}. 
\end{equation}

\begin{figure}[h!]
\centering
\includegraphics[width=0.5\textwidth]{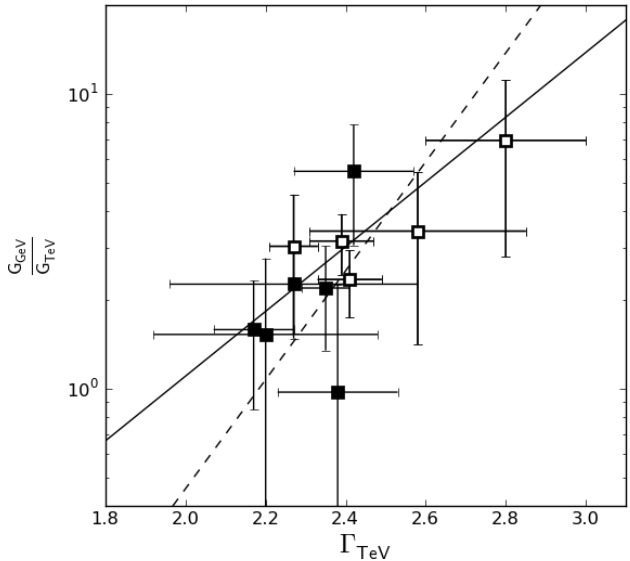}
\caption{Ratio of the LAT to the VHE energy flux as a function of the VHE spectral index for sources classified as ``PWN" or ``PWNc". The LAT flux is in the 10 GeV to 316 GeV energy range and the VHE flux is in the 1 TeV to 30 TeV energy range. The fluxes were obtained using power-law fits in each energy range. Sources classified in the TeVCat as ``PWN" are represented by full markers and sources classified as ``UNIDs" by hollow markers. The LAT fluxes were obtained subtracting any potential pulsar emission. The black line corresponds to a fit of a linear function to the points : $(1.10 \pm 0.48) \times \Gamma_{TeV} - (2.15 \pm 1.14)$ with $\chi^2/d.o.f = 11.2/9$. The dashed line corresponds to the correlation expected for a log-normal representation of $\beta=0.2$ and $E_0=300\,$GeV. HESS~J1303$-$631 and HESS~J1632$-$478 are not included in this figure, because their energy fluxes are known to be contaminated by neighboring sources in the LAT energy range (see Section~4). HESS~J1119$-$614 is not plotted since its TeV spectral index is not known.
\label{fig:EpeakETeV}}
\end{figure}

First, we checked if a log-normal spectral model was a possible representation of the spectrum of the sources observed at LAT and VHE energies. If the data were well-reproduced by a log-normal representation, then there should be a correlation between the energy-flux ratio and the VHE experiment spectral index. We fixed $E_0$ at 300\,GeV and $\beta$ at 0.2 (since $\beta$ corresponds to the typical curvature seen in the spectra) and fit the prefactor $N_0$ and the index $\alpha$ using our LAT and VHE spectral points. Figure~\ref{fig:EpeakETeV} plots the LAT to VHE energy flux ratios measured assuming a power-law spectrum in each energy range as a function of their spectral indices measured by VHE experiments and the relation expected for a log-normal model. To derive this relation, we randomly generated a thousand uniformly distributed sets of $\alpha$ providing a thousand different log-normal representations. For each log-normal model we derived the associated spectral points in $dN/dE$, assuming Poisson statistics with zero background counts, and used them to obtain the corresponding power-law index $\Gamma_{\text{TeV}}$ between 1 and 30 TeV. Then the log-normal model is used to compute the energy flux ratio and derive a relation between $\Gamma_{\text{TeV}}$ and this flux ratio. HESS~J1303$-$631 and HESS~J1632$-$478 are known to be contaminated by neighboring sources and were therefore removed from this figure. The energy flux ratio and the VHE index are correlated with a correlation coefficient of $+0.72\pm0.11$. Therefore, the log-normal model seems to be acceptable to reproduce the spectra between 10\,GeV and 30\,TeV.

\clearpage
 
We removed HESS~J1119$-$614 from the sample of sources since only the integrated flux above 1 TeV is available for this source and we required at least 3 spectral points to perform a fit. Then we defined the energy of the peak ($E_{peak}$) as the energy at which the modeled SED in $\nu F_{\nu}$ is maximal. This also corresponds to

\begin{equation}
\label{pos_epeak}
\alpha + 2\,\beta \times \log_{10}\left(\frac{E_{\text{peak}}}{E_0}\right) = 2.
\end{equation}

Since LAT spectral points typically suffer larger uncertainty and are less numerous than the VHE ones, this method is biased by the greater weight given to the VHE side of the log-normal representation. The fit results as well as the peak position are presented in Table~\ref{tab:Epeak}. 

\begin{deluxetable}{lcccl}
\tabletypesize{\scriptsize}
\tablecaption{IC peak fit results
\tabletypesize{\scriptsize}
\label{tab:Epeak}}
\tablewidth{0pt}
\tablehead{\colhead{Name} & \colhead{$\Gamma_{TeV}$} & \colhead{$\alpha$} & \colhead{$\text{E}_{\text{peak}}$} &  \colhead{$\chi^2$/d.o.f}\\
 \colhead{} & \colhead{} & \colhead{} & \colhead{(GeV)} & \colhead{}}
\startdata
HESS~J1023$-$575 & 2.58 $\pm$ 0.27 & $2.17 \pm 0.07$ & $113\pm45$ & 9.3/6\\
HESS~J1303$-$631\tablenotemark{a} & 2.44 $\pm$ 0.03 & $1.83\pm0.10$ & $798\pm459$ & 8.2/9 \\
HESS~J1356$-$645 & 2.20 $\pm$ 0.28 & $1.84\pm0.10$ & $754\pm434$ & 7.7/6\\
HESS~J1420$-$607 & 2.17 $\pm$ 0.10 & $1.88\pm0.06$ & $599\pm206$ & 8.2/10\\
HESS~J1514$-$591 & 2.27 $\pm$ 0.31 & $1.98\pm0.04$ & $337\pm178$ & 12.8/15\\
HESS~J1616$-$508 & 2.35 $\pm$ 0.06 & $2.11\pm0.06$ & $159\pm55$ & 4.4/7\\
HESS~J1632$-$478\tablenotemark{a} & 2.12 $\pm$ 0.20 & $2.18\pm0.12$ & $106\pm74$ & 11.6/5 \\
HESS~J1640$-$465 & $2.42 \pm 0.15$ & $2.27\pm0.08$ & $63\pm29$ & 4.9/6\\
HESS~J1825$-$137 & 2.45 $\pm$ 0.30 & $2.06\pm0.02$ & $212\pm24$ & 10.2/12\\
HESS~J1837$-$069 & $2.27 \pm 0.06$ & $2.03\pm0.08$ & $252\pm116$ & 11.5/13\\
HESS~J1841$-$055 & 2.41 $\pm$ 0.08 & $2.00\pm0.06$ & $300\pm103$ & 10.5/9\\
HESS~J1848$-$018 & $2.8 \pm 0.20$ & $2.23\pm0.15$ & $204\pm112$ & 6.9/8\\
HESS~J1857+026 & 2.39 $\pm$ 0.05 & $2.07\pm0.06$ & $201\pm69$ & 9.4/12\\
\enddata

\begin{flushleft}
a- Since these sources are known to be contaminated by neighboring sources the corresponding rows were not used in Section~5.
\end{flushleft}
\tablecomments{The fit parameters of the log-normal fit to the LAT and VHE data. Column 2 gives the spectral index of the power law fit by VHE experiments. Columns 3 and 4 give the index $\alpha$ at $E_0=300$ GeV and the peak position of the log-normal model as defined in Section~\ref{discussion}. Column 6 gives the reduced $\chi^2$ of the fit. The log-normal model was fit to the VHE and LAT SED with the pulsar's contribution subtracted.}
\end{deluxetable}
\normalsize
\noindent

\begin{figure}[h!]
\centering
\includegraphics[width=0.5\textwidth]{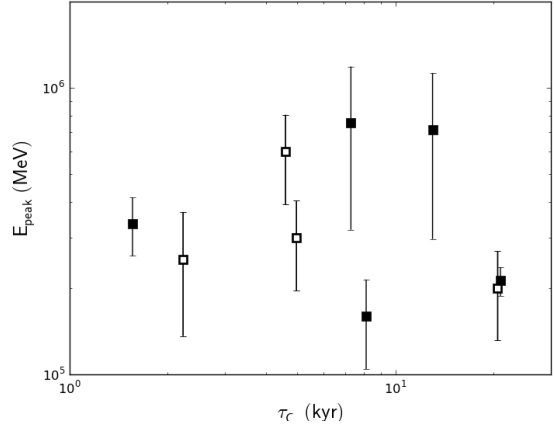}
\caption{Energy of the maximum of the IC peak as a function of the characteristic age of the pulsar for sources labelled as PWN or PWNc in Table~\ref{tab:Spat_results}. Full markers represent sources with a clear PWN association at VHE while hollow markers correspond to sources for which the association is less clear. The fitted SED corresponds to spectra in which contributions of pulsars summarized in Table \ref{tab:pulsars} are subtracted. HESS~J1640$-$465 and HESS~J1848$-$018 are not plotted since the age of their putative pulsar is not known. As in Figure~\ref{fig:EpeakETeV}, HESS~J1303$-$631, HESS~J1632$-$478 and HESS~J1119$-$614 are not plotted.
\label{fig:Epeakage}}
\end{figure}

We plotted the resulting $E_{\text{peak}}$ as a function of the age in Figure~\ref{fig:Epeakage}. These values of $E_{\text{peak}}$ and the age yielded a correlation coefficient of $-0.06 \pm 0.28$. Furthermore, a fit of the value assuming a linear function yielded $\log_{10}(E_{\text{peak}}/1\,\text{GeV})= (-0.13\,\pm\,0.09) \times \left(\log_{10}(\tau_C/1\,\text{kyr}) - \overline{\log_{10}(\tau_C/1 \text{kyr})} \right) + (2.41\,\pm\,0.04)$ with $\chi^2/dof=15.9/7$, while a simple model assuming a constant yielded $\log_{10}(E_{\text{peak}}/1\,\text{GeV})= 2.38\,\pm\,0.04$ with $\chi^2/dof=16.9/8$. This means that the linear function improves the fit at only the $1\,\sigma$ level. This coefficient and this comparison show that, contrary to the expectation \citep{2012arXiv1202.1455M}, no downward correlation is observed. However, this correlation might be obscured by the usage of the characteristic age of the pulsar which may not be a good age estimator for the PWN. For instance, MSH~15$-$52 is a known case for which two ages are proposed, either the characteristic age of the pulsar, 1.7\,kyr or an age between 20 and 40\,kyr as suggested by the size and general appearance of the SNR \citep{2001AA...374..259G}. 

\begin{figure}[h!]
\centering
\includegraphics[width=0.6\textwidth]{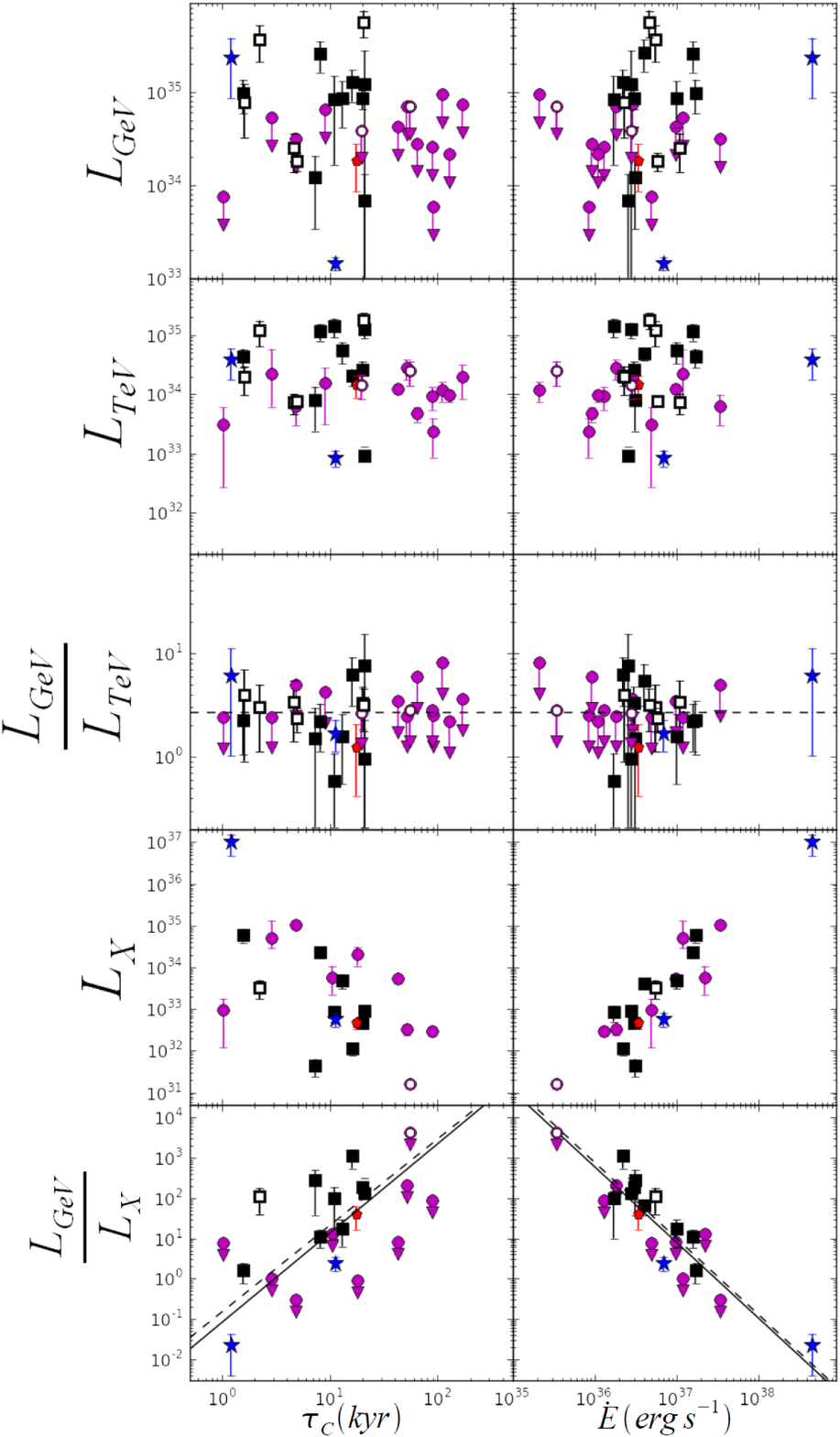}
\caption{From top to bottom: LAT luminosity, VHE luminosity, ratio of LAT to VHE luminosity, X-ray luminosity, and the ratio of LAT to X-ray luminosity as a function of the pulsar characteristic age (left) and the pulsar spin-down power (right). The X-ray, LAT, and VHE fluxes are integrated in the 2 keV to 10 keV, 10 GeV to 316 GeV, 10 TeV to 30 TeV energy ranges respectively.  Full markers correspond to sources with a clear PWN association at VHE energies while hollow markers correspond to sources for which the association is less clear. The black squares (\protect\includegraphics[scale=0.2]{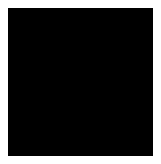}) represent the sources detected at LAT energies, the magenta circles (\protect\includegraphics[scale=0.2]{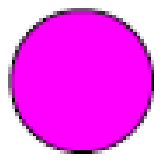}) show the upper limits, the red pentagon (\protect\includegraphics[scale=0.2]{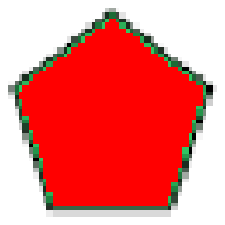}) is HESS~J1708$-$443 showing pulsar behavior in the LAT energy range and the blue stars (\protect\includegraphics[scale=0.2]{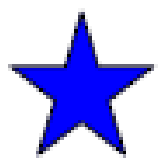}) represent the Crab nebula and Vela-X not studied in this work. The LAT luminosity corresponds to Table \ref{tab:table_luminosity2} and was computed after removing any potential emission from LAT-detected pulsars. The dashed and solid lines are explained in Section~\ref{discussion}.
\label{fig:rapportTeV}}
\end{figure}

Using the log-normal fit presented in Table~\ref{tab:Epeak} we derived the mean parameters $\bar{\alpha} = 2.06 \pm 0.02$ and $10^{\overline{\log_{10}(E_{\text{peak}})}}= 215 ^{+ 25}_{-23}\,\text{GeV}$. To find a putative correlation between luminosity ratio and the pulsar characteristics, we plot the GeV to TeV luminosity ratio as a function of the pulsar's characteristic age and spin-down energy in Figure~\ref{fig:rapportTeV}. To determine if the correlation of the luminosity ratio with the characteristic age and the pulsar spin-down power is due to one of the components, this figure also presents luminosities derived in the GeV, TeV and keV domains. We overlaid the mean ratio, $\bar{R}=10^{\overline{\log_{10}(R)}}=2.7^{+2.7}_{-1.4}$, found between the LAT and VHE energy ranges using the 14 sources flagged as PWNe and PWNc in Table~\ref{tab:Spat_results}. We included in this figure and the figures that follow all sources analyzed in this work including pulsar-like sources, providing an upper-limit on the PWN emission and sources flagged as ``O". This figure shows that no source is located at more than 2$\sigma$ from this mean ratio except HESS~J1804$-$216, which is not a PWN candidate as explained in section 4.3.  

\begin{figure}[h!]
\centering
\includegraphics[width=0.5\textwidth]{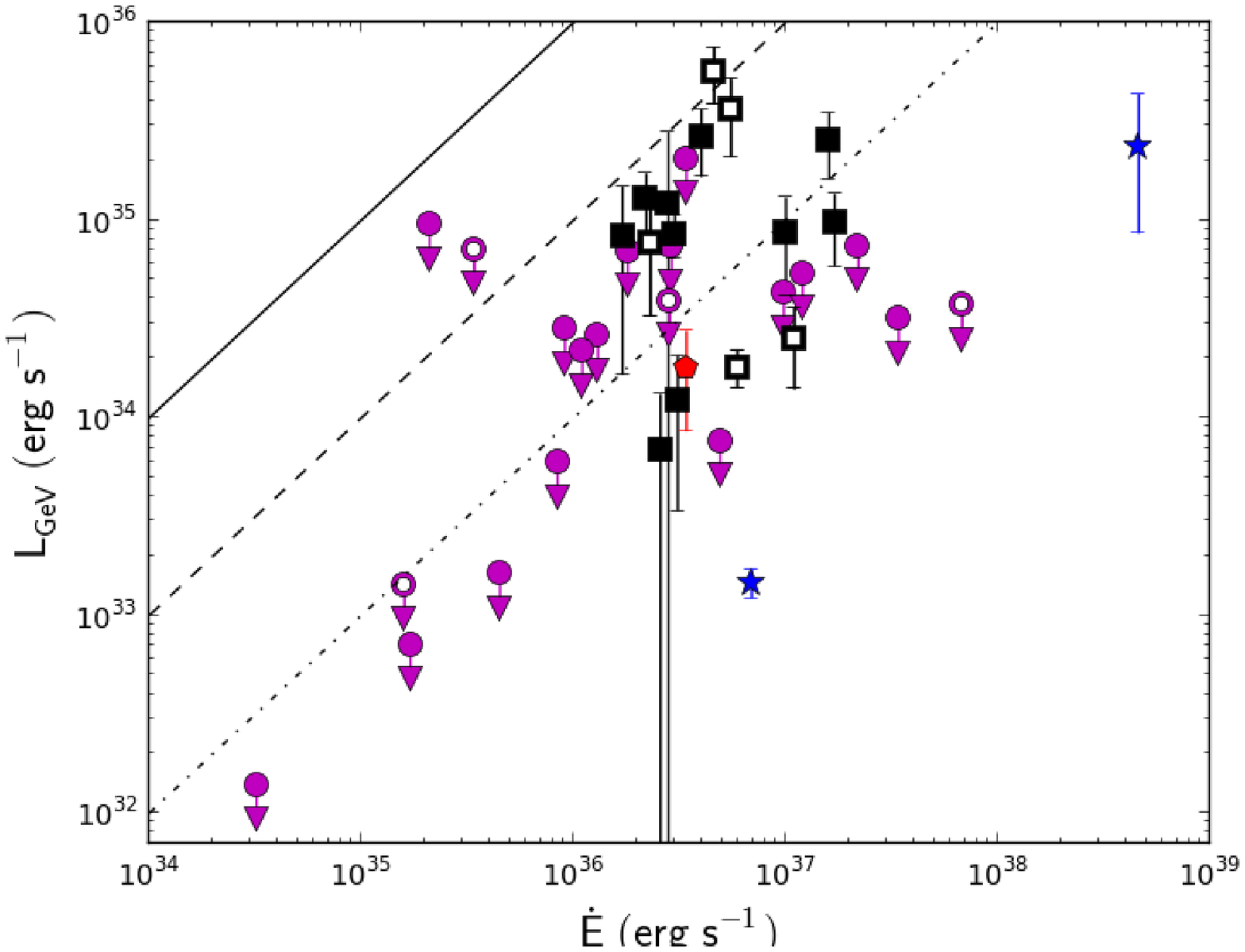}
\caption{LAT luminosity as a function of the pulsar's spin down power. The symbols are defined in Figure~\ref{fig:rapportTeV}. The LAT luminosity was computed after removing any potential emission from LAT-detected pulsars. The upper limit on HESS~J1818$-$154 is excluded from the plot because of its low value of $\dot{E}$.
\label{fig:dotelpwn}}
\end{figure}

As a second step, we investigated the relation between the spectra in the X-ray energy range and the LAT energy range. \cite{2009ApJ...694...12M} have shown a correlation between the X-ray flux of a PWN and the physical properties of the pulsar assuming the emission of the PWN to be composed of synchrotron emission in the keV domain. As described in \cite{2009ApJ...694...12M}, the $\gamma$-ray and the X-ray luminosities of PWNe are expected to decrease with time and the cooling time of electrons emitting synchrotron X-rays is smaller than the cooling time of the electrons emitting IC $\gamma$-rays. Therefore, the X-ray emission traces the recent evolution of a PWN while the $\gamma$-ray photons trace a longer period. Since the pulsar spin-down power driving the electron injection decreases with time, the ratio of the IC emission to the synchrotron emission is expected to increase over time.

Figure~\ref{fig:rapportTeV} and Table~\ref{tab:table_luminosity2} give the ratio of the $\gamma$-ray luminosity to the X-ray luminosity as a function of the age and as a function of the pulsar spin-down power. As reported for VHE experiments \citep{2009ApJ...694...12M}, no clear correlation is found between the $\gamma$-ray flux and spin-down power or between the $\gamma$-ray flux and the characteristic age. This can also be seen on Figure~\ref{fig:dotelpwn}. On the other hand, there is a correlation between the X-ray flux of the observed PWNe and the pulsars spin-down power and characteristic age. This causes the correlation between the ratio of the $\gamma$-ray flux to the X-ray flux and the pulsar properties. We also represented the relations derived in \cite{2009ApJ...694...12M} multiplied by $\bar{R}$ for the whole sample of sources and for the sources clearly identified with PWNe. The overall agreement with these relations is relatively good. 

In the plot showing the LAT to X-ray energy flux ratio as a function of the pulsar characteristic age, four upper limits are well below the correlation relations derived by \cite{2009ApJ...694...12M}. These outliers are, in increasing spin down power, HESS~J1833$-$105, HESS~J1554$-$550, HESS~J1849$-$000 and HESS~J1718$-$385. HESS~J1833$-$105 and HESS~J1718$-$385 are also included in \cite{2009ApJ...694...12M}, unlike HESS~J1554$-$550 and HESS~J1849$-$000. Table~\ref{tab:table_luminosity2} shows that the GeV over TeV upper limit is in each case larger than $\bar{R}$. Therefore, their small LAT/X-ray luminosity ratios do not come from an especially small LAT flux but instead from an especially large X-ray flux relative to other sources with a similar characteristic age.

These figures show that the \emph{Fermi}-LAT mainly detects young and middle-aged PWNe (1-30\,kyr) around energetic pulsars with spin-down powers between 10$^{36}$ and 10$^{39}$\,erg\,s$^{-1}$. One can see from these figures that the VHE sources older than 30\,kyr are not detected by the LAT, while \cite{2012arXiv1202.1455M} predict a higher flux in the LAT energy range than in the VHE experiments energy range. However it should be noted that among these nine VHE sources, at least 7 suffer one or several biases:
\begin{itemize}
\item Source misidentification: for example, HESS~J1912+101 has recently been proposed to be a shell-type SNR\footnote{This was proposed in a presentation at the AstroParticules et Cosmologie laboratory: \url{http://www.apc.univ-paris7.fr/$\sim$semikoz/CosmicRays/CosmicRays/Dec14/djannati-atai.pdf}}, but some of the emission could originate in the pulsar wind. Similarly, HESS~J1702$-$420 and HESS~J1646$-$458B cannot be clearly associated as PWNe and deeper multi-wavelength observations are needed to classify the emission.
\item Potential characteristic age overestimation. See, for example, HESS~J1809$-$193 \citep{2011ICRC....7..243S}.
\item HESS~J1026$-$582,  HESS~J1809$-$193, HESS~J1831$-$098 and HESS~J1849$-$000 are located in regions of strong diffuse emission where the LAT is less sensitive to any potential emission. 
\end{itemize}
Therefore, the lack of a population of old $\gamma$-ray-loud PWNe cannot be proven given the current statistics and systematics associated with this analysis. 

Finally, Figure~\ref{fig:dotelpwn} shows the $\gamma$-ray luminosity of the PWN as a function of the pulsar spin-down power assuming the distances listed in Table~\ref{tab:table_luminosity}. The $\gamma$-ray luminosity is, for all sources, computed by subtracting any potential emission from nearby LAT-detected pulsars. The distance uncertainties are calculated assuming a 20\% uncertainty on the dispersion measure (2PC). For most pulsars this yields a realistic uncertainty, for many however it leads to a large factor, for instance a factor of $6$ in \cite{2009ApJ...700.1059A}, between two distance estimates. Among the detected sources, eight show a $\gamma$-ray efficiency below 1\% and five are consistent within uncertainties of having an efficiency between 1 and 10\%. Six upper limits on the LAT luminosity of VHE sources associated to a PWN powered by an energetic pulsar ($\dot{E}\geq 10^{36}\,\text{erg}\,\text{s}^{-1}$) are well below an efficiency of 1\%. They are HESS~J1418$-$09, HESS~1813$-$178, HESS~J1833$-$105, HESS~J1849$-$000, MGRO~J2228$+$61 and VER~J1930$+$188.

\section{CONCLUSION}

We looked for LAT counterparts to the VHE sources potentially associated to PWNe using 45 months of \emph{Fermi}-LAT observations. We detected 30 of the 58 sources at LAT energies:

\begin{itemize}
\item{9 may be due to pulsar emission for energies above 10 GeV,}
\item{7 sources cannot be clearly associated to a PWN,}
\item{11 are PWNe candidates,}
\item{3 are clearly identified as PWNe: HESS~J1825$-$137 and HESS~J1514$-$591 (aka MSH~15$-$52) already detected and HESS~J1356$-$645 detected for the first time in this analysis.}
 \end{itemize}

Among these 30 sources, 23 were also detected in 1FHL and 15 in \cite{2012PhRvD..85h3008N}. We analyzed their morphology and found that 8 of them are significantly extended.

Adding the Crab Nebula and Vela$-$X to the 3 clearly identified PWNe, 5 PWNe are now detected at LAT energies. 11 sources are promising PWNe candidates such as HESS~J1420$-$607 and HESS~J1119$-$614. These 16 sources are associated with young (with an age between 1 and 30\,kyr) and powerful pulsars with a spin down power between 10$^{36}$ and 10$^{39}$\,erg\,s$^{-1}$ and typically have a conversion efficiency below 10\%. No correlation has been found between the LAT energy flux and the pulsar characteristic age. This work has not shown any evidence for a shift of the PWN IC emission towards the LAT energy range as a function of the characteristic age. However this can be due to large uncertainties on the systems' age. The correlation of the LAT to X-ray luminosities ratio with the pulsar characteristic age and its spin-down power is consistent with the work of \cite{2009ApJ...694...12M} derived for the VHE to X-ray luminosities ratio.

The \textit{Fermi} LAT Collaboration acknowledges generous ongoing support from a number of agencies and institutes that have supported both the development and the operation of the LAT as well as scientific data analysis. These include the National Aeronautics and Space Administration and the Department of Energy in the United States, the Commissariat \`a l'Energie Atomique and the Centre National de la Recherche Scientifique / Institut National de Physique Nucl\'eaire et de Physique des Particules in France, the Agenzia Spaziale Italiana and the Istituto Nazionale di Fisica Nucleare in Italy, the Ministry of Education, Culture, Sports, Science and Technology (MEXT), High Energy Accelerator Research Organization (KEK) and Japan Aerospace Exploration Agency (JAXA) in Japan, and the K.~A.~Wallenberg Foundation, the Swedish Research Council and the Swedish National Space Board in Sweden.

Additional support for science analysis during the operations phase is gratefully acknowledged from the Istituto Nazionale di Astrofisica in Italy and the Centre National d'\'Etudes Spatiales in France.

This research made use of pywcsgrid2, an open-source plotting package for Python\footnote{pywcsgrid2 can be obtained from: http://leejjoon.github.com/pywcsgrid2/.}. The authors acknowledge the use of the TeV catalog website\footnote{The TeV catalog can be obtained from: http://tevcat.uchicago.edu/} provided by the University of Chicago. The authors acknowledge the use of the \emph{Australia Telescope National Facility} pulsar catalog\footnote{The ATNF catalog can be obtained from: http://www.atnf.csiro.au/people/pulsar/psrcat/}.

\clearpage
\newpage
\bibliography{LAT_TeV_PWNe}


\clearpage
\appendix
\renewcommand{\thefigure}{A-\arabic{figure}}
\setcounter{figure}{0}
\appendix

\end{document}